\documentclass[reprint,aps,nofootinbib,superscriptaddress]{revtex4-2}

\usepackage{amsmath,amsthm,amssymb}
\usepackage{graphicx}
\usepackage{dcolumn}
\usepackage{bm}
\usepackage{braket}
\usepackage{mathtools}
\usepackage{hyperref}
\usepackage{suffix}
\usepackage{mathtools}

\DeclarePairedDelimiter\floor{\lfloor}{\rfloor} 

\newcommand{\e}{{\textrm e}} 

\DeclarePairedDelimiterX\MeijerM[3]{\lparen}{\rparen}%
{\begin{smallmatrix}#1 \\ #2\end{smallmatrix}\delimsize\vert\,#3}
\newcommand\MeijerG[8][]{%
	G^{\,#2,#3}_{#4,#5}\MeijerM[#1]{#6}{#7}{#8}}
\WithSuffix\newcommand\MeijerG*[7]{%
	G^{\,#1,#2}_{#3,#4}\MeijerM*{#5}{#6}{#7}}
\hypersetup{
    colorlinks = true,
    citecolor= black,
    linkcolor=black,
    urlcolor=cyan,
}

\begin{document}

    \author{Lucas Varela}
    \affiliation{Universidad de los Andes, Bogot\'a, Colombia}
    \affiliation{Universit\'e Paris-Saclay, CNRS, LPTMS, 91405, Orsay, France}	
    
    \author{Sergio Andraus}
    \affiliation{Graduate School of Physics, The University of Tokyo, 
    Tokyo 113-0033, Japan}
    
    \author{Emmanuel Trizac}
    \affiliation{Universit\'e Paris-Saclay, CNRS, LPTMS, 91405, Orsay, France}

    \author{Gabriel T\'ellez}
    \affiliation{Universidad de los Andes, Bogot\'a, Colombia}

	\title{Relaxation dynamics of two interacting electrical double-layers in a 1D Coulomb system}

	\begin{abstract}
	  We consider an out-of-equilibrium one-dimensional model for two electrical double-layers. With a combination of exact calculations and Brownian Dynamics simulations, we compute the relaxation time ($\tau$) for an electroneutral salt-free suspension, made up of two fixed colloids, with $N$ neutralizing mobile counterions. For $N$ odd, the two double-layers never decouple, irrespective of their separation $L$; this is the regime of like-charge attraction, where $\tau$ exhibits a diffusive scaling in $L^2$ for large $L$.
	  On the other hand, for even $N$, $L$ no longer is the relevant length scale for setting the relaxation time; this role is played by the Bjerrum length. This leads to distinctly different dynamics: for $N$ even, thermal effects are detrimental to relaxation, increasing $\tau$, while they accelerate
	  relaxation for $N$ odd.
	  Finally, we also show that the mean-field theory is recovered for large $N$ and moreover, that it remains an operational treatment down to relatively small values of $N$ ($N>3$).     
	\end{abstract}

	\maketitle
	\makeatletter
	\let\toc@pre\relax
	\let\toc@post\relax
	\makeatother 
	
\section{Introduction}
The contact between a charged surface or colloid particle with an ionic solution is known to create an electrical double-layer: a diffuse counterion cloud around the charged interface which extends a distance typically in the colloidal range, from molecular dimensions to the micrometer scale \cite{hunterbook,Levin2002}. The study of electric double-layers is an active research topic due to their importance to understand phenomena such as like-charge attraction \cite{Jonsson,Levin05}, ion transport in biological membranes \cite{Caffrey} or nanofluidics \cite{SchochHanRenaud,Jubin2018,Kavokine2019}. Furthermore, they play a significant role in the design and development of bio-chemical sensors \cite{Drummond2003}, super capacitors \cite{BurtRyanZhao,Simon2008,Sharma2010,Merlet2012}, electric-double-layer-gated transistors \cite{Xu2020}, water treatment \cite{AndersonCuderoPalma}, etc.

Electrical double-layers have been studied extensively \cite{hunterbook,Grahame1947,VerweyOverbeek,Ohshima}. 
They were first introduced in 1897 by Helmholtz while investigating electrodes in electrolytes subject to an external potential \cite{Helmholtz}. In this model, the counterions form a single layer close to the electrode. 
Then, Gouy \cite{Gouy} and Chapman \cite{Chapman} introduced the concept of a diffuse layer, where the counterion typical position results from the competition between entropic and electrostatic effects. 
This model was further improved by Stern \cite{Stern}, by assuming that the electrolyte system is made up of two parts: first comes a layer that is strongly bound to the electrode surface (practically immobile) and then follows the diffuse part, where counterions are loosely bound. 
The study of electrical double-layers displays numerous approaches that quickly increase in difficulty as the model restrictions are relaxed and ingredients are added. 
The Gouy-Chapman model resorts to a mean-field treatment that assumes weak electrostatic interactions and considers the ionic fluid as a continuum, discarding discreteness effects. The latter shortcoming can be partly addressed including ionic size effects within a mean-field framework   \cite{Bikerman,Freise,Kralj,Borukhov,Kornyshev2007} (see \cite{FrLe12} for a criticism on these approaches, and \cite{Huang2016} for a review on mean-field electric double-layers). Using mean-field models is convenient since they may yield analytic expressions. Interestingly, while mean-field techniques fail when the electrostatic coupling increases \cite{Guldbrand1984,Allahyarov1999,Moreira2002MC,MoreiraNetz2001,Boroudjerdi2005,Naji2005}, strongly-coupled systems lend themselves to analytical progress \cite{SaTr11PRL,SaTT18}. 
The remaining intermediate regime between weak and strong coupling is mostly accessible through numerics. So far, equilibrium properties mostly have been studied, and results are scarce for time dependent phenomena. The physics of out-of-equilibrium electrical double-layers is primarily described via mean-field and numerical approaches \cite{hunterbook,Golestanian2000,Bazant2004,Morrow2006,AlexeIonescu2006,Lim2007,Hojgaard2010,JanssenBier2018,Palaia2020,telles2021}, with few beyond mean-field contributions \cite{Bazant2009,Storey2012,Palaia2020}. 

\begin{figure}[htp]
	\begin{center}
	\includegraphics[width=0.48\textwidth]{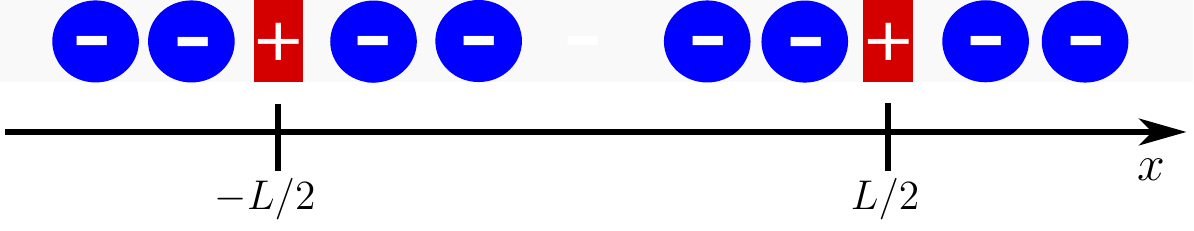}%
		\caption{Sketch of an electroneutral system consisting of two interacting electric double-layers, each made of a colloid (rectangle) and $4$ counterions (circles). The distance between colloids is $L$. The interaction between particles is mediated by the 1D Coulomb potential, linear in separation. The dynamics of the counterions is considered and the colloids are treated as static since their time scale is assumed to be much larger than that of the counterions. All particles are point-like and they can ``cross over'' each other.}
		\label{fig:intro_sketch}
	\end{center}
\end{figure}

Herein are reported exact and numerical results, within all coupling regimes, for the relaxation time towards equilibrium of two interacting double-layers in one dimension, at a distance $L$. Such systems have been studied at equilibrium \cite{Dean1998,dean,TellezTrizac2015,vtt2016}. We consider the dynamics of an electroneutral system made of two symmetrically-placed, permeable like-charged colloids and $N$ counterions as shown in Fig.~\ref{fig:intro_sketch}. Due to their larger mass, the colloids are assumed to exhibit a very large time scale for displacement compared to the counterions, hence we consider them to be fixed, and address the dynamics of the diffuse layer. The single counterion case ($N=1$) allows an analytical solution for the particle density and relaxation time. This subsumes the essential features of the dynamics for any odd number of counterions. Indeed, the parity of $N$ plays an important role, for both static and dynamics properties: if $N$ is even, the large $L$ regime features decoupled neutral entities, with $N/2$ counterions neutralizing each colloid. This is no longer possible for $N$ odd, for which there is always a misfit counterion \cite{TellezTrizac2015,Trizac2018}, that plays an important role in what follows. The misfit effect is accurately described by the mean passage of a free diffusing particle in a reduced length approximately given by the colloid separation minus the space taken by the double-layers between the colloids. For even $N$, the two double-layers form and completely neutralize each colloid. The decoupling between the two moieties, made possible for even $N$, explains why the relaxation process becomes $L$ independent for large $L$, in contrast with the odd case. 

It is relevant to point out that one-dimensional approaches may be insightful for more realistic systems, where theoretical results are scarce and simulations are computationally demanding. This is the case of like-charge attraction between two equally charged plates with counterions only \cite{Trizac2018}, where counterions interact pairwise by the 3D Coulomb potential in $1/r$. This phenomenon is exactly described in the strong coupling regime by a 1D system of two colloids and a single counterion 
\cite{MoreiraNetz2001,Netz2001,SaTr11PRL}. This regime arises when the distance between the plates is smaller than the typical counterion distance. 

The paper is structured as follows. The time evolution of the probability density for one counterion ($N=1$) is computed analytically in section \ref{sec:1_counterion}, parameterized by the colloid distance $L$. The density displays an exponential decay towards equilibrium, which naturally introduces a relaxation time. Then, a scheme is introduced to determine this quantity, from a simulation based on the corresponding Langevin dynamics. The results are in good agreement with exact values. 
In section  \ref{sec:Nrelaxation}, the numerical study of the relaxation time is generalized for the many-counterion case, $N>1$. The role of the parity of $N$ and the symmetry of the initial condition is discussed, with the former leading to two different behaviors for the relaxation time.
In Section~\ref{sec:meanfield}, we study the analytic mean-field treatment of a system at zero colloid separation. We find that this solution serves as a good approximation for the discrete charge system, for a large number of counterions as expected, but more surprisingly already for $N$ as small as 3. 
Treating the counterions as discrete charges results in exponential relaxation dynamics. This is in contrast to mean-field theory, that features a slower, algebraic decay \cite{hunterbook}. We explain below how the two regimes are matched, and show explicitly how mean-field's accuracy improves, upon increasing 
the number of counter-ions $N$.
Finally, we investigate numerically in section \ref{sec:middle_counterion_trans} the first passage time of the middle/misfit counterion between the two double-layers when $N$ is odd, to rationalize the behaviour of the relaxation time. We conclude in 
section \ref{sec:conclusions}.

\section{One counterion \texorpdfstring{$(N=1)$}{} } \label{sec:1_counterion}

In this section, we consider an electroneutral system made of a single counter-ion ($N=1$), of charge $e$ and two colloids, each with charge $-e/2$. The colloids have fixed positions:  $-L/2$ and $L/2$. These charges interact via the 1D Coulomb potential energy, which for two charges $q_1$ and $q_2$ at $x_1$ and $x_2$ is given by 
\begin{equation}
    V_{1\text{D}}(x_1,x_2) = -\frac{q_1q_2}{\epsilon}|x_1-x_2| ,
\end{equation}
where $\epsilon$ is the dielectric constant of the medium.
The counterion's position $x$ is ruled by the following over-damped Langevin equation:
\begin{equation}
	m \gamma \frac{dx}{dt} = -\frac{d\Phi(x)}{dx}+ \sqrt{ m \gamma k_{\text{B}} T_{\text{bath}}}\,\xi(t),
	\label{1c_lang}
\end{equation}
where $\gamma$ is the damping coefficient, $T_{\text{bath}}$ the temperature, $m$ the counterion mass, and ${\Phi}({x}) =V_{1\text{D}}(x,L/2) + V_{1\text{D}}(x,-L/2)$ is the electrostatic potential energy due to the colloids. The stochastic Langevin force  $\xi({t})$ is a Gaussian white noise characterized by zero mean $\braket{\xi(t)} = 0 $, and delta time correlation $\braket{\xi({t}_1)\xi({t}_2)} =  2\delta({t}_1-{t}_2)$. 

There is an important length scale related to the electrostatic interaction, namely the Bjerrum length, which in this one-dimensional context is defined as $l_{\text{B}} = k_{\text{B}} T_{\text{bath}} \epsilon/e^2$. When two $e$-charges are pushed closer to one another, an energy budget of $k_{\text{B}} T_{\text{bath}}$ corresponds to a relative displacement over a distance $l_{\text{B}}$.
Note that $l_{\text{B}}$ is exactly the inverse of its three dimensional counterpart: this stems from the fact that the 1D Coulomb potential is linear in distance between charges, while it goes like the inverse distance in 3D. 
Associated to the diffusive dynamics we consider here, we can define a characteristic time scale $\tau_{\text{B}}= l_{\text{B}}^2 / D$ where $D=k_{\text{B}} T_{\text{bath}}/(\gamma m)$ is the diffusion coefficient. In the following, it will prove useful to work with the rescaled units: $\widetilde{x} = {x}/l_{\text{B}}$, $ \widetilde{t} ={t}/\tau_{\text{B}}$ and $\widetilde{\Phi} = \Phi /(k_{\text{B}} T_{\text{bath}})$;
the length and time scales are 
such that the diffusion coefficient is set to unity in the dimensionless units.

Since the dynamics are given by a Markov process, the time evolution for the probability density function starting at any given time $t_0$ can be determined with the transition probability $p(x,t|x_0,t_0)$, where $p(x,t_0|x_0,t_0) = \delta(x-x_0)$, without knowledge of the preceding time evolution. In the following we will set $t_0=0$ and avoid writing it explicitly, $p(x,t|x_0) \equiv p(x,t|x_0,0) $.
Equation \eqref{1c_lang} has an associated Fokker-Planck equation that governs the transition probability $p$ (also known as the propagator, which is nothing but the density of ions):
\begin{equation}
	\frac{\partial p(\widetilde{x},\widetilde{t})}{\partial \widetilde{t}} =  \frac{\partial }{\partial \widetilde{x}} \Big(
	p(\widetilde{x},\widetilde{t})
	\frac{\partial\widetilde{\Phi}(\widetilde{x},\widetilde{L})}{\partial \widetilde{x}}\Big)+\frac{\partial^2 p(\widetilde{x},\widetilde{t})}{\partial \widetilde{x}^2} ,
	\label{1c_fp}
\end{equation}
where the rescaled 1D Coulomb potential $\widetilde{\Phi}$ is given by
\begin{equation}
	\widetilde{\Phi}(\widetilde{x},\widetilde{L}) = \begin{cases}
	\widetilde{L}/2 & \text{if }|\widetilde{x}|<\widetilde{L}/2\\
	|\widetilde{x}| & \text{if } |\widetilde{x}|>\widetilde{L}/2
	\end{cases}.
\end{equation}
Note that when $\widetilde{L} =0$, Eq.~\eqref{1c_fp} is formally the same Fokker-Planck equation that describes a Brownian motion with dry friction~\cite{Gennes2005, dryfriction}. We review in Appendix  \ref{appx:soa} the corresponding solution, which will be relevant in the following sections as a limiting case. 

Equation~\eqref{1c_fp} is a forward Fokker-Planck equation that features a piece-wise constant force. It is equipped with boundary condition $p(\widetilde{x} \to \pm \infty,\widetilde{t}|\widetilde{x}_0) = 0$. 
It can therefore be solved analytically using an eigenvalue expansion \cite{risken}:
 \begin{align}
	\label{1c_prop} p(\widetilde{x},\widetilde{t}|\widetilde{x}_0) =& p_{\infty}(\widetilde{x}) + \sum_{\alpha=o,e}\sum_{k} \frac{u_{k}^{\alpha}(\widetilde{x}) \nu_{k}^{\alpha}(\widetilde{x}_0)}{Z^{\alpha}_{k}} \e^{-\lambda_k^{\alpha} \widetilde{t}}   \nonumber\\
	&+ \sum_{\alpha=o,e}\int_{\frac{1}{4}}^{\infty} d\lambda   \frac{u^{\alpha}(\widetilde{x},\lambda)\nu^{\alpha}(\widetilde{x}_0,\lambda)}{Z^{\alpha}(\lambda)} \e^{-\lambda \widetilde{t}},
\end{align}
where $u_{k}^{\alpha}$ and $u^{\alpha}$ are the eigenfunctions of the Fokker-Planck operator (rhs of Eq.~\eqref{1c_fp}),  $\nu_{k}^{\alpha}$ and $\nu^{\alpha}$ are their adjoint eigenfunctions, and $Z_{k}^{\alpha}$ and $Z^{\alpha}$ the normalization constants. The equilibrium distribution $p_{\infty}$ is given by:
\begin{equation}
p_{\infty}(\widetilde{x})  =  \frac{\e^{-\widetilde{\Phi}(\widetilde{x})}}{Z_{\infty}} =\frac{\e^{-\widetilde{\Phi}(\widetilde{x})+\widetilde{L}/2}}{\widetilde{L} + 2} . \label{eq_dist}
\end{equation} 

The superscript $\alpha$ in Eq.~\eqref{1c_prop} indicates the parity of the eigenfunctions, which will be discussed in upcoming sections. The explicit expression of the functions and their derivation can be found in Appendix \ref{appendix:FokkerPlanck}.
Note that Eq.~\eqref{1c_prop} features  two different types of terms, corresponding either to discrete eigenvalues (with subscript $\lambda_k$) or to a continuous spectrum (the integral terms, involving functions of $\lambda$). This aspect will play a key role in the subsequent treatment.

\subsection{Analytic dynamics}	

We investigate here the dynamics of the counterion density $p(\widetilde{x},\widetilde{t})$, given that 
it starts with initial position $\widetilde{x}_0$.
There are two different possibilities for this initial condition: in the interstitial region
between the colloids ($|\widetilde{x}_0|\leq\widetilde{L}/2$) or outside ($|\widetilde{x}_0|>\widetilde{L}/2$). 

\begin{figure}[htp]
	\centering
	\includegraphics[width=0.48\textwidth]{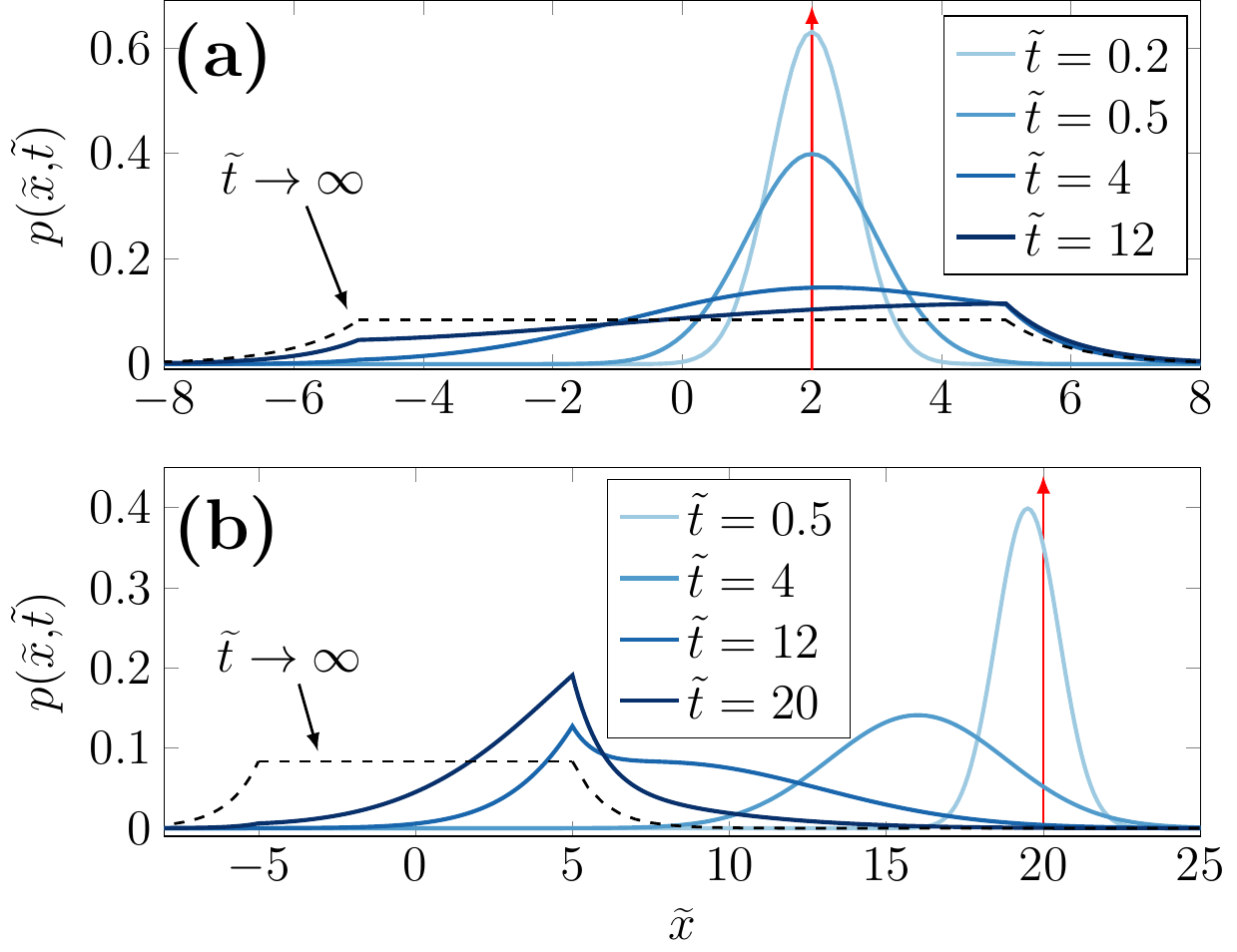}%
	\caption{Time evolution of the density $p(\widetilde{x},\widetilde{t})$ for a single counterion ($N=1$),  with colloid separation $\widetilde{L}=10$ and localized initial distribution centered at $\widetilde{x}_0$ (red arrows): (a) $\widetilde{x}_0 = 2$, and (b)
	$\widetilde{x}_0 = 20$. The equilibrium state ($\widetilde{t}\to \infty$) is given by the dashed lines. In panel (b), the cusp formed at $\widetilde{x} = 5$ for $\widetilde{t} = 20$ is understood in terms of the wedge potential: to the right of the colloid there is a constant force causing a greater  probability flow than from the left side where the counterion undergoes free diffusion.}%
	\label{fig:1c_p-xo-in}%
\end{figure}

\begin{figure}[htp]
	\centering
	\includegraphics[width=0.48\textwidth]{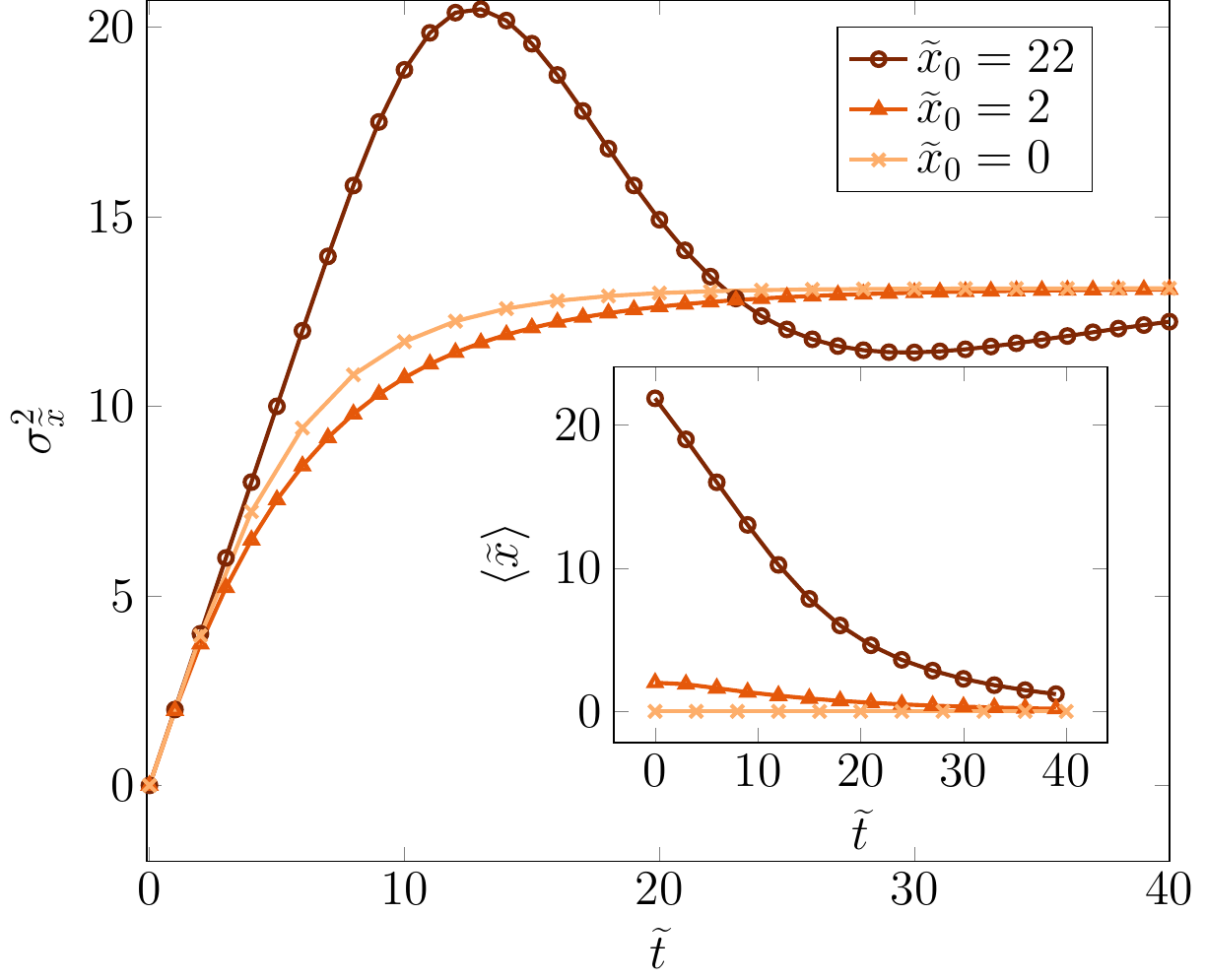}%
	\caption{Position variance $\sigma_x^2(\widetilde{t}\,)$ of a single counterion $N=1$ with a localized initial condition $\widetilde{x}_0 = 0,2,22$ and colloid separation $\widetilde{L}=10$. Note that for short times $\sigma_x^2(\widetilde{t}\,)$ is linear in time $\sigma_x^2(\widetilde{t}\,) \approx 2\widetilde{t}$. 
	For $\widetilde{x}_0 = 22$ the variance is not monotonous, which happens because the initial position is large enough to allow the regime of linear expansion to overshoot the equilibrium variance. The inset is for the corresponding 
	average position $\braket{\widetilde{x}(\widetilde{t}\,)}$ for each initial condition. 
	For large times the average position decays exponentially towards 0, except for the case $\widetilde{x}_0 =0$ where it identically vanishes.} %
	\label{fig:1c_x_var}%
\end{figure}

Let us start with the counterion in the space between the two colloids. The initial dynamics is that of a particle in free diffusion, lasting approximately until 
spread of the density (increasing in $\sqrt{2\widetilde{t}\,}$) reaches the nearest colloid. 
This behavior can be seen directly in Fig.~\ref{fig:1c_p-xo-in} where we observe that the 
density remains left-right symmetric  with respect to the counterion's initial position and it starts to become skewed once the nearest colloid is ``hit'' (located at 
$\widetilde{x}=5$). This is corroborated in Fig.~\ref{fig:1c_x_var}, which features the dynamics of the average position $\braket{x}$ (inset) and the position's variance $\sigma^2_{\widetilde{x}}$: 
the former is constant,
and the latter 
increases linearly with time until 
the density ``hits'' the colloid.

On the other hand, when the counterion starts outside the interstitial region ($|\widetilde{x}_0|>\widetilde{L}/2$) it initially experiences a constant drift towards the colloids. This causes the particle to move towards the colloids, with a mean position that travels at constant speed. This is seen in the inset of Fig.~\ref{fig:1c_x_var}, where for small times the mean position is a linear function of time when $\widetilde{x}_0> \widetilde{L}$. Besides, there is a constant diffusion which is manifested in a linear growth of the position variance ($\sigma_{\widetilde{x}}^2 \sim 2\widetilde{t}$, see Fig.~\ref{fig:1c_x_var}). 
The time during which the drift diffusion occurs lasts approximately until the mean position of the counterion and its nearest colloid are one  
standard deviation
apart. Upon reaching the central region, it has to ultimately accommodate to the asymptotic steady state given by Eq.~\eqref{eq_dist}; hence the non-monotonic behaviour of the position variance in Fig.~\ref{fig:1c_x_var}.
	
We now turn to the asymptotic equilibrium distribution; there is a closed form for the counterion position's average and variance:
\begin{align}
\lim_{\widetilde{t} \to \infty}	\braket{\widetilde{x}(\widetilde{t})}  &= 0, \\
\lim_{\widetilde{t} \to \infty}	\sigma_{\widetilde{x}}^2(\widetilde{t}) &= \frac{\widetilde{L}^2}{12}+\frac{\widetilde{L}}{3}+\frac{4}{3}+\frac{4}{3 (\widetilde{L}+2)}.
\end{align}
Averages are taken here with respect to the equilibrium distribution $p_{\infty}$ (Eq.~\eqref{eq_dist}). Note that for large $\widetilde{L}$, the dominant term is $	 \widetilde{L}^2/12$ which recovers the result of a Brownian particle inside a 1D box of length $\widetilde{L}$.

\subsection{Eigenvalues and relaxation time}\label{sec:1c_eigen}

 Equation \eqref{1c_prop} shows that there is an exponential relaxation towards the equilibrium distribution $p_{\infty}$. Therefore, the relaxation time $\widetilde{\tau}$ can be defined as the inverse of the decay rate associated to the dominant term in Eq. \eqref{1c_prop} for large times. 
 To determine this rate, it is necessary to know the eigenvalue structure of the Fokker-Planck equation (Eq.~\eqref{1c_fp}) and more precisely the spectral gap (first non-vanishing eigenvalue), which is the inverse of $\widetilde{\tau}$. In this section, we explore the $\widetilde{L}$-dependence of the spectrum and consequently of $\widetilde{\tau}$. 

\begin{figure}[htp]
	\centering
	\includegraphics[width=0.48\textwidth]{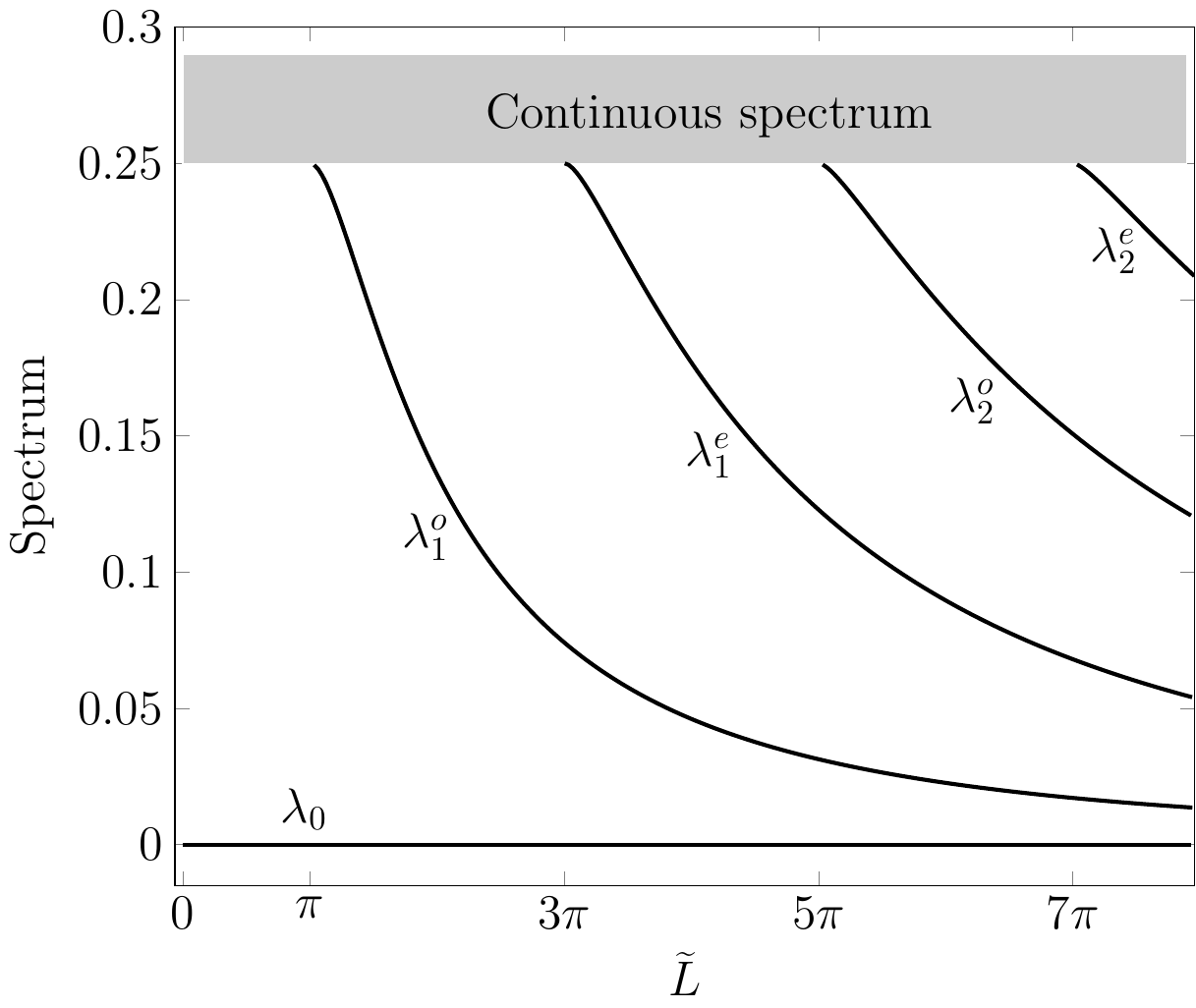}%
	\caption{Spectrum of the Fokker-Planck operator as a function of the system size $\widetilde{L}$. The continuous spectrum's range is  $1/4<\lambda<\infty$, regardless of $\widetilde{L}$. On the other hand, the number of discrete eigenvalues increases with $\widetilde{L}$ in a sequence that alternates odd ($\lambda_k^o$) and even eigenvalues ($\lambda_k^o$). The equilibrium eigenvalue $\lambda_0=0$ is always present.}%
	\label{spectralgap}%
\end{figure}

The spectrum is made of two parts, one discrete and one continuous, which stem from enforcing vanishing boundary conditions at infinity to the eigenfunctions of the Fokker-Planck operator (more details are presented in appendix \ref{appendix:FokkerPlanck}). Since the Fokker-Planck equation is formally equivalent to a Schr{\"o}dinger
equation \cite{risken}, we can also understand the spectrum in terms of quantum mechanics: the discrete and continuous parts correspond to the bounded and unbounded equilibrium states of a single nonrelativistic particle in one dimension, subject to a square well-like potential \cite{schrodinger}.
The discrete set of eigenvalues $\{\lambda_k\}_k \in [0,1/4)$ 
is  $\widetilde{L}$-dependent, while the continuous spectrum is the interval $[1/4,\infty)$.    
The discrete eigenvalues are obtained from solving the following equations subject to the restriction $0\leq \lambda_k < 1/4$:
\begin{align}\label{eodd}
	\big(1-\sqrt{1-4 \lambda_k^{o}}\,\big) \tan \big(\widetilde{L} \sqrt{\lambda_k^{o}}/2\big)&=2 \sqrt{\lambda_k^{o}} , 
	\\
	\label{eeven}\big(\sqrt{1-4 \lambda_k^{e}}-1\big) \cot \big(\widetilde{L} \sqrt{\lambda_k^{e}}/2\big)&=2 \sqrt{\lambda_k^{e}} ,
\end{align}
where Eq.~\eqref{eodd} and \eqref{eeven} are for odd and even eigenfunctions respectively. 
Note that the discrete set is non-empty because  $\lambda_0=0$ is always a solution to Eq. \eqref{eeven}. 
From Eqs.~\eqref{eodd}-\eqref{eeven}, it follows that the number of odd ($N_{\lambda^o}$) and even  ($N_{\lambda^e}$) discrete eigenvalues for a given length are: 
\begin{align}
	N_{\lambda^o} &=  \floor*{\frac{\widetilde{L}-\pi}{4\pi}} + 1, \label{eq:noddeigvl}\\
	N_{\lambda^e} &=  \floor*{\frac{\widetilde{L}+\pi}{4\pi}} + 1\label{eq:neveneigvl} .
\end{align}
From $N_{\lambda^o}$ and $N_{\lambda^e}$ we see that 
discrete eigenvalues emerge as $\widetilde{L}$ increases: odd and even eigenvalues appear in an alternating sequence as a function of $\widetilde{L}$ with period $2\pi$, such that the second eigenvalue $\lambda_2$ is odd and present for $\widetilde{L}>3\pi$, and so on, as sketched in Fig.~\ref{spectralgap}. There is no degeneracy in the discrete spectrum; an eigenvalue cannot solve Eqs.~\eqref{eodd} and \eqref{eeven} simultaneously. This allows a strict ordering:  $\lambda_0<\lambda_1^{o}<\lambda_1^{e} < \lambda_2^{o} <\cdots $.

Further analysis of equations \eqref{eodd}-\eqref{eeven} yields the large $\widetilde{L}$ behavior for the discrete eigenvalues:
\begin{align}
	\lambda^{o}_k &\underset{\widetilde{L}\gg 1}{\sim} \frac{4\pi^2}{\widetilde{L}^2 } \Big(k -  \frac{1}{2} \Big)^2, & 1\leq k \leq N_{\lambda^o} ,\label{eq:oddeigvlasymptotics}\\
	\lambda^{e}_k &\underset{\widetilde{L}\gg 1}{\sim}   \frac{4\pi^2 }{\widetilde{L}^2}(k-1)^2, &1\leq k \leq N_{\lambda^e}. \label{eq:eveneigvlasymptotics}
\end{align}
This implies that for large colloid separations the relaxation time is quadratic in $\widetilde{L}$, regardless of $\widetilde{x}_0$.
In other words, the large $L$ limit yields, quite expectedly, the same
spectrum as a free diffusion in a box of size $L$.

We conclude with the explicit expression for the relaxation time. For this purpose, we identify the dominant term of Eq.~\eqref{1c_prop} at large times, which is associated to the minimum nonzero eigenvalue among the non-vanishing projection of the ionic density onto the eigenbasis. This means that, unlike the spectrum, the relaxation time does depend on the initial conditions. Indeed, there are two distinct behaviors depending on the symmetry of the initial condition: $\widetilde{x}_0 = 0$ (symmetric) and $\widetilde{x}_0 \neq 0$ (asymmetric).
For the former, we have $\nu_k^o(0) =0$ and therefore only the even branches in Fig.~\ref{spectralgap} do matter in Eq.~\eqref{1c_prop}. Then, the relaxation time is given by 
\begin{equation}
    \widetilde{\tau} = \begin{cases}
     \max\{4, \frac{1}{\lambda_1^o}\}, & \widetilde{x}_0 \neq 0,\\
     \max\{4, \frac{1}{\lambda_1^e}\}, & \widetilde{x}_0 = 0.
    \end{cases}\label{tau}
 \end{equation}

\subsection{Simulations}\label{sec:1C_sims}

This section introduces a method to estimate the relaxation time using the counterion density. The scheme is tested on the exact density and  approximation obtained from a computer simulation. Then, these results are compared to the theoretical spectral gap. In this way, the single counter-ion case, which provides us with reference analytic results, is used as a test bench for the method which is later employed for $N>1$ counterions. 

\subsubsection{Relaxation time estimation scheme}
\label{sec:extrapolationmethod}

First, the Kullback-Leibler Divergence (KLD) \cite{kullback1951} is introduced, as it will be used in the relaxation time estimation scheme. Also known as relative entropy and widely used in information theory, this function is defined as:
\begin{equation}
    D_{\text{KL}}(\rho_1||\rho_2) = \int_{\mathbb{R}} dx\; \rho_1(x) \log \frac{\rho_1(x)}{\rho_2(x)} \label{def_KLD},
\end{equation}  
where $\rho_1$ and $\rho_2$ are probability densities and $D_{\text{KL}}(\rho_1||\rho_2)$ is defined as the Kullback-Leibler divergence from $\rho_2$ to $\rho_1$. The discrete definition of the KLD follows from replacing probability densities for probabilities, and performing a summation instead of integrating. The relative entropy is bounded from below $D_{\text{KL}}(\rho_1||\rho_2)\geq 0$, with equality satisfied when $\rho_1=\rho_2$, provided that $\rho_1$ and $\rho_2$ are both normalized. The Kullback-Leibler divergence is not a metric because it is neither symmetric nor does it obey the triangular inequality.  Nonetheless, it is conveniently related to the relaxation time when used with the appropriate distributions: $p(\widetilde{x},\widetilde{t}|\widetilde{x}_0)$ and $p_{\infty}(\widetilde{x})$.

For large times, the Kullback-Leibler divergence of the equilibrium distribution $p_{\infty}$ to the instantaneous density $p$ decreases exponentially to zero. To see this, consider the ionic density;  from \eqref{1c_prop} it follows that
	\begin{equation}
	p(\widetilde{x},\widetilde{t}|\widetilde{x}_0) = p_{\infty}(\widetilde{x}) + \delta p(\widetilde{x},\widetilde{t},\widetilde{x}_0),
	\label{prho*}
	\end{equation}
where $\delta p(\widetilde{x},\widetilde{t},\widetilde{x}_0) / p_{\infty}(\widetilde{x}) \ll 1 $ at long times.
Then, the Kullback-Leibler divergence from $p_{\infty}$ to $p$ has the following large time behavior:
	\begin{align}
	D_{\text{KL}}(p||p_{\infty}) &\sim	\int_{\mathbb{R}}  \delta p\, d\widetilde{x} +	\int_{\mathbb{R}} \frac{(\delta p)^2}{p_{\infty}} d\widetilde{x} + \mathcal{O}\Big(\left({\delta p}/p_{\infty}\right)^2\Big)\nonumber\\
	&\sim		\int_{\mathbb{R}} \frac{(\delta p)^2}{p_{\infty}} d\widetilde{x} + \mathcal{O}\left(\left({\delta p}/p_{\infty}\right)^2\right),
	\label{DKL_epsilon}
	\end{align}
where the integral of $\delta p$ vanishes in view of Eq.~\eqref{prho*} and the fact that both $p$ and $p_{\infty}$ are normalized. Likewise, the associated divergence with transposed arguments ($D_{\text{KL}}(p_{\infty}||p)$) exhibits the same asymptotic behavior as that given by Eq.~\eqref{DKL_epsilon}. From Eq.~\eqref{DKL_epsilon} we observe that the KLD has an exponential decay constant $2/\widetilde{\tau}$, which is twice the value for the ionic density itself.    

\begin{figure}[htp]
	\centering
	\includegraphics[width=0.48\textwidth]{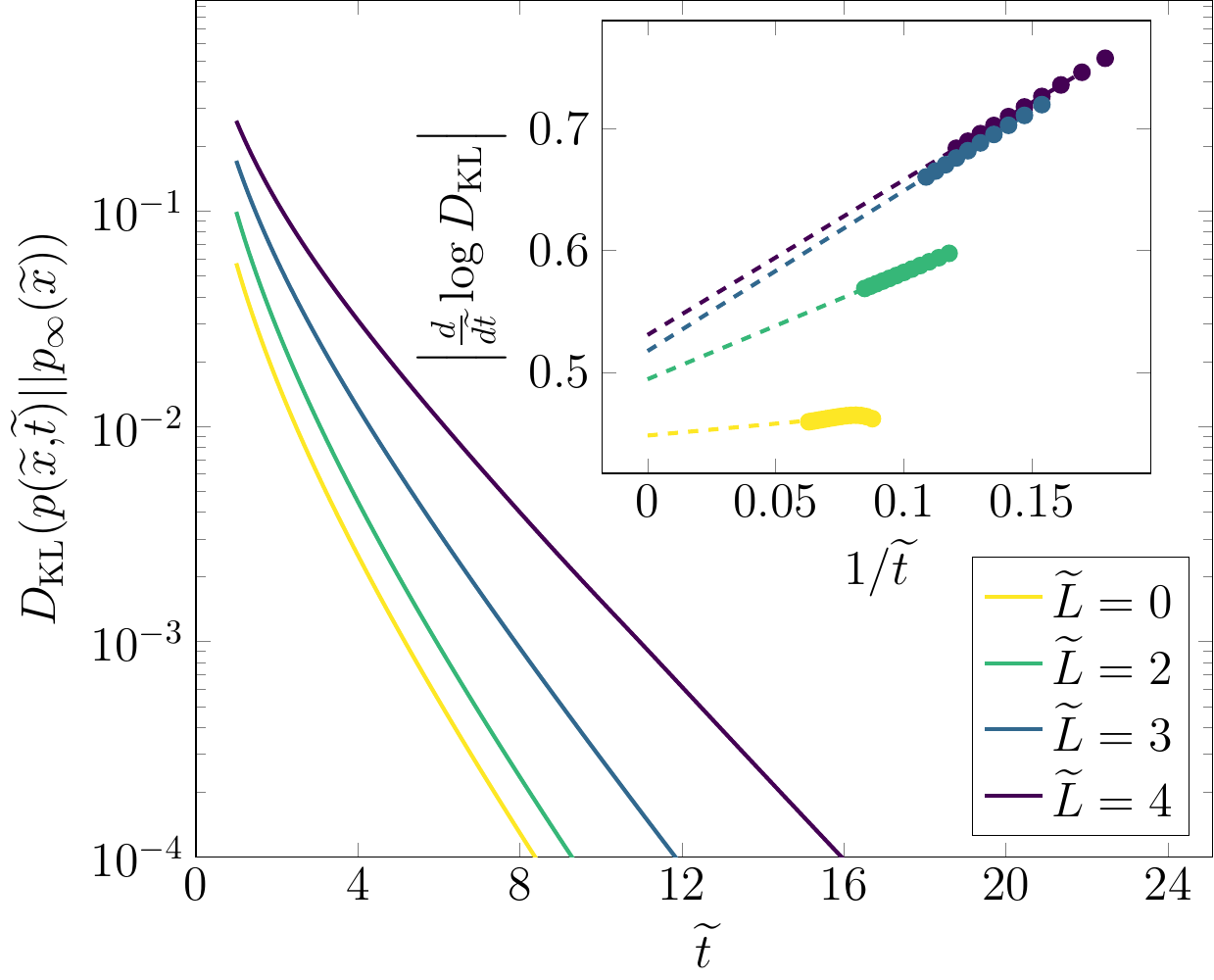}%
	\caption{Kullback-Leibler divergence for one counterion $N=1$ with various colloid separations $\widetilde{L} = 0,2,3,4$ and initial counterion position $\widetilde{x}_0 = \widetilde{L}/4$. Note that for $\widetilde{L}< \pi$, 
	the log KLD is still convex-up for $D_{\text{KL}} >10^{-4}$ 
	due to the subdominant power law $\widetilde{t}^{-s}$ in Eq.~\eqref{KLD_relaxtime}. Contrarily, for $\widetilde{L}> \pi$  the curvature vanishes quickly since the subdominant term decreases exponentially. The inset is a sketch of the scheme to estimate the relaxation time $\widetilde{\tau}$ for a given $\widetilde{L}$. Each same color set of points is the numerical derivative of the logarithm of  $D_{\text{KL}}(p(\widetilde{x},\widetilde{t}|\widetilde{x}_0)||p_{\infty}(\widetilde{x}))$ as a function of the inverse time. The dashed lines are the corresponding minimum square regressions, 
	used to extrapolate the behavior of the dots to $1/\widetilde{t}= 0$. According to Eq.~\eqref{KLD_relaxtime}, this yields twice the decay rate, $2/\widetilde{\tau}$.}%
	\label{fig:KLD_analytic}%
\end{figure}

The term $(\delta p)^2$ induces two distinct behaviors for $D_{\text{KL}}(p||p_{\infty})$ depending on the symmetry 
of the initial condition: $\widetilde{x}_0 = 0$ (symmetric) or $\widetilde{x}_0 \neq 0$ (asymmetric). In the former case, the projection of the ionic density onto odd eigenfuctions vanishes and therefore only the even branches of the
spectrum in Fig.~\ref{spectralgap} do matter. We next describe the asymmetric case, which in Appendix \ref{appx_KLD} is found to be:
\begin{equation}
D_{\text{KL}}(p||p_{\infty})
\sim \begin{cases}
    \widetilde{t}^{-s}\e^{-2\widetilde{t}/4} =\widetilde{t}^{-s}\e^{-2\widetilde{t}/\widetilde{\tau}}, & \widetilde{L} <\pi \\
    \e^{-2\widetilde{t}/\lambda_1^o}=\e^{-2\widetilde{t}/\widetilde{\tau}}, & \widetilde{L} >\pi,
    \end{cases}
    \label{KLD_relaxtime}
\end{equation}
where $1/2<s<5/2$. Although for very large times the 
algebraic term  $\widetilde{t}^{-s}$ is negligible, this is not the case for the times available in the simulations,
which may cause a difficulty in extracting the decay rate. Take for example the case $\widetilde{x}_0 = \widetilde{L}/4$ (Fig.~\ref{fig:KLD_analytic}): notice the curve's concavity for $\widetilde{L} <\pi$, which stems from the subleading term  of order $\log \widetilde t$ 
in the logarithm of the KLD. On the contrary, for $\widetilde{L}>\pi$ this term is always negligible. The symmetric case $\widetilde{x}_0=0$ is similar to  Eq.~\eqref{KLD_relaxtime},
provided the substitutions $\lambda_1^o \to \lambda_1^e$ and $\pi \to 3\pi$
are performed. The inset of Fig.~\ref{fig:KLD_analytic} presents the 
method used for extracting the decay rate $\tau$ from the dynamical data.

\subsubsection{Numerical Integration}

We now proceed to introduce the simulation used to compute the numeric density profiles. We integrate the following stochastic Langevin equation
\begin{equation}
\dot{\widetilde{x}} = \widetilde{F}(\widetilde x, \widetilde{L}) + \xi(\widetilde{t}),
\label{LangevinN1}
\end{equation}
where $\widetilde{F}(\widetilde{L}) = -\partial \widetilde \Phi / \partial \widetilde x$ is the dimensionless force. Equation \eqref{LangevinN1} is Eq.~\eqref{1c_lang} in rescaled units. We record $10^8$ positions per time. Each temporal step was set to $4\times 10^{-4}$ and a histogram was recorded every 200 time steps, using bins of size 0.2. 
These histograms give the numeric estimation of the ionic density, $p(\widetilde{x},\widetilde{t})$.

\begin{figure}[htp]
\centering
\includegraphics[width=0.48\textwidth]{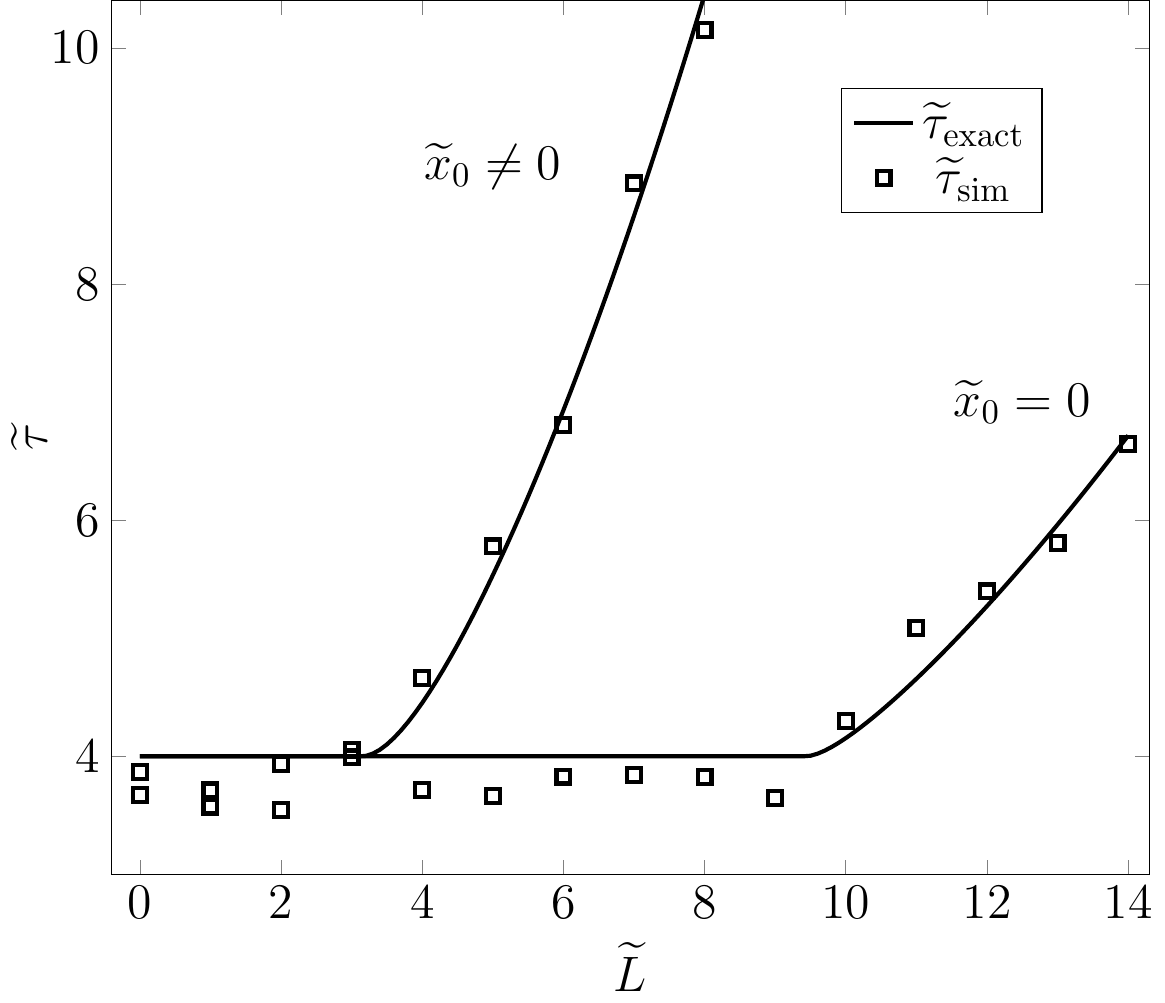}%
\caption{Relaxation time for a system made of two colloidal particles at a distance $\widetilde{L}$ and one counterion ($N=1$), from the exact calculation (solid) given by Eq.~\eqref{tau} and an estimation using a simulation (squares). Two cases for the localized initial distribution should be distinguished: asymmetric ($\widetilde{x}_0 \neq 0$) and symmetric ($\widetilde{x}_0 = 0$).}%
\label{fig:1_counterion_texp_KLD}%
\end{figure}

Then, the discrete Kullback-Leibler divergence $D_{\text{KL}} (p(\widetilde{x},\widetilde{t})||p_{\infty}(\widetilde{x}))$ is calculated. The relaxation times follow from the long-time behavior of
$D_{\text{KL}}$, as explained above;
the results are shown in Fig.~\ref{fig:1_counterion_texp_KLD}, where the numerical scheme is seen to be in agreement with the analytical curve. Finally, note that the relaxation time $\widetilde{\tau}_{\text{sim}}$  extracted from the simulations for a symmetric initial condition exhibits a slight non-monotonous behavior in the vicinity of $\widetilde{L} = 9$. This 
non-monotonicity appears to be an artifact of the numerical procedure used.
We come back to this in section \ref{sec:Nrelaxation}.

We conclude with some additional remarks on our numerical study.
It is based on simulations of the Brownian dynamics that describe the time evolution of the system. For our purposes, the Euler-Maruyama method \cite{Maruyama} is both the simplest and most efficient technique to simulate the counterions' paths. The reason for this is that the coefficients of the white noise terms in the Langevin equations (defining the diffusion coefficients) are all constant in time and the electrostatic force is bounded; these are sufficient conditions for numerical convergence of weak order $1.0$ \cite{Milstein,comment10}. 
Higher-order methods do not provide any computational advantages, as they usually rely on derivatives of the white noise term coefficients \cite{KloedenPlaten}, so the only real improvement in this direction is the use of computation in parallel. As it is necessary to collect $n$ samples to obtain results with a precision of order $n^{-1/2}$ from simulations, computation in parallel provides a way to obtain results with high precision at a reasonable computation cost.


\subsection{First passage time}
\label{sec:firstpassage}

At equilibrium, the counterion plays an important role in the pressure: the colloids share this particle, which for large enough $\widetilde{L}$, induces them to attract each other  \cite{TellezTrizac2015}. It is of interest to investigate how the counterion is shared, and how much time it spends in the vicinity of one colloid before reaching the other. 

The first passage time $\widetilde{T}$, with initial and final position on each colloid, is an appropriate quantity to measure the crossing time between colloids. Due to the symmetry of the system we can choose the particle starting on either side and ending in the opposite side. Taking the initial position of the counterion to be at the left colloid $-\widetilde{L}/2$ and the final position at the right colloid $\widetilde{L}/2$,  the first passage time distribution $w_{\widetilde{L}}(\widetilde{T})$ follows from:
\begin{equation}
w_{\widetilde{L}}(\widetilde{T}) = -\int_{-\infty}^{\widetilde{L}/2} \frac{\partial}{\partial \widetilde{T}} P_{a}(\widetilde{x}, \widetilde{T}|-\widetilde{L}/2) \;  d\widetilde{x},
\end{equation}
where $P_{a}(\widetilde{x}, \widetilde{T}|\widetilde{x}_0)$ is the probability density function of an initially localized counterion  at $\widetilde{x}_0$ with an absorbing wall at $\widetilde{L}/2$ (see \cite{risken}). Since $P_a$ follows from solving  Eq.~\eqref{1c_fp} with different boundary conditions than those for $p$ (Eq.~\eqref{1c_prop}),  the Fokker-Planck operator for $P_a$ features a different spectrum. It has both a continuous and a discrete part, and their boundaries are: $\lambda^{a} \in (1/4,\infty)$ and $0<\lambda^{a}_k< 1/4$ respectively. The discrete part of the spectrum follows from solving
\begin{equation} 
1-\sqrt{1-4\lambda^{a}_k}  =2\sqrt{\lambda^{a}_k} \cot(\sqrt{\lambda^{a}_k}\widetilde{L})
\label{eigenabsortion},
\end{equation}
which has no solutions for  $\widetilde{L}< \pi/2$.  Therefore, the spectrum is completely continuous when $\widetilde{L}< \pi/2$, leaving aside the gapped steady-state eigenvalue $\lambda=0$.     

In a similar fashion as for $P$, we obtain the analytic solution for $P_a$ and consequently for the first passage time distribution $w_{\widetilde{L}}(\widetilde{T})$:
\begin{equation}
\begin{split}
&w_{\widetilde{L}}(\widetilde{T}) = \int_{1/4}^{\infty}\e^{-\lambda^{a}  \widetilde{T}}\frac{\sqrt{4 \lambda^{a} -1}  \sin \left(\sqrt{\lambda^{a} } \widetilde{L}\right)}{2 \pi  \sqrt{\lambda^{a} }-\pi  \sin \left(2 \sqrt{\lambda^{a} } \widetilde{L}\right)} d\lambda^{a}
\\
+& \sum_{k} \frac{ \e^{-\lambda^{a}_k  \widetilde{T}} \big[1+2 \sqrt{\lambda^{a}_k } \tan \Big(\frac{\sqrt{\lambda^{a}_k } \widetilde{L}}{2}\Big)-\sqrt{1-4 \lambda^{a}_k }\,\,\big]}{ 2(1-4 \lambda^{a}_k)^{-\frac{1}{2}}+ \widetilde{L} \csc ^2(\sqrt{\lambda^{a}_k } \widetilde{L})-\cot (\sqrt{\lambda^{a}_k } \widetilde{L})/\sqrt{\lambda^{a}_k}} \label{w0}
\end{split}
\end{equation}
where $\lambda^{a}_k<1/4$ is a solution of Eq.~\eqref{eigenabsortion}. From the spectrum follows the asymptotic behavior which decays as $\e^{-\widetilde{T}/4}$ for $\widetilde{L}<\pi/2$ and as  $\e^{-\lambda^{a}_{1}\widetilde{T}}$ for $\widetilde{L}>\pi/2$ where $\lambda^{a}_{1}$ is the smallest discrete eigenvalue. 

The previous behavior bears a resemblance to the first passage time distribution of a free diffusing particle in a box of length $\widetilde{L}$. This situation consists of a free Brownian motion $\widetilde{X}(\widetilde{t})$ with diffusion constant equal to one and started at the origin. We place a pair of fixed walls, a reflecting one at the origin and an absorbing one at a distance $l>0$. Then, we define the transport time as the first hitting time of $\widetilde{l}$, that is,
\begin{equation}\label{eq:tau_l}
\tau_{\widetilde{l}}=\inf\{\widetilde{t}>0:\widetilde{X}(\widetilde{t})\geq \widetilde{l}\}.
\end{equation}
We derive the distribution $q_{\widetilde{l}}(\widetilde{t})$ of $\widetilde{\tau}_l$ in appendix  \ref{appx:transport_time}:
\begin{equation}\label{eq:trtimedist}
q_{\widetilde{l}}(\widetilde{t})=\sum_{n=0}^\infty(-1)^n\frac{\pi(2n+1)}{\widetilde{l}^2}\exp\Big(-\frac{\pi^2(2n+1)^2}{4\widetilde{l}^2}\widetilde{t}\Big).
\end{equation}

\begin{figure}[htp]
\centering
\includegraphics[width=0.48\textwidth]{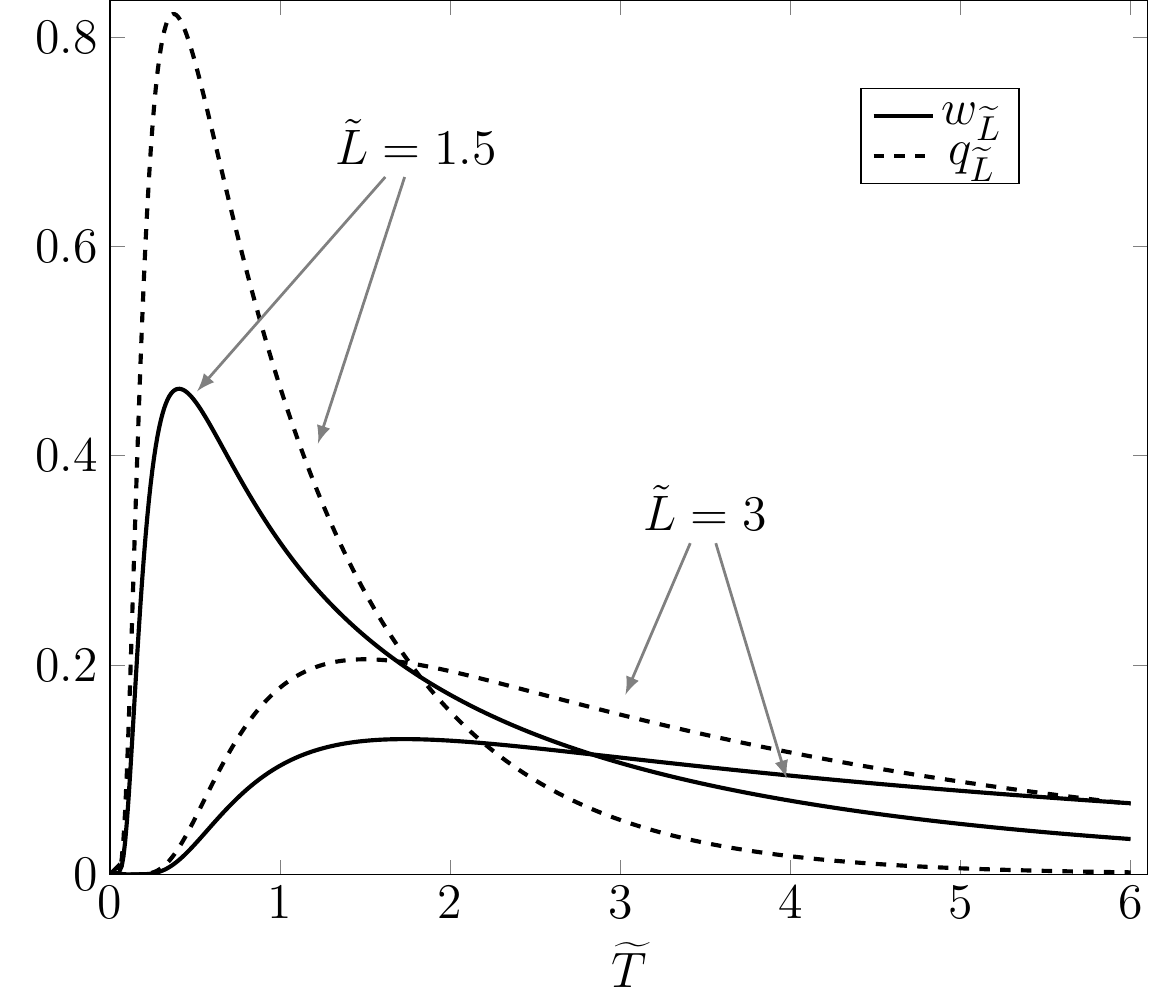}%
\caption{First passage distribution for a single counter-ion traveling between the colloids ($w_{\widetilde{L}}$, solid) and a free diffusing particle in a box ($q_{\widetilde{L}}$, dashed) with a colloid/box length of $\widetilde{L} =1.5$ and $3$. The former is seen to have a larger mean first passage time and variance: in that case the counter-ion has the possibility to make excursions with $\widetilde{x} <0$ outside the region delimited by the colloids unlike the box situation where it is limited to a finite space.}%
\label{fig:1counterion_FPT}%
\end{figure}

The confined Brownian particle exhibits a smaller average first passage time and variance than the one counter-ion colloid (see Fig.~\ref{fig:1counterion_FPT}). This is expected since it does not have an infinite region available to wander off. In the following section the boxed particle model will allow us to understand the behavior of the crossing time of the $M$th particle when $N=2M-1$ (the middle/misfit counter-ion), from one colloid double-layer to another. In this sense, the case $N=1$ is different from all the odd $N>1$ where the colloids have screening particles that prevent the middle counter-ion from scouting the exterior regions. In fact, we will see that the effective length $l$ is smaller than $L$. 

\section{Relaxation Time For Multiple Counterions}
\label{sec:Nrelaxation}

We move on to describe the relaxation time for the multiple  counterion case, $N>1$. Each counterion has fixed charge $e$, and the colloid charge $-Ne/2$ ensures electroneutrality.   
An analytical or numerical solution of the Fokker-Planck equation is impractical since it involves a partial differential equation in $N+1$ dimensions. Instead, we perform simulations of the corresponding Langevin equation to compute the time evolution of the density profile $n(\widetilde{x},\widetilde{t};N)$. To obtain the relaxation time, we analyze the evolution of the KLD between the density at time $\widetilde{t}$ and the equilibrium one by extending the scheme introduced in section \ref{sec:1C_sims}. This is done by replacing the single particle counterion density with the normalized counterion density profile $n/N$. This normalization is necessary to use the KLD, which is defined for probability density functions. Besides from the time evolution, we require the equilibrium density profile $n_{\infty}(\widetilde{x};N)$, which has the following analytical expression:
\begin{equation}
\begin{split}
	n_{\infty} = 
	\sum_{N_{\ell} = 0}^{N}\sum_{N_{r} = 0}^{N-N_{\ell}} \frac{z_{N_{\ell},N_r}(N,\widetilde{L}) }{Z(N,\widetilde{L})} {n}_{N_{\ell},N_r}\Big(\widetilde{x}+\frac{\widetilde{L}}{2},N,\widetilde{L}\Big),
\end{split}
\label{N_n}
\end{equation}
where $Z(N,\widetilde{L}) =\sum_{N_{\ell} = 0}^{N}\sum_{N_{r} = 0}^{N-N_{\ell}} z_{N_{\ell},N_r}(N,\widetilde{L})$. The functions  $z_{N_{\ell},N_r}(N,\widetilde{L})$ refer to partition functions of a system with a fixed number of particles at the left ($N_{\ell}$) and right ($N_{r}$) sides and are given in \cite{vtt2016} (Eqs.~(52)-(55)), and $n_{N_{\ell},N_r}$ are the density functions with fixed number of particles at each side and are given by equations (59)-(61) of Ref.~\cite{vtt2016}. Those results are for a system of two symmetric impermeable colloids and $N$ counterions distributed in the three possible regions: $N_{\ell}$, $N_r$ and $N-N_{\ell}-N_r$ are in the regions $x<0$, $\widetilde{x}>\widetilde{L}$ and $0<\widetilde{x}<\widetilde{L}$ respectively. These results must be translated to coincide with the present situation $\widetilde{x}\to \widetilde{x} + \widetilde{L}/2$, and then the sum over all the possible impermeable configurations weighted by $z_{N_{\ell},N_r}(N,\widetilde{L})  /Z(N,\widetilde{L}) $ yields the permeable result.   

The time evolution of the density profile  $n(\widetilde{x},\widetilde{t};N)$ is computed using a numerical simulation of the Langevin equation. The advantage of this scheme is that it is easily extended to an $N$ counter-ion system. The system of Langevin equations to be simulated is
\begin{equation}
\dot{\widetilde{x}}_j = \widetilde{F}_j(\widetilde{x}_1,\dots,\widetilde{x}_N) + \xi_j\left(\widetilde{t}\,\right),
\label{NLangevin}
\end{equation}
where  $\braket{\xi_i(\widetilde{t}') \xi_j(\widetilde{t})} = 2 \delta_{ij}\delta(\widetilde{t}-\widetilde{t}')$,  $\braket{\xi_i(\widetilde{t})} = 0$ and $\widetilde{F}_j = F_j \epsilon kT/e^4$ is the dimensionless Coulomb force on the particle at $\widetilde{x}_j$:
\begin{equation}
\begin{split}
\widetilde{F}_j(\widetilde{x}_1,\dots,\widetilde{x}_N) &=  
\sum_{\substack{i=1\\i\neq j}}^N  \text{sgn}(\widetilde{x}_j-\widetilde{x}_i)-\frac{N}{2}\,\text{sgn}\Big(\widetilde{x}_j+\frac{\widetilde{L}}{2}\Big)\\
&-\frac{N}{2}\,\text{sgn}\Big(\widetilde{x}_j-\frac{\widetilde{L}}{2}\Big),
\label{NForce}
\end{split}
\end{equation}
where $\text{sgn}(x)$ is the sign function. Eq.~\eqref{NLangevin} with a force given by Eq.~\eqref{NForce} is very close to a
one-dimensional Brownian particle system with rank-dependent drifts \cite{pal2008}, with the difference that the fixed colloids disable the rank-dependence because the drift of the counterion also depends on its position relative to the fixed points $-\widetilde{L}/2$ and $\widetilde{L}/2$. 
With these two ingredients, we can compute $D_{\text{KL}}(p(\widetilde{x},\widetilde{t};N)||p_{\infty}(\widetilde{x};N))$, where $p(\widetilde{x},\widetilde{t};N) = n(\widetilde{x},\widetilde{t};N)/N$ is the normalized density profile and $p_{\infty}(\widetilde{x};N)$ the corresponding equilibrium state (Eq.~\eqref{N_n}).

\begin{figure}[htp]
	\begin{center}
	\includegraphics[width=0.48\textwidth]{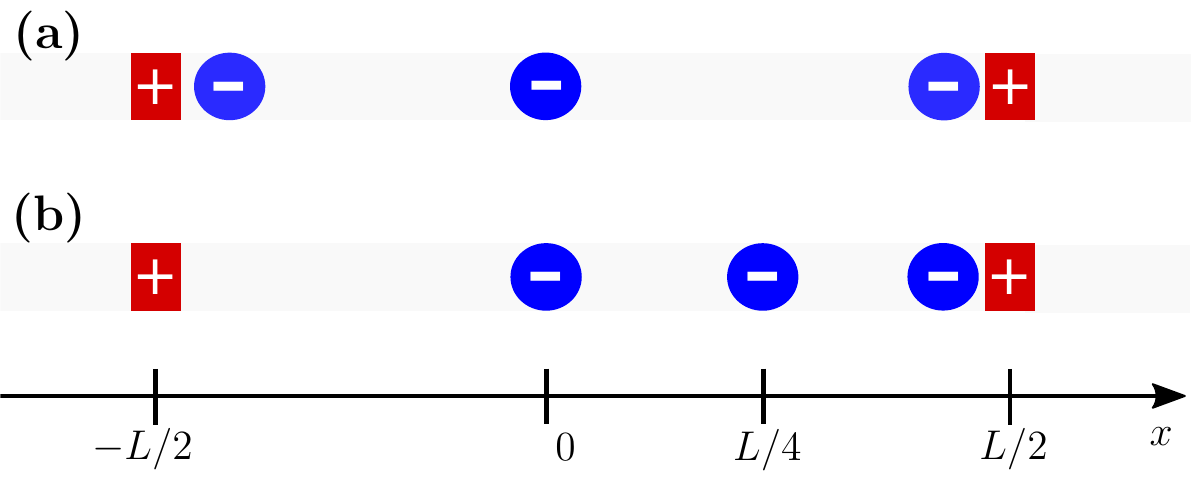}%
		\caption{Sketch of initial conditions (Eq.~\eqref{eq_IC} used for the colloid simulation (here $N=3$): 
		(a) Symmetric IC and (b) Asymmetric IC.}
		\label{fig:sketch_IC}
	\end{center}
\end{figure}

The simulations for the many counterions case use the same time and position discrezation as for for $N=1$. 
The number of positions recorded per particle is of order $10^7$. For $N=1$, a symmetric initial condition annihilates the projection of the ionic density onto the first non-zero eigenvalue ($\lambda_1^o$), and therefore changes the relaxation time. In view of this phenomenon, we implement two initial conditions (IC) with different parity, as presented in Fig. \ref{fig:sketch_IC} and specified by
\begin{equation}
\begin{split}
\text{Symmetric IC} &=
\Big\{-\frac{\widetilde{L}}{2},\dots,\frac{(2k-1-N)\widetilde{L}}{2(N-1)},\dots, \frac{\widetilde{L}}{2} \Big\}  
\\
	\text{Asymmetric IC} &=
\Big\{0,\dots, \frac{(k-1)\widetilde{L}}{2(N-1)},\dots, \frac{\widetilde{L}}{2}\Big\} 
,
\end{split}
\label{eq_IC}
\end{equation}
where $k\in\{1,\dots,N\}$.

\begin{figure}[htp]
	\begin{center}
	\includegraphics[width=0.48\textwidth]{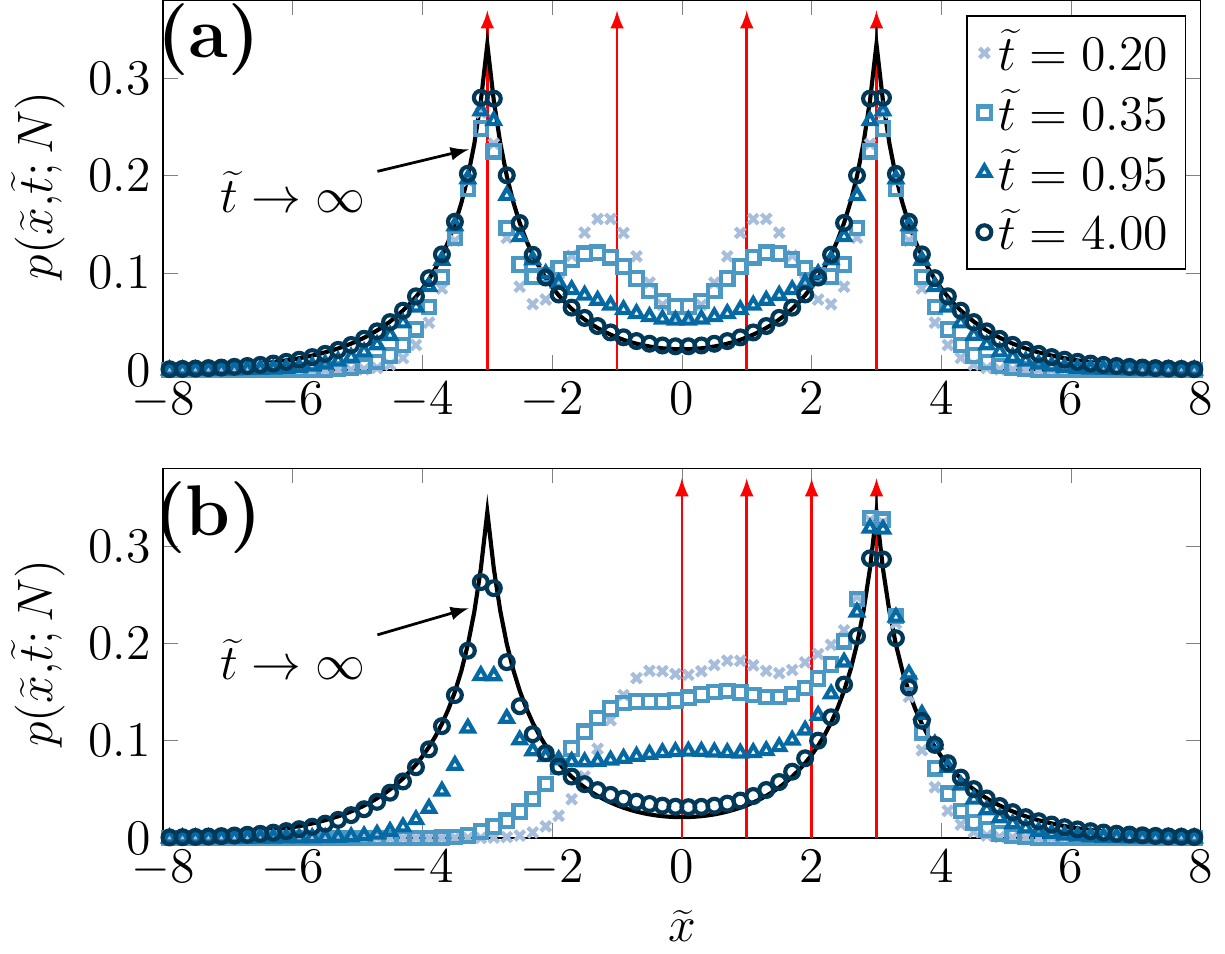}%
		\caption{Time evolution of the normalized density profile $p(\widetilde{x},\widetilde{t};N)$ for $N=4$ counterions and colloids at distance $\widetilde{L}=6$. The red arrows represent the localized initial conditions: (a) symmetric and (b) asymmetric. The solid line is the exact equilibrium distribution. Note that for $\widetilde{t}\approx4$ the density has almost reached its steady state.}
		\label{fig:N4L6}
	\end{center}
\end{figure}

The dynamics in the many counterion case is determined by two parameters: colloid separation and number of counterions. However, there are a few general properties that we proceed to describe using a concrete example: $N=4$ and $\widetilde{L}=6$ which is plotted in Fig.~\ref{fig:N4L6}. As a consistency check, note that the simulation approaches the exact equilibrium result (solid line). From $p_{\infty}(\widetilde{x};N)$ we know that the first moment vanishes at infinite times, which is readily seen from the Fig.~\ref{fig:N4L6} at $\widetilde{t} \approx 4$.
The position variance is seen to be monotonic as the density profile expands to its equilibrium state, just as for $N=1$ with the counterion initially in the region between the colloids. Both the first moment and the variance decay exponentially to their respective terminal values. The following part of the paper moves on to describe this process.

\subsection{Odd Number of Counterions}

\begin{figure}[htp]
\begin{center}
\includegraphics[width=0.48\textwidth]{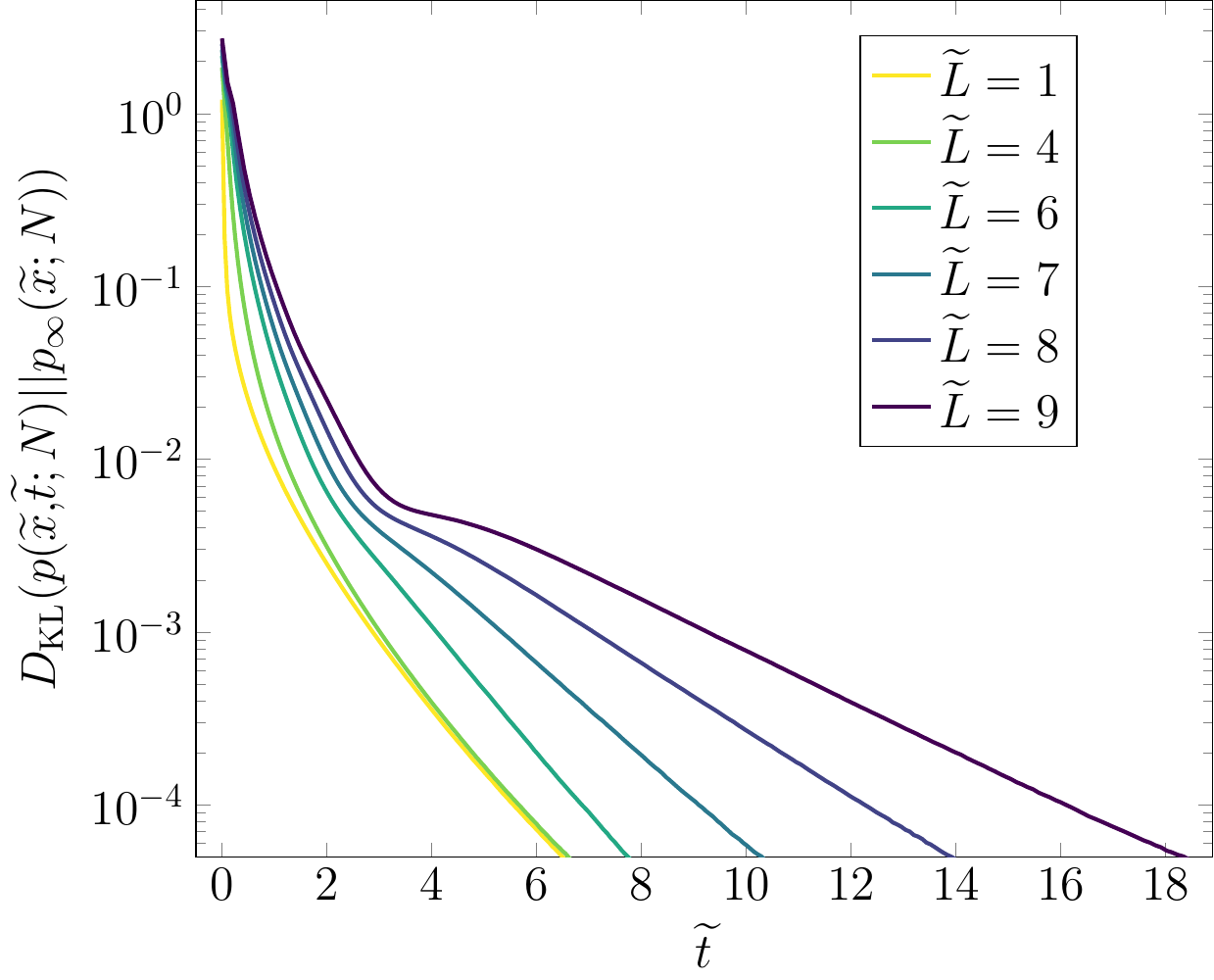}%
	\caption{Numeric Kullback-Leibler divergence for $N=15$ in the asymmetric IC case. Note the strong positive concavity for $\widetilde{L}=4$ in the vertical range between $10^{-3}$ and $10^{-4}$.}
	\label{fig:AsymKLDN15}
\end{center}
\end{figure}

\begin{figure}[htp]
\begin{center}
\includegraphics[width=0.48\textwidth]{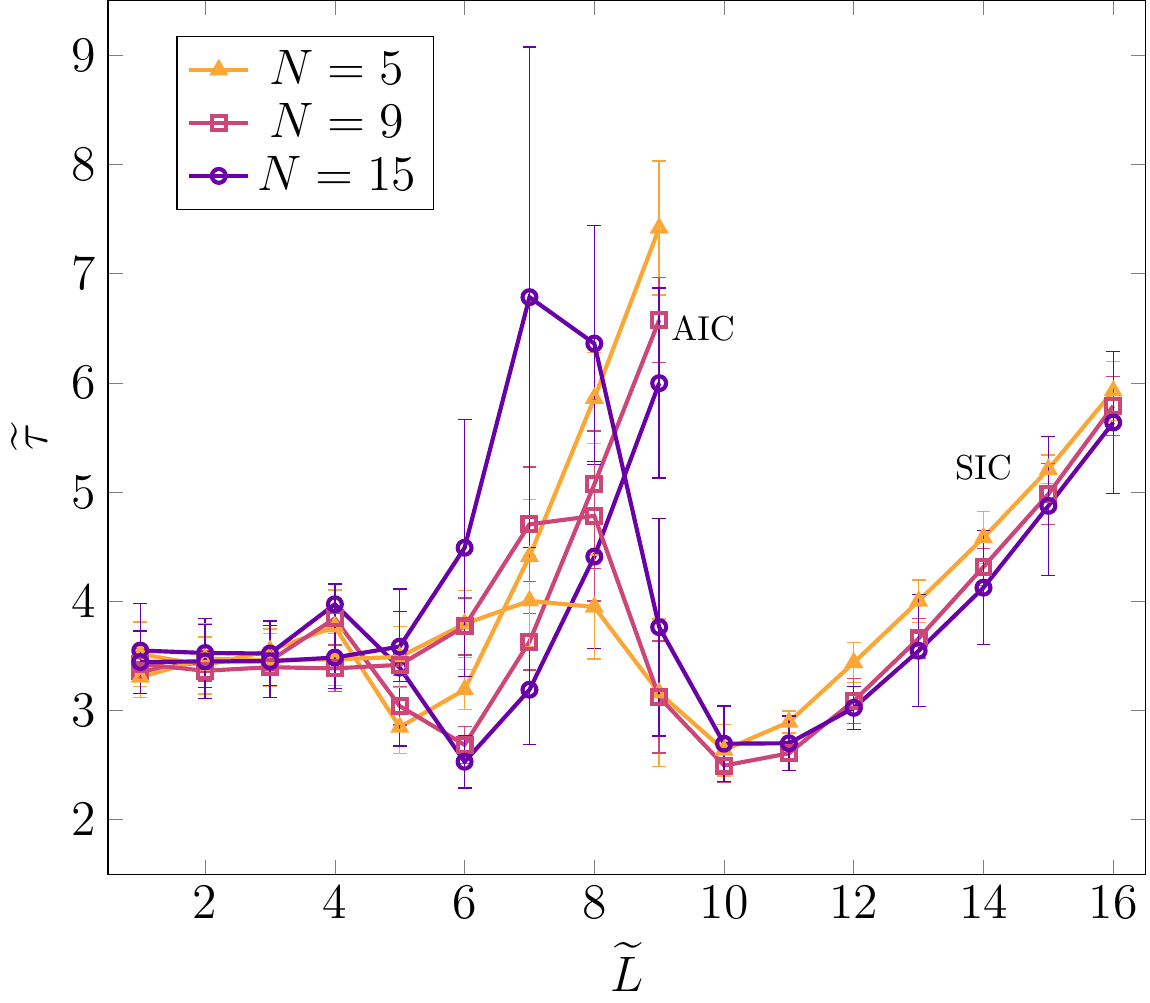}%
\caption{Numerical estimation of the relaxation time $\widetilde \tau$ as a function of $\widetilde{L}$ for $N = 5,9,15$ with asymmetric (AIC) and symmetric (SIC) initial conditions (Eq.~\eqref{eq_IC}). The straight lines are guides to the eye. Note the overall resemblance with the case $N=1$ (Fig.~\eqref{fig:1_counterion_texp_KLD}): a constant region at short colloid separations is followed by a quadratic increase of $\widetilde{\tau} \sim \widetilde{L}^2$ at large $\widetilde L$.  The error bars denote one standard deviation, obtained from the spread in the linear fits outlined in Sec.~\ref{sec:extrapolationmethod} performed over a moving window of 30 consecutive data points.
}
\label{fig:OddAsymmTime}
\end{center}
\end{figure}

Let us begin this part with the asymmetric IC case. Fig.~\ref{fig:AsymKLDN15} shows the time evolution of the KLD obtained from simulations with $N=15$ counterions. In order to carry out the extrapolation scheme described in Sec.~\ref{sec:extrapolationmethod}, we take the data points where the KLD lies between $10^{-3}$ and $10^{-4}$. It is clear from Fig.~\ref{fig:AsymKLDN15} that the log KLDs for $1\leq \widetilde{L}\leq 4$ are convex-up in this region, but no concavity is visible for $\widetilde{L}\geq 6$.
In the following discussion, it will appear that the behavior for odd $N>1$ 
can be mapped onto the $N=1$ case; the middle particle indeed acts as for $N=1$, but in a reduced length: the whole colloid-colloid length $L$ is no longer accessible, 
given the presence of the ions localized in the vicinity of the colloids.
An effective length, $L$ minus the two double-layer sizes, turns out to 
make the above mapping operational.
We come back to this question in section \ref{sec:middle_counterion_trans}.
In the one-counterion case with asymmetric IC, the first nonzero discrete eigenvalue appears when $\widetilde{L}\geq \pi$ according to Eq.~\eqref{eq:noddeigvl}; in the present situation with $N>1$, taking into account the double-layer size of the ion clouds, we may surmise that this eigenvalue appears at a slightly larger $\widetilde{L}$. Indeed, this is confirmed in Fig.~\ref{fig:OddAsymmTime}, where the characteristic time seems to be close to $\widetilde{\tau}=4$ for $\widetilde{L}\leq 4$, and increases for $\widetilde{L}\geq 6$.

When imposing the symmetric IC, the projection of $p$ onto any odd eigenfunction vanishes. Therefore, the dominating term becomes $\lambda^e_1$, which for $N=1$
separates from the continuum at $\widetilde{L}\geq 3\pi$, according to Eq.~\eqref{eq:neveneigvl}. 
If we take the double-layer size into account, we expect the separation of this eigenvalue to occur at around $\widetilde{L}=10$. The corresponding dynamics of the KLDs for $N=15$ are very similar in shape to those shown in Fig.~\ref{fig:AsymKLDN15}, in that the log KLDs show a noticeable convex behavior in the range between $10^{-3}$ and $10^{-4}$ for small $\widetilde{L}$. This convexity disappears for large $\widetilde{L}$, a phenomenon that seems to occur at $\widetilde{L}>10$, and that is reflected in Fig.~\ref{fig:OddAsymmTime}. As expected from our considerations in the single counterion case, we observe that the characteristic time is compatible with $\widetilde{\tau}\simeq 4$ for $\widetilde{L}\leq 6$, and that it increases for $\widetilde{L}> 10$. 

\begin{figure}[htp]
	\centering
	\includegraphics[width=0.48\textwidth]{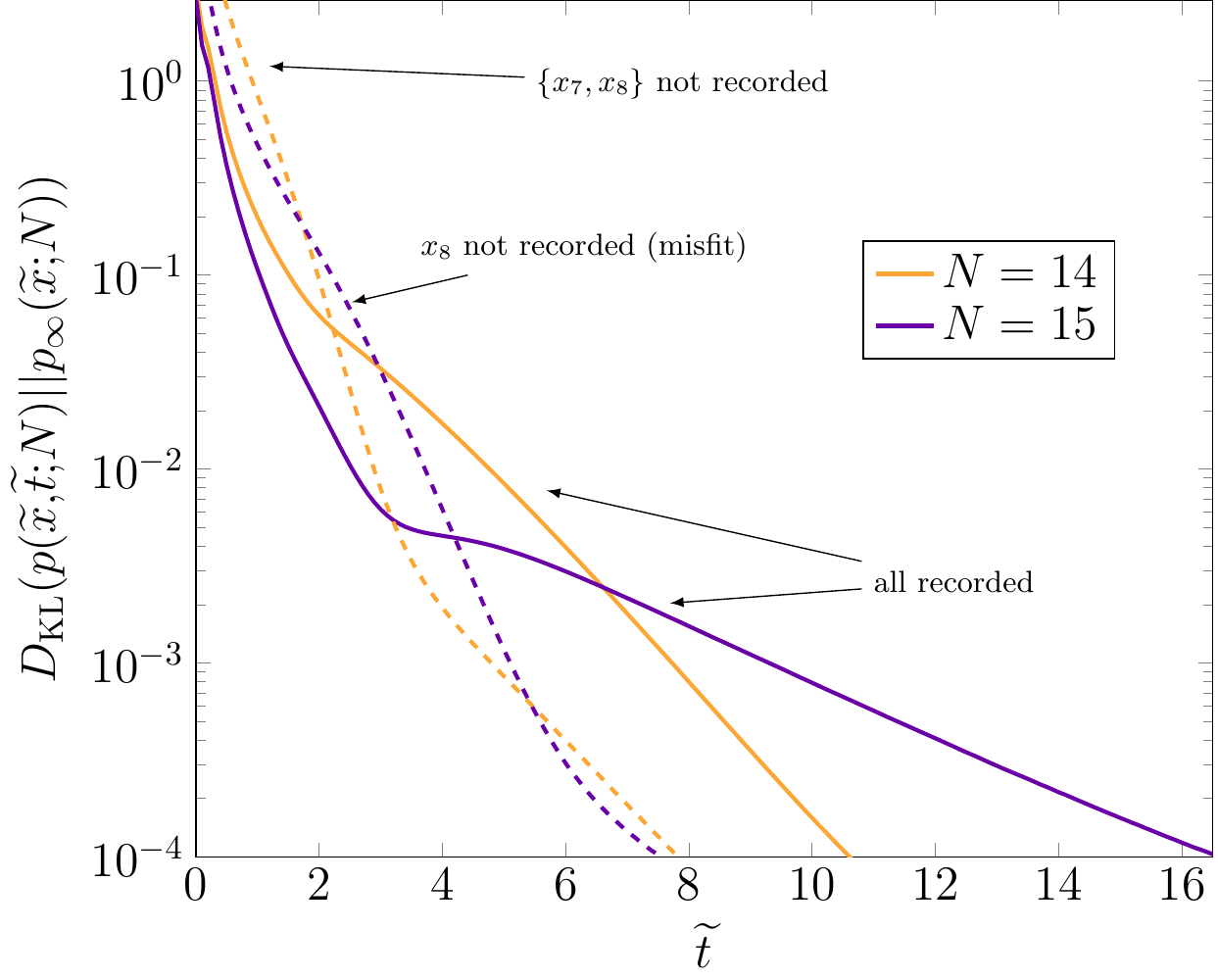}%
	\caption{Kullback-Leibler divergence of a modified ionic density (dashed) at a given time and its equilibrium distribution, for $\widetilde L = 9$.  For $N=15$, the simulation does not record the misfit's position ($x_8$). Likewise, for $N=14$ the middle particles $x_7,x_8$ are not recorded. The solid curves correspond to the case where the true counterion density profile is used. 
	}%
	\label{fig:misfitNR}%
\end{figure}

While the diffusive behaviour of the relaxation
time for $N$ odd is due to the misfit, one may wonder how fast all
other ions do relax, and surmise that they presumably do so on a much smaller 
time scale than $\widetilde\tau$. We show now that this indeed is the case
and examine the effect of the middle particle in the double-layer relaxation.
To this end, we have considered the density of the counterions
discarding the misfit, and investigated how it departs from its equilibrium distribution, through the corresponding KLD.
In Figure \ref{fig:misfitNR}, the resulting dashed curve, for $N=15$, does not
exhibit a pure exponential behaviour, but its slope yields a relaxation time
close to $1.3$, which is significantly smaller than the $\widetilde\tau$ value,
here close to $6.0$, and given by the large time slope of the continuous 
$N=15$ curve. For completeness, we also report 
a benchmark calculation for $N=14$. There is then no misfit, but
this calculation allows to estimate the effect of removing 
ions for the KLD considered. We note that the two curves for $N=14$ 
quickly become parallel, and thus yield the same decay rate.
This is at variance with the situation at $N=15$. In the following section, the case where $N$ is even is explored in detail.

\subsection{Even Number of Counterions}

\begin{figure}[htp]
	\centering
	\includegraphics[width=0.48\textwidth]{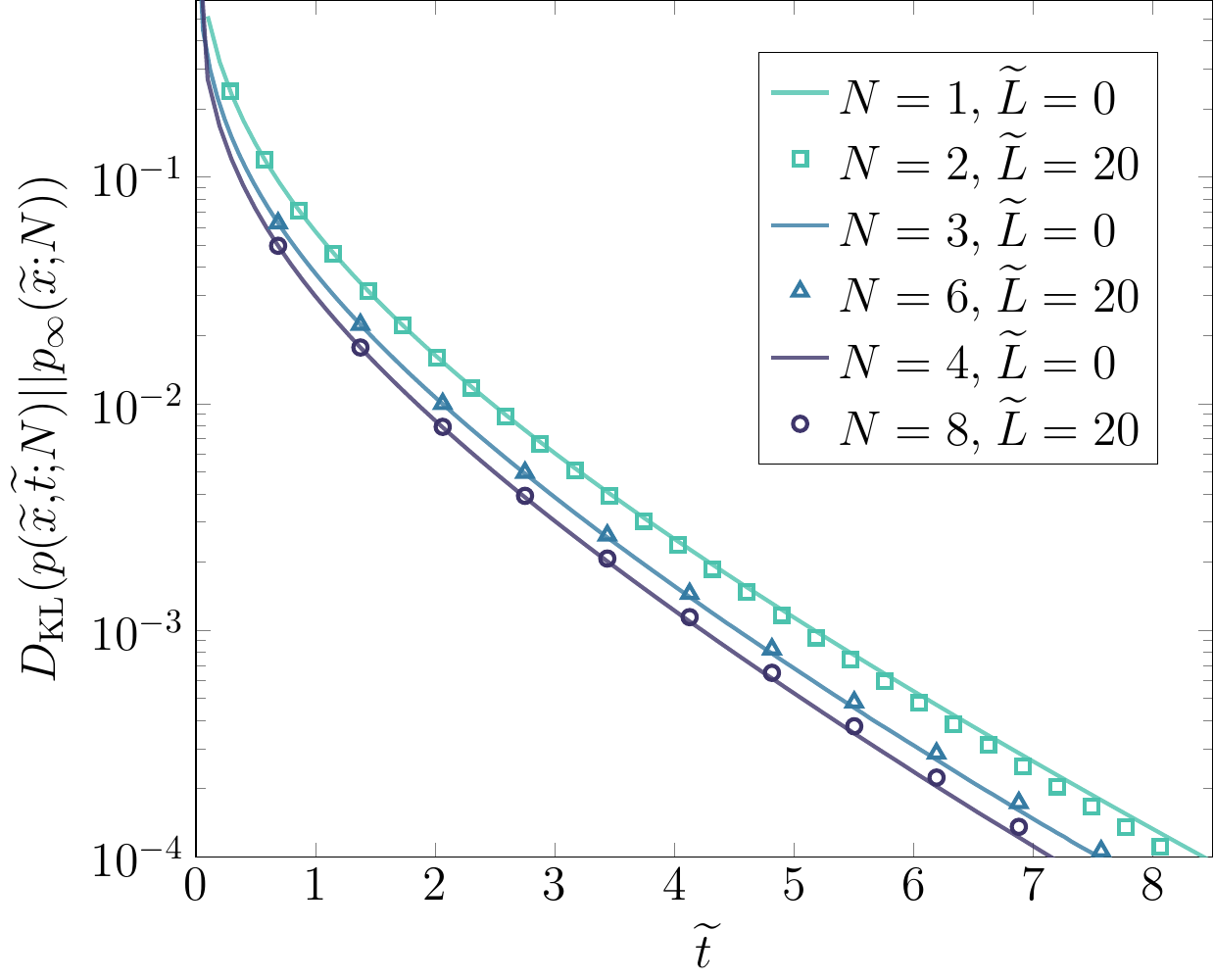}%
	\caption{Kullback-Leibler divergence for one counter-ion (analytic) and for $N=2,3,4,8$ counter-ions (numerical). Solid lines have $\widetilde{L}=0$ and marks $\widetilde{L}=20$ (large colloid separation). Data for $N/2$ and $N$ counterions do overlap. The agreement of each pair of curves shows that at large distances, a system with even $N$ counterions effectively behaves as two decoupled neutral subsystems of $N/2$ counterions.}%
	\label{fig:N_Nhalf}%
\end{figure}

We now consider an even number of counterions. For this case we do not posses any analytical results. However, we expect that for two counterions ($N=2$) and $\widetilde{L} \gg 1$, there is correspondence to the system with a single counterion ($N=1$) and zero colloid separation ($\widetilde{L} =0$). The rationale behind this argument is that at large colloid separations, the $N$ counterions split into two groups of $N/2$ particles which completely screen each colloid. In other words, two neutral subsystems are formed. Moreover, the screened colloids do not exert any force onto each other. Hence, there is an effective decoupling of the whole system into two non-interacting screened colloids, with $N/2$ counterions each. Consequently, we expect the case $N=1$ and $\widetilde{L}=0$ to have the same relaxation towards equilibrium as $N=2$ and large $\widetilde L$.
In Fig.~\ref{fig:N_Nhalf}, we see that both systems share the same KLD. Note that the mapping $(N,\widetilde L \gg 1) \to (N/2,\widetilde L = 0) $  holds for any even $N$, as seen in Fig.~\ref{fig:N_Nhalf} for a couple of cases $(N=2,6,8)$.

\begin{figure}[htp]
	\centering
	\includegraphics[width=0.48\textwidth]{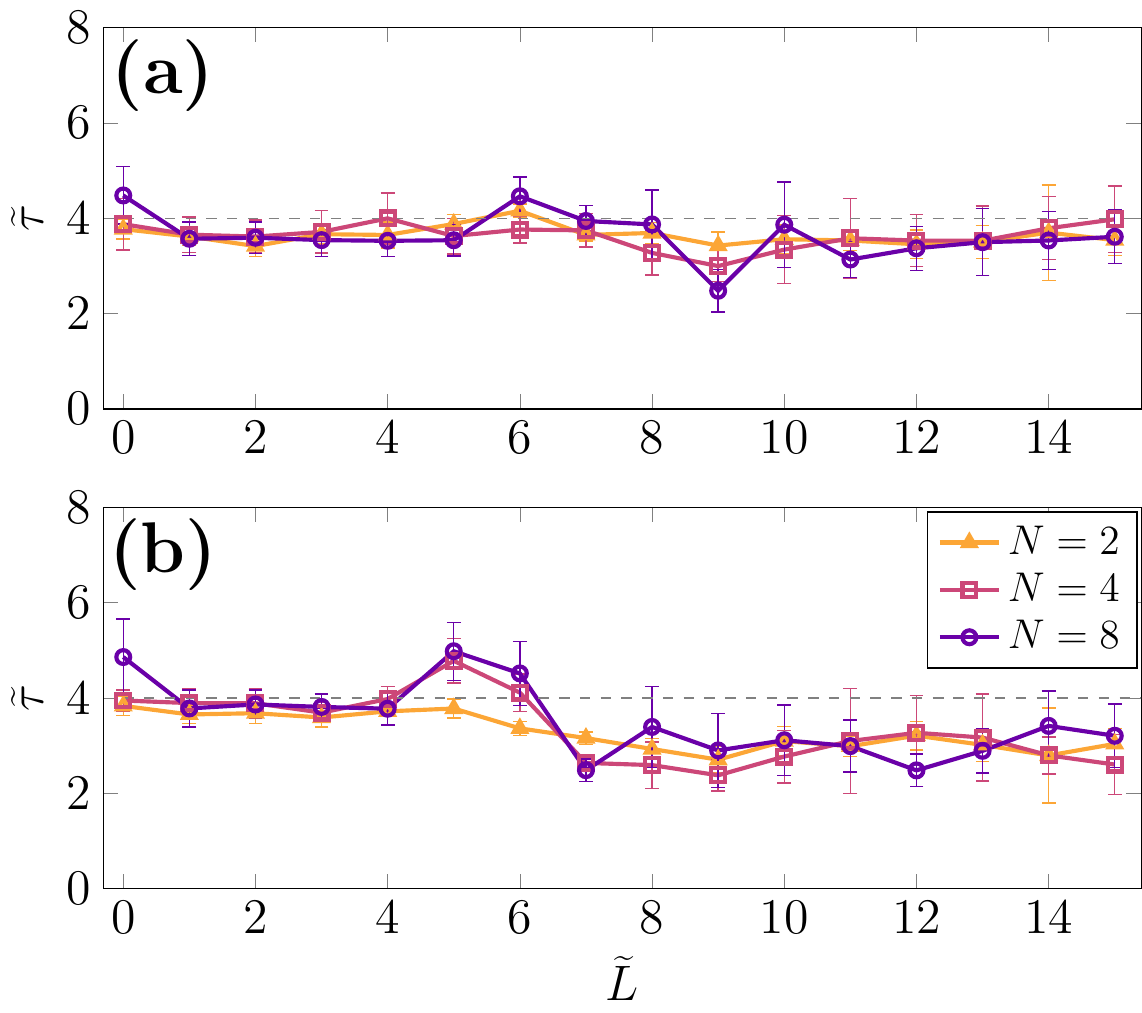}%
	\caption{Relaxation time $\widetilde{\tau}$ estimation for $N = 2,4,8$ counterions as a function of colloid distance $\widetilde{L}$ for two initial conditions (Eq.~\eqref{eq_IC}): (a) symmetric and (b) asymmetric. The straight lines are guides to the eye.
	}%
	\label{fig:Neven_sym_asym}%
\end{figure}

Figure \ref{fig:Neven_sym_asym} shows the simulation based estimation for the relaxation time as a function of $\widetilde{L}$, for several even $N$. 
The results for $N\geq 2$ give evidence that the relaxation time has constant value $\widetilde{\tau} = 4$ for any $\widetilde{L}$. This is seen for both initial conditions (Eq.~\eqref{eq_IC}) and suggests that $\widetilde{\tau}$ is independent of both $N$ and $\widetilde{L}$. Additionally, physical arguments lead to the same conclusion in limiting cases,
such as the aforementioned mapping $(N,\widetilde L \gg 1) \to (N/2,\widetilde L = 0) $.
In Figs. \ref{fig:OddAsymmTime} and \ref{fig:Neven_sym_asym} we observe that for every $N$ considered, the relaxation time is $\widetilde{\tau} = 4$ when the colloids are together ($\widetilde{L} = 0$). Therefore,  it is plausible that when $N$ is even and the colloids are sufficiently separated, the relaxation time is also $4$.
	
An argument to understand the $\widetilde L$ independence phenomenon goes as follows: for even $N$, the left and right moieties of the system, each being neutral, are decoupled. The only $L$ dependence in the problem
therefore arises through the initial condition, that does not affect the relaxation rate measured.
Of course, initial conditions with counterions more distant from their native colloid will collectively take longer to relax than if all ions would start, say, from a typical equilibrium double-layer distance from the colloid. Yet, this difference will manifest itself in the short time evolution, 
and leave unaffected the large-time decay rate. Indeed, ions starting far away from their native colloid will undergo a ballistic motion on average. Such behavior occurs in a finite time, whereas the equilibration of the double-layer is an exponential decay.
The $N$ independence is traced to the double-layer's length, which is  practically constant in $N$ when expressed in units of $l_{\text{B}}$, as done here \cite{TellezTrizac2015}. Then, for any number of counterions, the space to be probed is the same.

Our results are limited by the minimum value obtained for $D_{\text{KL}}$, which is mainly determined by the binning of positions used to compute the histograms at each time. As $\widetilde t$ increases, so does the number of simulations needed. This especially affects the asymmetric initial condition that requires a larger $\widetilde{t}$ to be in the regime where $\log D_{\text{KL}}$ is ruled by a linear term in time and with a subdominant (yet relevant) term $\log \widetilde{t}$. In that regime the method for determining the relaxation time is very effective, as previously seen in $N$ odd and for the analytic solution when $N=1$.

To summarize, two essentially different scenarios are identified depending on the parity of $N$. The odd $N$ case has a qualitative behavior identical to the single counter-ion case. When $\widetilde{L} < \widetilde{L}^*(N)$ then $\widetilde{\tau} = 4$, where $\widetilde{L}^*(N) \geq \pi$ is some length dependent on $N$. For large $\widetilde{L}$ the relaxation time is quadratic on the distance: $\widetilde{\tau} \propto \widetilde{L}^2$, showing typical diffusive behaviour. On the other hand, when $N$ is even, we gave evidence that the relaxation time is both $N$- and $\widetilde{L}$-independent.  
In all cases, the relaxation dynamics follow the exponential decay outlined in Eq.~\eqref{KLD_relaxtime}, which is fundamentally different from the long-time behavior shown by similar systems with weak-coupling, that are amenable to a mean-field treatment. We now establish the connection between these two behaviours.


\section{Mean-Field dynamics}\label{sec:meanfield}

In this section, we consider the dynamics within a mean-field treatment. This approach is justified in the weak-coupling regime; in our case, this corresponds to taking the limit $e\to 0$ and $N\to \infty$ while keeping $Ne$ fixed (i.e. the colloids charge). This was shown analytically using different formulations in \cite{dean,TellezTrizac2015}, for an 
equilibrium system with counterions limited to remain in the inter-colloidal space. We generalize here to the dynamics. 

\begin{figure}[htp]
	\begin{center}
	\includegraphics[width=0.48\textwidth]{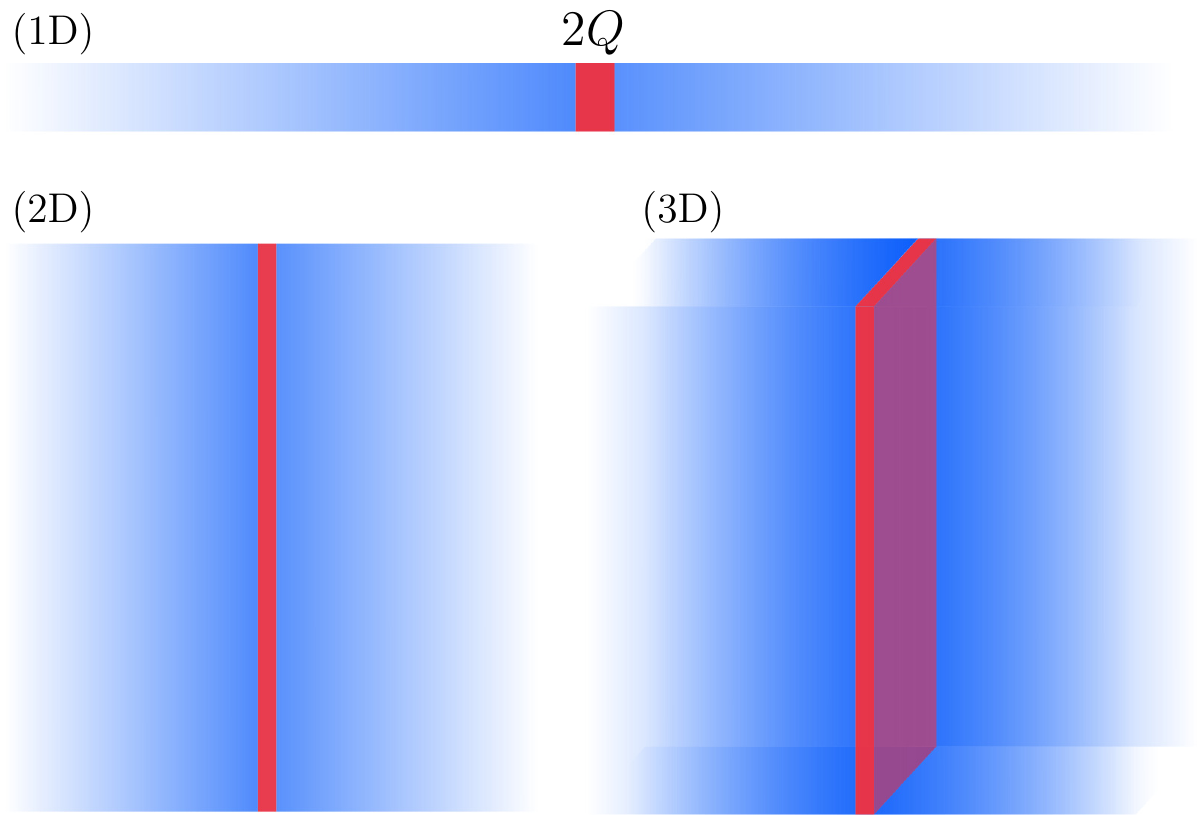}%
		\caption{Sketch for three mean-field systems described by the same equation: a point charge in 1D (top), a uniformly charged line in 2D (bottom left) and a uniformly charged plate in 3D (bottom right).  In red are the colloids (point/line/plate) and in blue the counterions which have a continuous charge distribution.  
		At equilibrium, exactly half of the distribution lies within a Gouy-Chapman length from its colloid.
		Under the appropriate time and length rescaling, the equations in (PB) and 
		out of equilibrium (PNP) are identical for these three systems.
		}
		\label{fig:mf_sketch}
	\end{center}
\end{figure}

To keep the discussion within reasonable bounds, we focus on the case $L=0$. The mean-field problem then admits a simple exact solution, be it at equilibrium or out of equilibrium, that allows to draw conclusions from analytic expressions and to assess how a discrete system approaches the mean-field regime. We will now treat the counterions as a continuous charge distribution rather than a discrete set of point-particles. 
We consider a system made of two colloids at the origin, each of charge $Qe$, together with the counterion distribution. The top illustration in Fig.~\ref{fig:mf_sketch} envisions the mean-field system, were we take the distance between colloids to be zero. Although our discussion is centered in a 1D Coulomb gas, the mean-field results in this section also describe systems in 2D and 3D, as depicted in Fig.~\ref{fig:mf_sketch}. Since our system is symmetric with respect to the origin, we restrict to $x>0$.  
First, we analyze the equilibrium behavior, which follows from solving the Poisson-Boltzmann equation \cite{andelman2006}:
\begin{equation}
\phi_{\text{PB}}''(x)  = \frac{2n_0e}{\epsilon} \exp \{e \beta\phi_{\text{PB}}(x)\}, 
\label{PB}
\end{equation}
where $n_0$ is a normalization constant. The previous equation is complemented with the boundary condition $\phi_{\text{PB}}'(0^+) = -2Qe/\epsilon$, which accounts for the colloidal charge. Equation \eqref{PB} is obtained by assuming the counterions have a Boltzmann distribution $n_0 \exp(-e \beta\phi_{\text{PB}})$ and then inserting this density into Poisson equation.
The solution to Eq.~\eqref{PB} is
\begin{equation}
n_{\text{PB}}(x) = \frac{Q}{b(1+x/b)^2},
\label{cPB}
\end{equation}
where $b=\epsilon/\beta Q e^2$ is the Gouy-Chapman length. In 1D, this quantity is proportional to the Bjerrum length $b = l_{\text B}/Q$. 

\begin{figure}[!htp]
\centering
\includegraphics[width=0.48\textwidth]{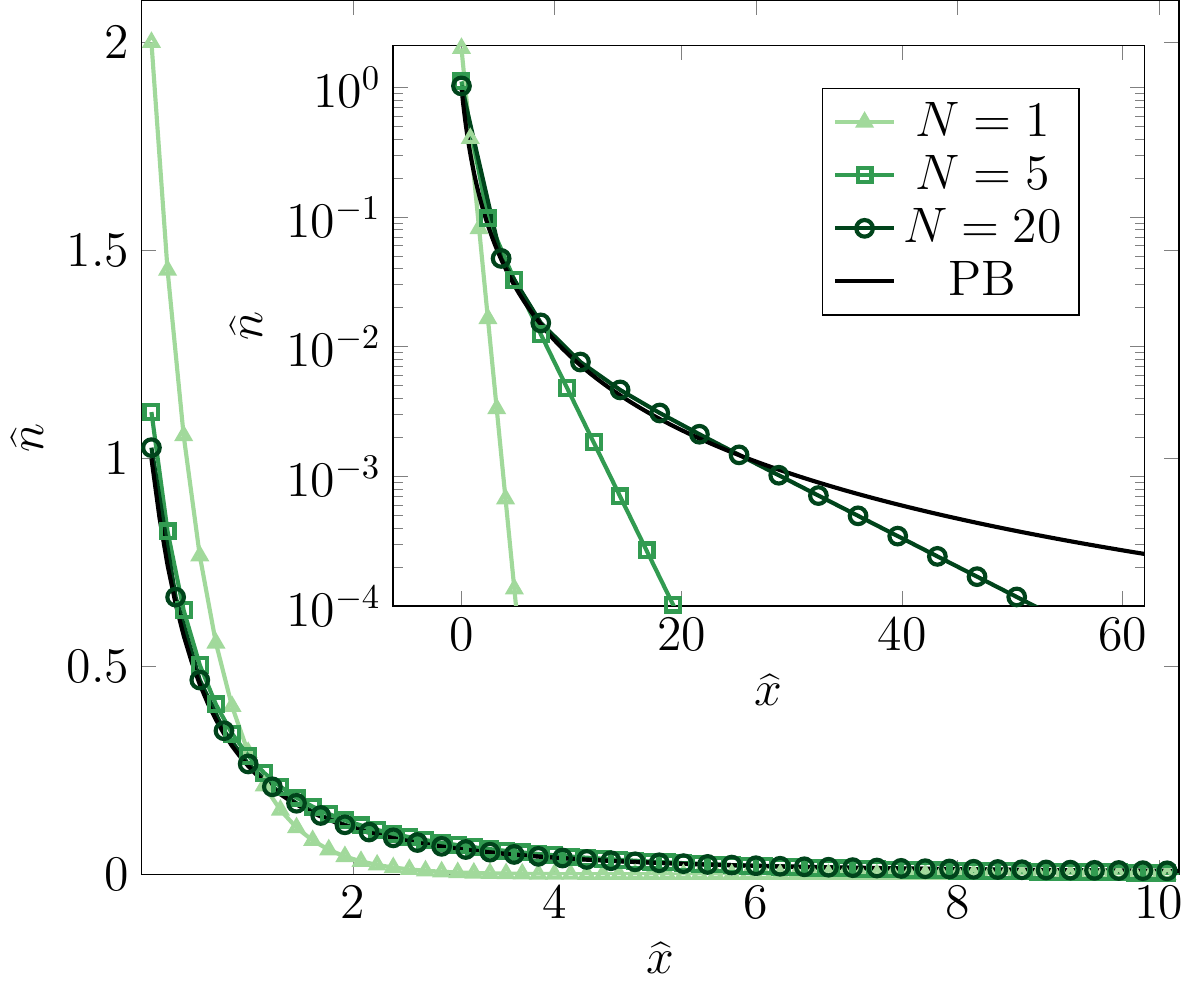}%
\caption{Equilibrium density profile $\widehat{n}_{\infty}(\widehat{x})$ (Eq.~\eqref{N_n}) with $L=0$ for $N=1,5,20$. The solid line is for the Poisson-Boltzmann solution   $\widehat{n}_{\text{PB}}$. Note that for as few as $N=5$ counterions, the mean-field theory (i.e. $\widehat{n}_{\text{PB}}$) is a good approximation,
except in the tail.
The inset shows the same plot in logarithmic scale to emphasize the tails: increasing $N$ (while keeping $Ne$ fixed) augments the region of overlap between the discrete and mean-field models. }%
\label{fig:cPBvsN}%
\end{figure}

In order to discuss the mean-field results, we rescale the position by the $b$ length: $\widehat{x} = x/b$ and introduce the reduced density $\widehat{n} =  (b/Q)n$. Figure \ref{fig:cPBvsN} features the Poisson-Boltzmann density and three exact profiles for $N=1,5$ and $20$, computed using Eq.~\eqref{N_n}. The plot shows that by increasing $N$, the discrete result approaches the Poisson-Boltzmann density. Beyond a few particles (e.g. 5), the effect of increasing $N$ mostly affects the tail behavior.       

We now turn to the mean-field dynamics. For this purpose, 
we solve the dynamical generalization of the Poisson-Boltzmann equation, the Poisson–Nernst–Planck (PNP) equations \cite{hunterbook}:
\begin{equation}
\begin{split}
-\partial^2_x\phi_{\text{\tiny PNP}}  &= -2\,e\,n_{\text{\tiny PNP}}/\epsilon, \\
\partial_t n_{\text{\tiny PNP}}  &= D\,\partial^2_x n_{\text{\tiny PNP}}  - \mu\,\partial_x \left(n_{\text{\tiny PNP}}\,\partial_x \phi_{\text{\tiny PNP}} \right),
\end{split}\label{electrokinetic_u}
\end{equation}
where  $D$ is the diffusion constant and $\mu$ the electric mobility of the counterions. These constants are related by the Einstein relation $D= (\mu/e) k_{\text{B}}T$. Again, we account for the colloidal particle by fixing the electric field at the origin for all times:
\begin{equation}
\partial_x  \phi_{\text{\tiny PNP}}(0^+,t) = -{2 Q e}/{\epsilon}.
\label{interface_bc}
\end{equation}

We proceed to use the ``hat'' rescaling, which was already defined for position and ionic density. The remaining time and potential units are given by $ \widehat{t} = t/({b^2/D})$ and $\widehat{\phi}_{\text{\tiny PNP}} = e\beta \phi_{\text{\tiny PNP}}$ respectively. In \cite{Golestanian2000}  it was shown that integrating twice the second equality in Eq.~\eqref{electrokinetic_u} and assuming a symmetric initial condition, the electrokinetic equations become
\begin{equation}
    \partial_{\widehat t} \widehat{\phi}_{\text{\tiny PNP}}(\infty,\widehat t) -\partial_{\widehat t} \widehat{ \phi}_{\text{\tiny PNP}}(\widehat x,\widehat t) = \partial_{\widehat x}^2 \widehat{ \phi}_{\text{\tiny PNP}}(\widehat x,\widehat t) +\frac{[\partial_{\widehat x}\widehat{ \phi}_{\text{\tiny PNP}}(\widehat x,\widehat t)]^2}{2} 
    \label{electrokinetic2}
\end{equation}
In \cite{Golestanian2000} 
a Cole-Hopf transformation was used. It
amounts to introducing $W$ such that:
\begin{equation}
\widehat{\phi}_{\text{\tiny PNP}}({\widehat{x}},{\widehat{t}}) -\widehat{\phi}_{\text{\tiny PNP}}(\infty,{\widehat{t}}) = -2 \ln W({\widehat{x}},{\widehat{t}}) ,
\label{eq:CH}
\end{equation}
which brings interesting simplifications, see below.
Note that up to a constant, we have $\widehat{\phi}_{\text{\tiny PNP}}(\widehat{x},{\widehat{t}}\to \infty) =\widehat{\phi}_{\text{PB}}(\widehat{x}) = 2 \ln (1+\widehat{x})$. As a result, $\widehat{\phi}_{\text{\tiny PNP}}(\infty,{\widehat{t}})$ is divergent for
$\widehat{t} \to \infty$. This nevertheless does not lead to any physical
difficulty nor any ill-posedness in \eqref{eq:CH}: an additive $x$-independent 
term in the potential $\phi_{\text{\tiny PNP}}$ does not change the charge density. 
What we may gather from the above remark is that
\begin{equation}
    \lim_{\widehat x\to \infty}  W({\widehat{x}},{\widehat{t}}) \,=
    \, 1 \quad \hbox{while}\quad \lim_{\widehat t\to \infty}  W({\widehat{x}},{\widehat{t}}) \,=\, 0.
    \label{eq:limitsnoncumuting}
\end{equation}
We then substitute the Cole-Hopf transformation into Eq.~\eqref{electrokinetic2}, and it doing so it becomes
\begin{equation}
\partial_{\widehat{t}} W = \partial_{\widehat{x}}^2 W,
\label{W}
\end{equation}
which is equipped with the following Robin boundary condition:
\begin{equation}
\partial_{\widehat{x}} W(0^+,{\widehat{t}}) - W(0^+,{\widehat{t}}) = 0,
\end{equation} 
as follows from Eq.~\eqref{interface_bc}. 

In order to find $W$, we can use the reflection method where we introduce a function $W_0$, defined from a modification of the initial condition $W(\widehat x,0)$.
The idea is to restrict to $\widehat x \geq 0$, and to continualize 
$W$ to $\widetilde x <0$ in a convenient fashion. 
In order to satisfy the Robin boundary condition, the function $W_0$ is taken such that $\partial_{\widehat{x}} W_0({\widehat{x}}) - W_0({\widehat{x}}) $ is an odd function, and consequently vanishes at the origin. We also require $W_0$ to coincide the initial condition for $\widehat{x}>0$. The function that satisfies the previous requirements is given by:
\begin{equation}
W_0({\widehat{x}}<0) = W_0(0) \e^{{\widehat{x}}}+\e^{{\widehat{x}}} \int_0^{|{\widehat{x}}|} \e^{y}\left[W_0'(y) -W_0(y) \right] dy,
\end{equation}
where $W_0(\widehat{x}\geq 0) = W(x,0)$. Then, the solution to Eq.~\eqref{W} reads
\begin{equation}
W({\widehat{x}},{\widehat{t}}) = \int_{\mathbb{R}} dy\, W_0(y) \times \frac{1}{\sqrt{4\pi {\widehat{t}}\,\, }} \exp\Big\{-\frac{({\widehat{x}}-y)^2}{4{\widehat{t}}}\Big\} ; \label{Wsol}
\end{equation}
this expression is used for $\widehat{x}>0$. Once we have computed $W$, we obtain the ionic density through 
\begin{equation}
\widehat{n}_{\text{\tiny PNP}}({\widehat{x}},{\widehat{t}}) = \partial^2_{\widehat{x}} \big[-\ln W({\widehat{x}},{\widehat{t}})\big],
\end{equation}
which follows from the Cole-Hopf transformation and 
Poisson's equation.
The asymptotic behavior as $\widehat{t}\to \infty$ of the PNP density profile is given by the Poisson-Boltzmann solution: $\widehat{n}_{\text{\tiny PNP}}({\widehat{x}},{\widehat{t}}\to \infty) = \widehat{n}_{\text{PB}}({\widehat{x}})$.

We now focus on an initially localized density profile $\widehat{n}_{\text{\tiny PNP}}({\widehat{x}},0) = \delta({\widehat{x}})$, which leads to the following $W$ function:
\begin{equation}
W(\widehat{x},\widehat{t})=  \text{erf}\Big(\frac{\widehat{x}}{2 \sqrt{\widehat{t}}}\Big)- \e^{\widehat{t}+\widehat{x}}\, \text{erf}\Big(\frac{2 \widehat{t}+\widehat{x}}{2 \sqrt{\widehat{t}}}\Big)+ \e^{\widehat{t}+\widehat{x}}. \label{Wdelta}
\end{equation}
The previous expression has to be handled with caution since in general spatial and time limits cannot be exchanged, see Eq.~\eqref{eq:limitsnoncumuting}. 
We can then analyze the asymptotic time behavior of Eq.~\eqref{Wdelta} by computing the large time expansion:
\begin{equation}
W(\widehat{x},\widehat{t}) \underset{\widehat{t}\to \infty}{\sim} \frac{ 1+\widehat{x}}{\sqrt{\pi \widehat{t} \,\,}} + \mathcal{O}(\widehat{t}^{-3/2}).
\end{equation}

\begin{figure}[!htp]
\centering
\includegraphics[width=0.48\textwidth]{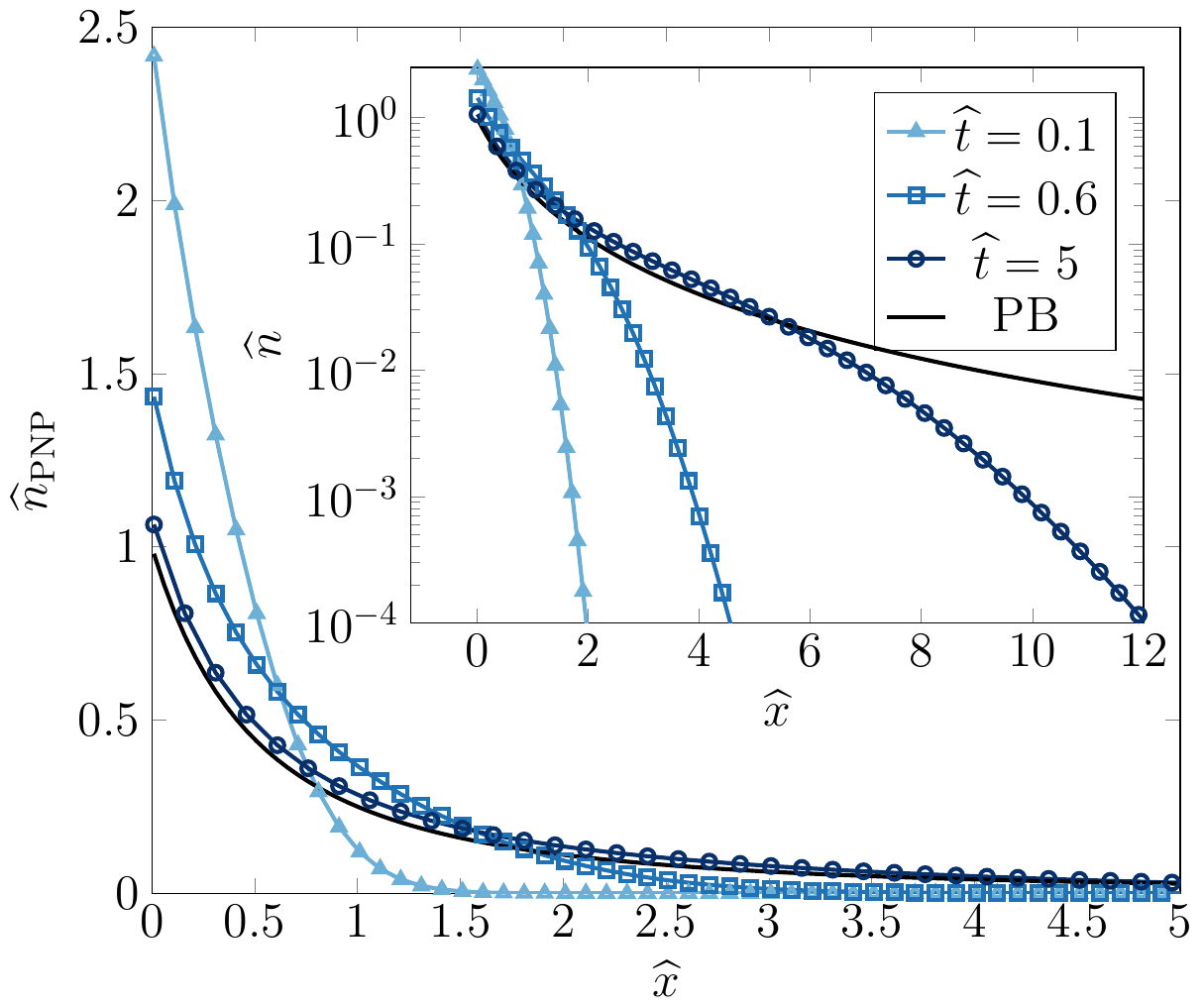}%
\caption{Ionic density at times  $\widehat{t}=0.1,0.6,5$ for the Poisson-Nernst-Planck (PNP) mean-field dynamics (markers) and the equilibrium Poisson-Boltzmann (PB) solution (solid). The initial condition is $\widehat{n}_{\text{\tiny PNP}}(\widehat{x},0)= \delta(\widehat{x})$. For $\widehat{t} = 5$, the dynamical PNP solution is close to its equilibrium counterpart. The inset shows the same plot in linear-log scale to emphasize the tails of the distributions, where lies the largest difference between PB and PNP at large times.}%
\label{Fig:PNP_t}%
\end{figure}

The mean-field density profile follows, as
\begin{equation}
\begin{split}
\widehat{n}_{\text{\tiny PNP}}({\widehat{x}},{\widehat{t}})  = \frac{\e^{-\widehat{x}^2/4\widehat{t}} }{\sqrt{\pi\widehat{t}\,} \Big[\text{erf}\Big(\frac{{\widehat{x}}}{2 \sqrt{{\widehat{t}}}}\Big)+\e^{{\widehat{t}}+{\widehat{x}}}\, \text{erfc}\Big(\frac{2 {\widehat{t}}+{\widehat{x}}}{2 \sqrt{{\widehat{t}}}}\Big)\Big]}\\
-\frac{\e^{{\widehat{t}}+{\widehat{x}}} \,\text{erf}\Big(\frac{{\widehat{x}}}{2 \sqrt{{\widehat{t}}}}\Big) \text{erfc}\Big(\frac{2 {\widehat{t}}+{\widehat{x}}}{2 \sqrt{{\widehat{t}}}}\Big) }{ \Big[\text{erf}\Big(\frac{{\widehat{x}}}{2 \sqrt{{\widehat{t}}}}\Big)+\e^{{\widehat{t}}+{\widehat{x}}}\,\text{erfc}\Big(\frac{2 {\widehat{t}}+{\widehat{x}}}{2 \sqrt{{\widehat{t}}}}\Big)\Big]^2},
\end{split}
\label{ntPNP}
\end{equation}
which admits the following expansion at large times
\begin{equation}
\widehat{n}_{\text{\tiny PNP}} \underset{{\widehat{t}}\to \infty}{\sim} \frac{1}{ (1+{\widehat{x}})^2}\Big[1+\frac{{\widehat{x}}^3+3 {\widehat{x}}^2+3 {\widehat{x}}+3}{6  ({\widehat{x}}+1)\widehat{t}}+\mathcal{O}\Big(\frac{1}{{\widehat{t}}^{3/2}}\Big)\Big], 
\end{equation}
where the equilibrium distribution $\widehat{n}_{\text{PB}}({\widehat{x}})$ is ultimately reached, as anticipated.  The next leading order decays as inverse
time, and reveals that the corrections to Poisson-Boltzmann are of order $\sim \widehat{x}^2/\widehat{t}$ at large distances;
this indicates that the tails take longer to converge towards equilibrium. Furthermore, it implies that it takes a time $\widehat{t} \sim \widehat{x}^2$ for the distribution to reach equilibrium at distance $\widehat{x}$ from the colloid. Fig.~\ref{Fig:PNP_t} shows how the PNP solution approaches the equilibrium distribution: starting by the region close to the colloid and then spreading outwards. Equation \eqref{ntPNP} also admits an expansion when $\widehat{x} \to \infty$:
\begin{equation}
    \widehat{n}_{\text{\tiny PNP}} \underset{{\widehat{x}}\to \infty}{\sim} \frac{\e^{-\widehat{x}^2/4\widehat{t}}}{\sqrt{\pi \widehat{t}\,\,\,}} \Big[1-\frac{2 \widehat{t}}{\widehat{x}}+\mathcal{O}\Big(\frac{1}{{\widehat{x}}^{2}}\Big)\Big], 
\end{equation}
which shows that for a finite time, the distribution has Gaussian tails. Then, the algebraic decay is featured exclusively at equilibrium.  

\begin{figure}[!htp]
\centering
\includegraphics[width=0.477\textwidth]{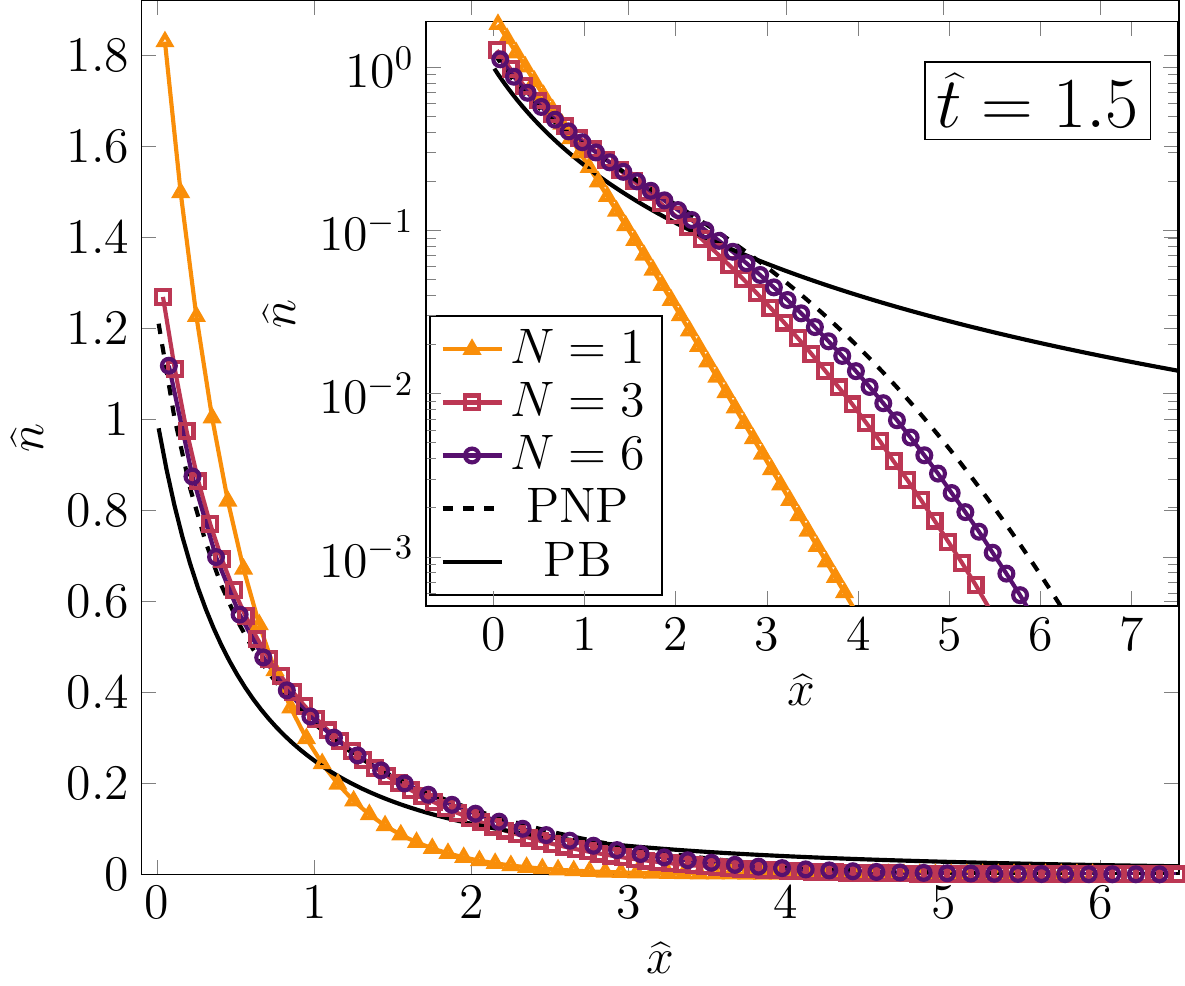}%
\caption{Ionic density at time  $\widehat{t}=1.5$, for the Poisson-Nernst-Planck (PNP) mean-field dynamics (dashed) and the discrete $N$ counterion  simulation with $N=1,3,6$ (markers). The initial condition for the different cases consists of all the particles localized at ${\widehat{x}}=0$. The equilibrium Poisson-Boltzmann (PB) mean-field solution $\widehat{n}_{\text{PB}}$ (solid) is given for reference. The insets show the ionic density in logarithmic scale to magnify the behavior on the tails, where the discrete case departs from the PNP solution.
}%
\label{fig:PNP_N_t-1p5}%
\end{figure}

\begin{figure}[!htp]
\centering
\includegraphics[width=0.477\textwidth]{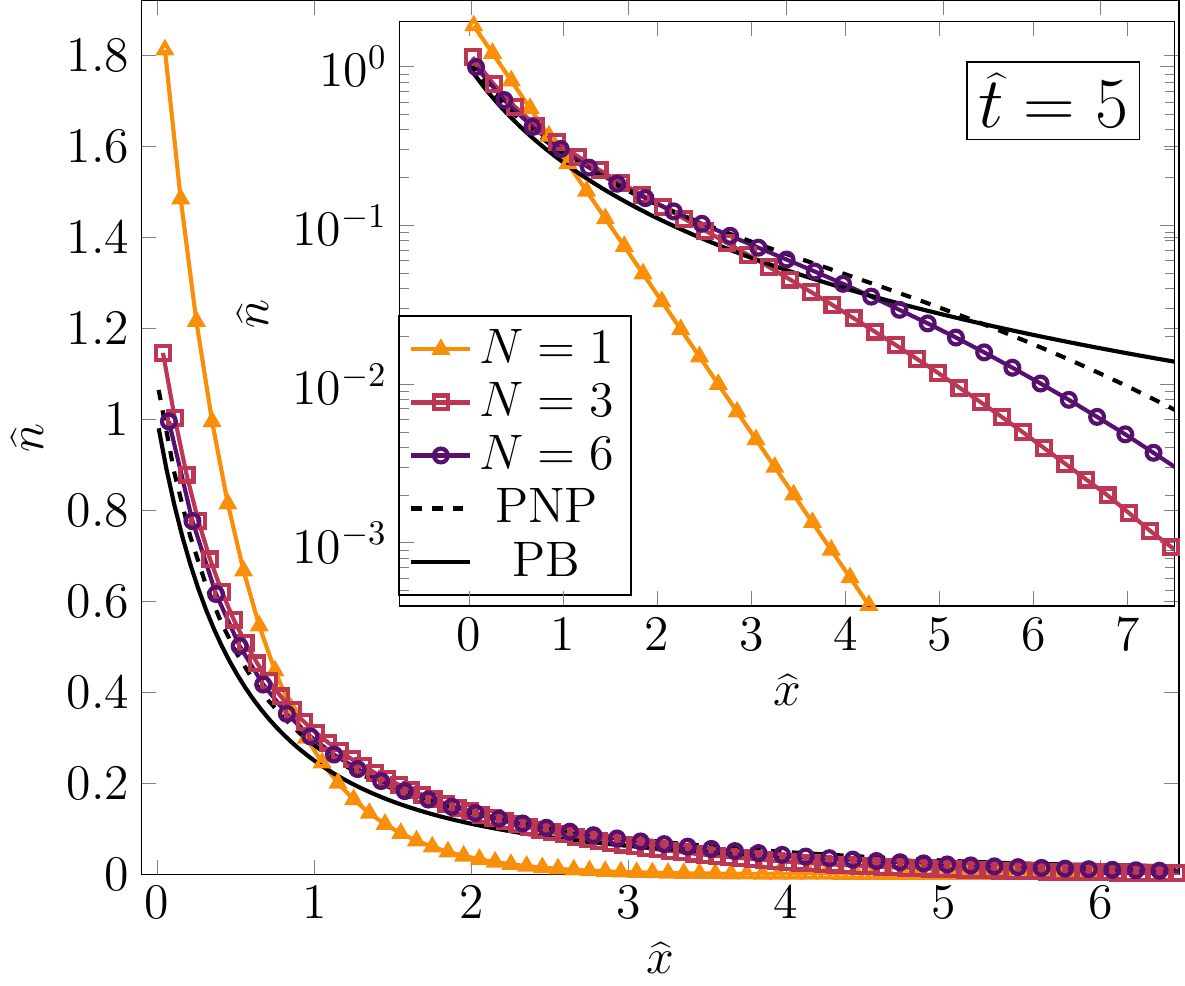}%
\caption{Same as Fig.~\ref{fig:PNP_N_t-1p5} with $\widehat{t}=5$.}%
\label{fig:PNP_N_t-5}%
\end{figure}

We move on to compare the dynamics of the discrete counterion model and the PNP solution. In Figs.~\ref{fig:PNP_N_t-1p5} and \ref{fig:PNP_N_t-5},
the exact discrete results are plotted at different times.
For as few as $N=3$ counterions, the mean-field theory becomes operational,
if we exclude the tail. Besides, the larger $N$, the closer to mean-field the
tail behaves.


\begin{figure}[!htp]
\centering
\includegraphics[width=0.48\textwidth]{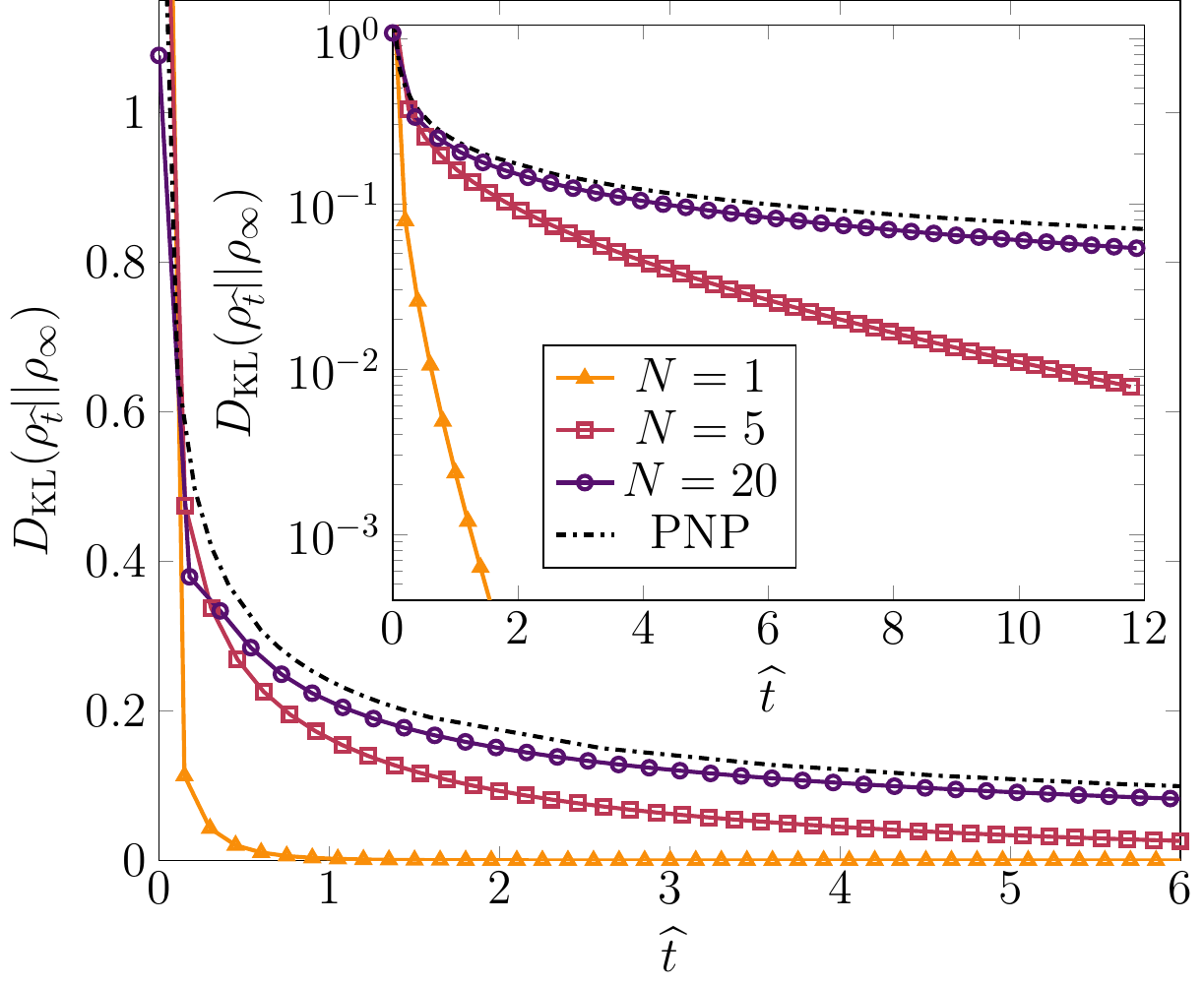}%
\caption{Kullback-Leibler divergence for $N=1,5,20$ (marks) and for the Poisson-Nernst-Planck (PNP) mean-field dynamical solution (dashdotted). The Kullback-Leibler divergence is taken between the dynamical distribution and its corresponding equilibrium state. Note how this observable converges to the PNP curve as the number of counterions increases. 
}%
\label{fig:PNP_DKL}%
\end{figure}

Finally, we discuss the characteristic relaxation time. The absence of an exponential decay means, that the PNP dynamics is ruled by an infinite characteristic time. To see how this matches with the finite $N$ exact results,
we examine how the characteristic time $\widehat{\tau}$ behaves when 
$N$ increases while keeping the colloid charge $Ne$ ($2Q=N$ in the discrete case) fixed. For $L=0$, we have found $\widetilde{\tau} \sim 4$, for even and odd values of $N$. This means that $\widehat{\tau} \sim  N^2$. Therefore, in the mean-field limit $N\to \infty$ and $e\to 0$ with fixed $Ne$, the characteristic time diverges, 
to yield the PNP result of a diverging scale. Figure \ref{fig:PNP_DKL} illustrates how an observable, the Kullback-Leibler divergence, approaches the mean-field as $N$ increases while keeping $Ne$ fixed. 

\section{Misfit Counterion Transport Time }
\label{sec:middle_counterion_trans}

We have seen that the relaxation dynamics depends on whether the number of ions is even or odd, with the slowest relaxation occurring in the odd case. In order to investigate this further, let us revisit the case where we have an odd number of counterions $N=2M-1$ with $M> 1$, and focus on the misfit ion. If we label the ions from $1$ to $N$ according to their increasing position, the misfit ion is the middle one or $M$th ion. Due to the nature of 1D Coulomb interactions and the charge neutrality of the system, there is no net drift force acting on the $M$th ion whenever it lies between the two colloids. As a result, it undergoes free Brownian motion until it collides with either of its neighbours.

\subsection{Transport time distribution}

We consider the time and distance between collisions of the $M$th ion with its neighbours as follows. Suppose that the system is in equilibrium and that the misfit ion collides with, say, the $(M-1)$th ion (its left neighbor). We record the time and position of this collision and let the system evolve until the $M$th ion collides with its right neighbor (the $(M+1)$th ion). We then calculate the time interval and distance between collisions. Finally, we record the position and time of this new collision and repeat the process for the next collision with the $(M-1)$th ion. In this manner, we record $10^6$ samples of times and distances, obtaining the corresponding sample averages and standard deviations. Note that we do not record data from successive collisions with the same neighbor. For example, after recording a collision with the left neighbor, we do not record any new collisions with the left neighbor until a collision with the right neighbor has occurred. In doing so, we define the transport time distribution between double-layers.

\begin{figure}[htp]
\begin{center}
    \includegraphics[width=0.48\textwidth]{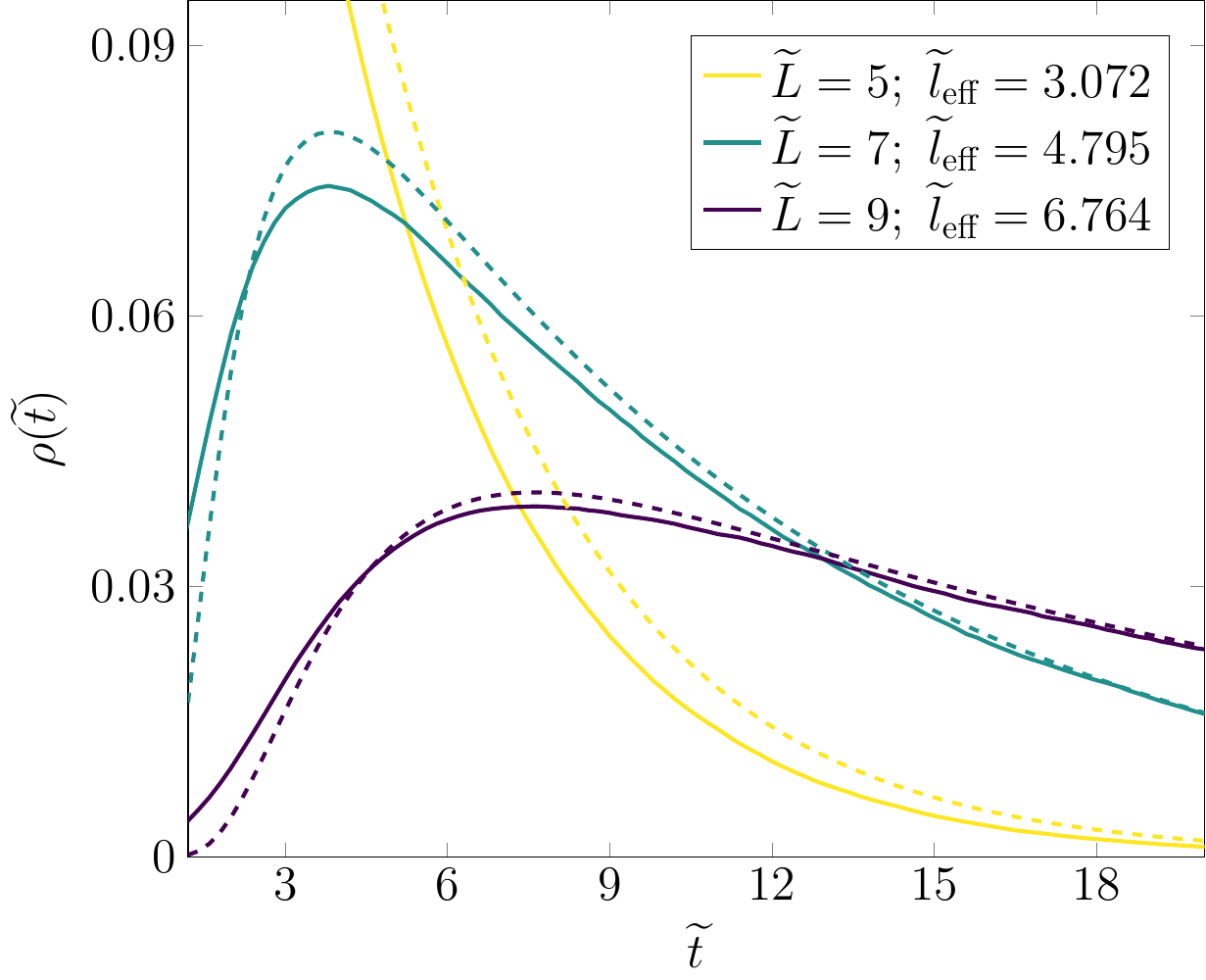}%
	\caption{Transport time distribution for the misfit ion (solid lines) and 
	fitted fixed wall model time distribution $q_{\widetilde{l}_\text{eff}}(\widetilde{t})$ (dashed lines), for $N=25$.}
	\label{fig:TimeN25}
\end{center}
\end{figure}

With $N=25$, the time between collisions is distributed as shown in Fig.~\ref{fig:TimeN25}. We observe that for $\widetilde{L}=3$ the collision times take mostly small values, while for $\widetilde{L}\geq7$ the distributions show noticeably longer tails, indicating that the $M$th ion requires a much longer time to go from one of its neighbors to the other.
Besides, it is seen in Fig.~\ref{fig:TimeN25} that
the short-time collision probability diminishes as $\widetilde{L}$ increases. This is due to the minimum distance the counterion needs to find another particle:
at large $\widetilde{L}$, all ions except the misfit are located in the vicinity of the colloids. The misfit thus needs to travel a distance of order $\widetilde{L}$
to collide with a new partner (see more details below).
We then calculate the average transport time; 
the results, shown in Fig.~\ref{fig:DiffL}a
do exhibit the expected diffusive $\widetilde{L}^2$
scaling for large $\widetilde{L}$.


\subsection{Effective model for the misfit's free space}

We are interested in estimating the length of the available free space for the misfit. This is expected to be given by $\widetilde{L}$ minus two double-layer sizes (one on the right, another on the left hand side). 
While we do not have the explicit form of the collision time distribution as a function of $\widetilde{L}$, we can resort to the boxed particle model 
introduced in Sec.~\ref{sec:firstpassage}. We then estimate the transport time as a function of the distance between walls $\widetilde{l}$, namely $\tau_{\widetilde{l}}$, as given in \eqref{eq:tau_l}. We present a derivation of its distribution in Appendix~\ref{appx:transport_time}. 
Using this object, we can investigate the effective size of the system between the ion clouds using the $M$th ion as a probe. We perform a parameter fit to find the value of $l$ for which $q_{\widetilde{l}}(\widetilde{t})$, which is the transport distribution (Eq.~\eqref{eq:trtimedist}) for a Brownian particle in a 1D box of length $\widetilde{l}$, reproduces the collision time distributions most closely, and we denote it by $\widetilde{l}_{\text{eff}}$. The result of the parameter fit is shown in  Fig.~\ref{fig:TimeN25}; as expected, the fixed wall model time distribution seems to reproduce the transport time distribution more closely as $\widetilde{L}$ grows.

\begin{figure}[htp]
\begin{center}
	\includegraphics[width=0.48\textwidth]{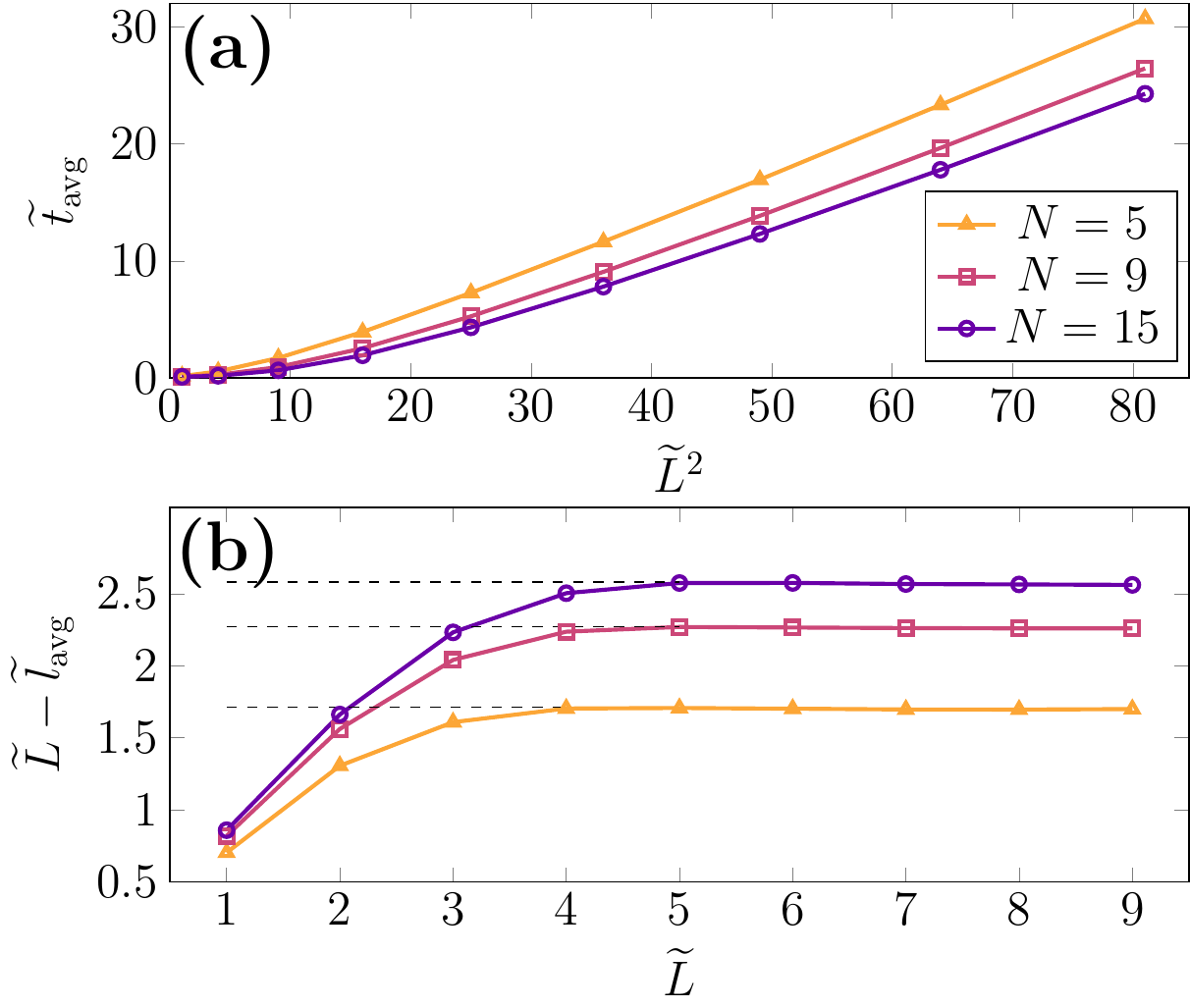}%
	\caption{(a) Average transport time as a function of $\widetilde{L}^2$. The 
	lines are guides to the eye.
	(b) Difference $\widetilde{L}-\widetilde{l}_\text{avg}$ for $N=5, 9$ and 15. The 
	lines are guides to the eye and the horizontal dashed lines show the asymptotic value as $\widetilde L \to \infty$. The equivalent plot for $\widetilde{L}-\widetilde{l}_\text{eff}$ (not shown) yields a similar behavior: a bounded monotonic increase to a slightly different terminal value, yet reaching it nearly for the same $\widetilde L$. 
	}
	\label{fig:DiffL}
\end{center}
\end{figure}

The first observation to be made about $q_{\widetilde{l}}(\widetilde{t})$ as expressed in Eq.~\eqref{eq:trtimedist} is that it obeys the scaling relation
\begin{equation}
q_{\widetilde{l}}(\widetilde{t})=\widetilde{l}^{-2} \, {\cal Q} \Big(\frac{\widetilde{t}}{\widetilde{l}^2}\Big),
\end{equation}
where $\mathcal{Q}(t)$ is a scaling function. This means time scales like $\widetilde{l}^2$.
Indeed,
\begin{equation}
\int_{0}^{\infty}\widetilde{t} \, q_{\widetilde l}(\widetilde t) \, d\widetilde t = \frac{\widetilde l^2}{2}.
\end{equation}
For large enough $\widetilde L$, this quadratic behavior is observed in Fig.~\ref{fig:DiffL}a.


For smaller values of $\widetilde{L}$, the boxed particle model fails to describe the situation. The behavior of $\widetilde{L}-\widetilde{l}_{\text{eff}}$ as a function of $\widetilde{L}$, resembles the one featured for $\widetilde{l}_{\text{avg}}$ (Fig.~\ref{fig:DiffL}(b)).
After calculating the coefficient of determination $R^2$ of each fit, we find that its value is quite low (less than 0.5 in several cases) for $\widetilde{L}<5$, while it lies consistently between 0.75 and 1 and shows a monotonically increasing behavior when $\widetilde{L}\geq 5$, that is, the fits become more reliable as $\widetilde{L}$ grows. Moreover, with increasing $\widetilde{L}$, $\widetilde{L}-\widetilde{l}_{\text{eff}}$ tends to a value that depends only on $N$, and as $N$ grows this value converges to a limit close to 2; this limit is roughly twice the double-layer size, which in the non-permeable case is exactly 1 \cite{TellezTrizac2015}.
To see this, we examined $\widetilde{L}-\widetilde{l}_{\text{eff}}$ as a function of $N^{-1}$. The $\widetilde{L}-\widetilde{l}_{\text{eff}}$ behaves as a linear function of $N^{-1}$ as $N\to\infty$,
which allows to extrapolate the asymptotic behavior
\begin{equation}
\lim_{N\to\infty}(\widetilde L-\widetilde{l}_\text{eff})=2.46\pm 0.01.
\end{equation}

\begin{figure}[htp]
\begin{center}
	\includegraphics[width=0.48\textwidth]{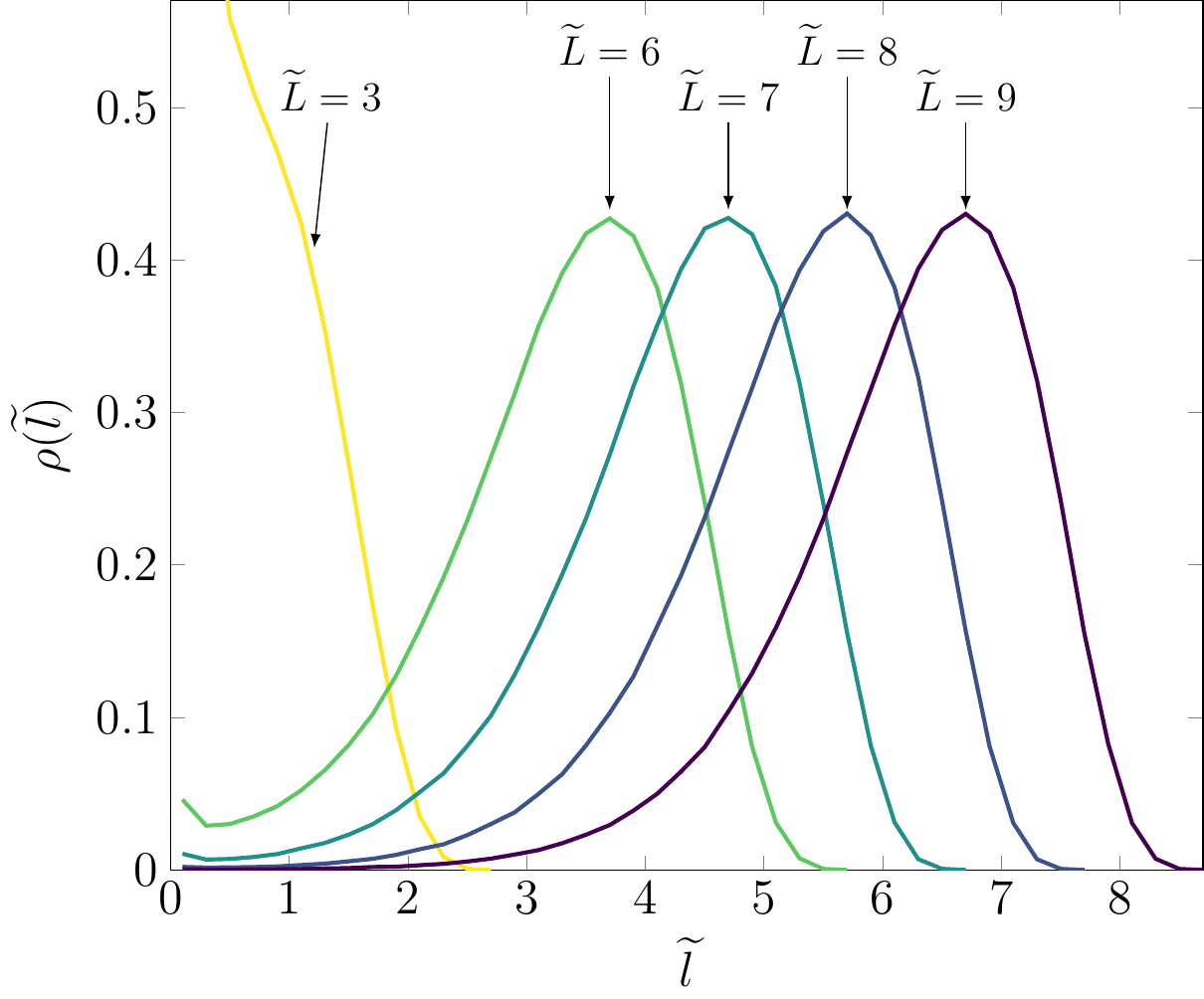}%
	\caption{Distribution of distances traveled by the misfit ion between the double-layers, for $N=25$ and various $\widetilde{L}$. 
	}
	\label{fig:DistN25}
\end{center}
\end{figure}

Finally, let us analyze directly the distribution of distances travelled by the misfit ion between collisions. In Fig.~\ref{fig:DistN25}, we show the results for a system of 25 counterions. When $\widetilde L=3$, a large fraction of the collisions occur at very small distances (less than 0.5) which results in a peak at the origin. For $\widetilde L \geq 7$, the distribution has no such peak. We also observe that that as $\widetilde L$ increases, the distribution shifts to the right without significant changes in the shape. 
We define $\widetilde{l}_{\text{avg}}$ as the average of the collision distance data, that is, the mean
of the collision distance distributions depicted in Fig.~\ref{fig:DistN25}. 
Performing a similar analysis as for $\widetilde{l}_{\text{eff}}$ yields similar asymptotic results for growing $\widetilde{L}$, see Fig.~\ref{fig:DiffL}(b). 
By considering  the difference $\widetilde L-\widetilde{l}_\text{avg}$ for $\widetilde L =5$ as a function of $1/N$, we observe that $\widetilde{l}_{\text{avg}}$ converges to a well-defined value, as $N$ tends to infinity. Similar to the behavior of $\widetilde{L}-\widetilde{l}_{\text{eff}}$, we see that $\widetilde L-\widetilde{l}_\text{avg}$ is a linear function of $N^{-1}$ as $N\to\infty$. From the fit we obtain
\begin{equation}
\lim_{N\to\infty}[\widetilde L-\widetilde{l}_\text{avg}(N,\widetilde{L})]=3.05\pm 0.01.
\end{equation}
As before, we expect this value to only vary slightly and converge as $\widetilde L\to\infty$ in the same way as $\widetilde L-\widetilde{l}_\text{eff}$.

\subsection{Misfit's role in the relaxation time}

\begin{figure}[htp]
\begin{center}
	\includegraphics[width=0.48\textwidth]{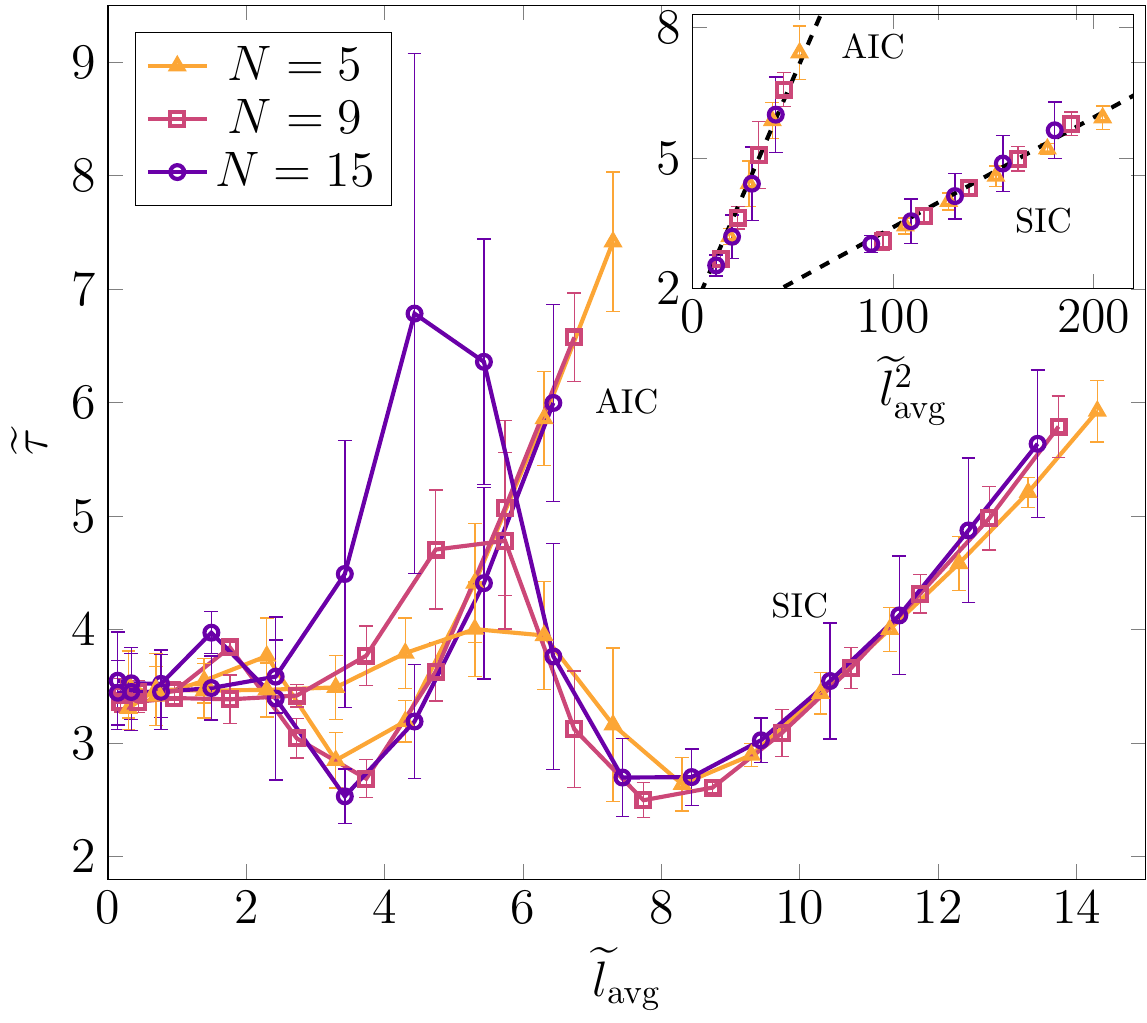}%
	\caption{Characteristic time for the asymmetric (AIC) and symmetric (SIC) initial conditions as a function of $\widetilde{l}_\text{avg}$. The straight lines are guides to the eye, and the error bars denote one standard deviation. In the inset, we plot the characteristic time as a function of $\widetilde{l}_\text{avg}^2$; the dashed lines correspond to linear fits.}
	\label{fig:ShiftOddSymmTime}
\end{center}
\end{figure}

In view of the analysis in Sec.~\ref{sec:1_counterion}, we expect the characteristic time to be a quadratic function of $\widetilde{L}$ when the latter is sufficiently large. In addition, Eq.~\eqref{eq:noddeigvl} indicates that the smallest discrete eigenvalue appears when $\widetilde{L}\geq \pi$ for $N=1$. In the present case, we may expect a similar behavior, whenever the free space available to the
$M$th counterion exceeds $\pi$; 
this length can be estimated by $\widetilde{l}_\text{avg}$.
As can be seen in Fig.~\ref{fig:ShiftOddSymmTime}, the behaviour of 
$\widetilde\tau$
in $\widetilde{l}_\text{avg}^2$ is well obeyed.
More specifically, we get for asymmetric initial conditions
\begin{equation}
\widetilde{\tau}(N,\widetilde{L})=A+B\ \widetilde{l}_\text{avg}(N,\widetilde{L})^2\label{eq:chartl2},
\end{equation}
with $A=1.47\pm 0.18$, $B=0.107\pm 0.005$, and $\widetilde{L}\geq 7$. Note that the coefficient $B$ is very close to the value predicted by Eq.~\eqref{eq:oddeigvlasymptotics}, which is $1/\pi^2\approx 0.1$. The curves in Fig.~\ref{fig:OddAsymmTime} plotted as functions of $\widetilde{l}_\text{avg}$ are also shown in Fig.~\ref{fig:ShiftOddSymmTime}, and we see that the characteristic time follows a single curve that does not depend on the number of counterions, provided it is odd. We observe that with the asymmetric initial conditions, for all curves, the characteristic time is close to $4$ when $\widetilde{l}_\text{avg}\leq \pi-1$, and that $\widetilde{\tau}$ follows a quadratic growth when $\widetilde{l}_\text{avg}\geq \pi+1$. 

We perform the same analysis for the symmetric IC, reaching similar conclusions,
see Fig.~\ref{fig:ShiftOddSymmTime}  \cite{lavgforlargeL}. Eq.~\eqref{eq:chartl2} is still obeyed, with $A=0.89\pm0.18$ and $B=0.0252\pm0.0012$. Again, the coefficient $B$ is consistent with the asymptotics given by Eq.~\eqref{eq:eveneigvlasymptotics}, which indicate a value of $1/(4\pi^2)\approx 0.0253$. Assuming in addition that the single counterion situation subsumes the key effects, we expect from Eq.~\eqref{eq:neveneigvl} an increasing behavior when $\widetilde{l}_\text{avg}\geq 3\pi$;
this is confirmed by the figure.
Moreover, when $\widetilde{l}_\text{avg}\leq 3$ the characteristic times seem to be independent of the number of ions, and very close to the completely-continuous spectrum value of 4 found for $N=1$.

The non-monotonicity of the characteristic time shown in Fig.~\ref{fig:ShiftOddSymmTime} for asymmetric IC with $2\leq \widetilde{l}_\text{avg}\leq 4$, and for symmetric IC with $3\leq \widetilde{l}_\text{avg}\leq 9$ may be the consequence of the finite number of samples taken during our simulations as well as the finite simulation time. As seen in Fig.~\ref{fig:1_counterion_texp_KLD}, the characteristic time estimated from simulations is quite close to the exact curves when $N=1$, but it can be observed that $\widetilde{\tau}$ is underestimated at small $\widetilde{L}$; also, estimations made for $N=1$ in a time interval and a noise level close to those used for $N>1$ reveal a similarly non-monotonic behavior in the intermediate $\widetilde{L}$ region, which we know is nonexistent in the exact $N=1$ results.
We conclude that these non-monotonicities are numerical artifacts, see also the end of
Appendix~\ref{appx_KLD}.

\section{Conclusion}\label{sec:conclusions}

We have determined the relaxation time $\tau$ of an overdamped electroneutral two-colloid system as a function of the colloid separation $L$ and the number of counterion $N$. 
The parity of $N$ determines whether $\tau$ depends on the distance between colloids. For $N$ odd, we found a behavior that mirrors the single counterion case:
$\tau \propto L^2/D$, where $D$ is the diffusion coefficient. From the Stokes-Einstein relation, $D$ grows linearly with temperature ($T_{\text{bath}}$), so that here, $\tau$ decreases upon 
increasing $T_{\text{bath}}$. 
On the other hand, for $N$ even, $\tau \propto l_{\text{B}}^2/D$ where $l_{\text{B}}$ is the 
Bjerrum length, which provides a measure of the extension of the equilibrium 
double-layer in 1D \cite{TellezTrizac2015}. Since $l_{\text{B}} \propto T_{\text{bath}}$,
we conclude that $\tau$ increases when increasing $T_{\text{bath}}$: this is due to the
enhancement of the double-layer size, under the influence of thermal 
agitation. The quasi-independence of double-layer size on $N$ 
--ionic charges being fixed, and therefore at varying colloidal charge--
is at the root
of the rather striking independence of $\tau$ on the number of 
counterions, when this quantity is an even integer. The irrelevance of $N$,
interestingly, is also observed for $N$ odd, stemming from a distinct mechanism.
There, what matters is the presence of a misfit counterion, that will be,
by and large, the dynamical limiting factor. By symmetry, this central ion 
does not experience any force, while all other ions are subject to
a non-vanishing electric field. For large $L$, this ion diffuses in a domain 
of size $L$; hence the scaling in $L^2/D$ for the characteristic time. Leaving aside
the misfit ion, odd-$N$ systems behave much like even ones, and equilibrate 
over a common time scale $l_{\text{B}}^2/D$.



We showed that the analytical solution for the mean-field dynamics (namely, the Poisson-Nernst-Planck electrokinetic equations) provides a reasonable approximation for a system with as few as $N=3$ counterions (see Figs.~\ref{fig:PNP_N_t-1p5},\ref{fig:PNP_N_t-5}). We can surmise that the mean-field framework becomes exact 
in the limit $N\to \infty$. While the exact equilibrium 
density profiles at finite-$N$ feature an exponential tail at large distances, their mean-field expressions are longer range, with an algebraic decay.
This translates into an infinite characteristic time for equilibration
at mean-field level. We have shown that the finite-$N$ finite-$\tau$ 
results did approach this limit as $\tau \propto l_{\text{B}}^2/D \propto e^{-2}$ where $e$
is the charge of the counterions. Since the mean-field limit, for a  
colloidal object of charge $Q$, is met for $N\to \infty$,  electroneutrality
$Q=2 Ne$ requires that $e\to 0$. Thus, $\tau \propto e^{-2}$ becomes infinite 
in the mean-field limit.



\section*{Acknowledgments}
We would like to thank A. Chepelianskii, S. Majumdar and L. \v{S}amaj for useful discussions.
This work was supported by an ECOS-Nord/Minciencias C18P01 action of Colombian and French cooperation. 
L.V. and G.T. acknowledge support from Fondo de Investigaciones, Facultad de Ciencias, Universidad de los Andes, Program INV-2019-84-1825 and Exacore HPC Uniandes for providing high performance computing time. L.V. acknowledges support from Action Doctorale Internationale (ADI 2018) de l'IDEX Universit\'e Paris-Saclay. S.A. is supported by the JSPS Kakenhi Grant number JP19K14617.

\appendix
\section{Dry friction in a wedge potential}
\label{appx:soa}

This section deals with the limit $L\rightarrow0$, where the system is described by the same equation as that ruling the velocity distribution of Brownian motion with dry friction \cite{Gennes2005,dryfriction}.  
The latter describes a particle under the influence of a Langevin force $\xi(t)$ with Gaussian distribution characterized by $\braket{\xi(t)} = 0$ and $\braket{\xi(t)\xi(0)} = m^2 \Gamma \delta(t)$. There is also a dry friction force term of magnitude $\Delta_F$. For our purpose, we only mention the results obtained in the ``partly stuck'' regime, in which the friction coefficient is small enough to avoid getting stuck but large enough to differ from free Brownian motion. The equation of motion in that case is given by:
\begin{equation}
m \dot{v} + m\gamma v = - \Delta_F \text{ sgn}(v) + \xi(t).
\end{equation} 
If the viscous damping is neglected ($\gamma\rightarrow 0$), this equation becomes:
\begin{equation}
\dot{v}  = - \Delta \text{ sgn}(v) + \frac{1}{m}\xi(t),
\end{equation} 
where $\Delta = \Delta_F/m$ and $\text{sgn}(x)$ is the sign function. The previous equation has an associated Fokker-Planck formulation, which we write in terms of the dimensionless variables 
$\widetilde{x} =2 \Delta v/\Gamma $  and $ \widetilde{t} = 2\Delta^2 t/\Gamma$:
\begin{equation}
\frac{\partial p(\widetilde{x},\widetilde{t}|\widetilde{x}_0,0)}{\partial \widetilde{t}} = \frac{\partial }{\partial \widetilde{x}}\left(p(\widetilde{x},\widetilde{t}|\widetilde{x}_0,0) \text{sgn}(\widetilde{x})\right) +  \frac{\partial^2 p(\widetilde{x},\widetilde{t}|\widetilde{x}_0,0)}{\partial \widetilde{x}^2} ,
\end{equation}
where $p(\widetilde{x},\widetilde{t}|\widetilde{x}_0,0)$ is the  propagator for the velocity distribution with initial dimensionless velocity $\widetilde{x}_0$ at a time $ \widetilde{t}_0 = 0$. Note that the analogy with  Eq.~\eqref{1c_fp} for $\widetilde{L} = 0$. This Fokker-Planck equation has been solved using an eigenfunction expansion in \cite{dryfriction}, which is the same treatment we used for an arbitrary $\widetilde{L}$. Simplifications ensue for 
$\widetilde{L}=0$ and a closed
form  expression is available:
\begin{align}
p(\widetilde{x},\widetilde{t}|\widetilde{x}_0,0) = \frac{1}{2\sqrt{\pi \widetilde{t}}}\e^{-\widetilde{t}/4 } \e^{-(|\widetilde{x}|-|\widetilde{x}_0|)/2} \e^{ - (\widetilde{x}-\widetilde{x}_0)^2/4\widetilde{t}} \nonumber\\+ \frac{\e^{-|\widetilde{x}|}}{4} \left[ 1 + \text{erf}\left(\frac{\widetilde{t}-|\widetilde{x}|-|\widetilde{x}_0|}{2\widetilde{t}^{1/2}}\right)\right],
\end{align}
where $\text{erf}(x)$ is the error function. This density distribution explicitly shows the time scale $\tau = 2 m^2 \Gamma/ \Delta_F^2$ (or $\widetilde{\tau} = 4$ in dimensionless units). 

The constant drift diffusion described in previous sections is observed directly from the average position:
\begin{align}
\braket{\widetilde{x}(\widetilde{t})} = \text{sgn}(\widetilde{x}_0) \left[  \left(\frac{ |\widetilde{x}_0|-\widetilde{t}}{2}\right) \text{erfc}\left(\frac{\widetilde{t}-|\widetilde{x}_0|}{2\widetilde{t}^{1/2}}\right) \right. \nonumber\\
\left.+ \e^{|\widetilde{x}_0|}\left(\frac{ |\widetilde{x}_0|+\widetilde{t}}{2}\right)  \text{erfc}\left(\frac{\widetilde{t}+|\widetilde{x}_0|}{2\widetilde{t}^{1/2}}\right) \right],
\end{align}
where $\text{erfc}(x)$ is the complementary error function. 
The previous expression can be shown to follow a ballistic behavior  $\braket{\widetilde{x}(t)} \approx\widetilde{x}_0 - \widetilde{t} \;\text{sgn}(\widetilde{x}_0)$ during the period $\widetilde{t} < \widetilde{x}_0$ 
($t <v_0/\Delta$). After this, the average velocity decays exponentially to $0$. 

\section{Fokker-Planck Equation for \texorpdfstring{$N=1$}{} }
\label{appendix:FokkerPlanck}

The eigenfunctions $u$ of the Fokker-Planck operator follow from solving:
\begin{equation}
\label{efde}
\frac{d^2 u(\widetilde{x},\lambda)}{d \widetilde{x}^2} + \frac{d}{d \widetilde{x}} \left(\widetilde{\Phi}'(\widetilde{x}) u(\widetilde{x},\lambda)\right) + \lambda u(\widetilde{x},\lambda) = 0,
\end{equation} 
where the eigenvalue $\lambda$ is real and positive. This result stems from the fact that the 1D Fokker-Planck equation can be transformed to a Schr\"{o}dinger equation, which involves a Hermitian operator and therefore a real spectrum \cite{risken,schrodinger}. Furthermore, the potential is confining enough to allow for a localized steady state. 


The eigenproblem equation (Eq.~\eqref{efde}) is piece-wise in three regions: $\widetilde{x}< -\widetilde{L}/2$, $-\widetilde{L}/2<\widetilde{x}<\widetilde{L}/2$
and $\widetilde{x}> \widetilde{L}/2$. For that reason, Eq.~\eqref{efde} is solved for each region, with
\begin{align}
\label{cef}u((\widetilde{L}/2)^+,\lambda) &=u((\widetilde{L}/2)^-,\lambda),\\
\label{cpc}u'((\widetilde{L}/2)^+,\lambda) + u((\widetilde{L}/2)^+,\lambda) &=u'((\widetilde{L}/2)^-,\lambda),
\end{align}
where these equations express the continuity of the eigenfunction and of the probability current $j(\widetilde{x}, \lambda) = \widetilde{\Phi}'(\widetilde{x}) u(\widetilde{x}, \lambda) + u'(\widetilde{x}, \lambda)$ at $\widetilde{x}= \widetilde{L}/2$, respectively. The interface conditions at $-\widetilde{L}/2$ are analogous. 
Eqs.~(\ref{cef}) and (\ref{cpc}) only have non-trivial solutions for a discrete set of eigenvalues $\{\lambda_k\}_k$ in the domain $[0,1/4)$. This set is $\widetilde{L}$-dependent and non-empty since the equilibrium distribution, given by $\lambda_0 = 0$,  is always present. On the other hand, the spectrum is continuous in $[1/4,\infty)$, independent of $\widetilde{L}$. We use the notation $u_{\lambda_k}(\widetilde{x})$ and $u(\widetilde{x},\lambda)$ for the eigenfunctions of each case respectively. The eigenfunctions of the adjoint problem $v(\widetilde{x},\lambda)$  follow from $v(\widetilde{x},\lambda) = u(\widetilde{x},\lambda) \e^{\widetilde{\Phi}(\widetilde{x})}$, where the discrete case is obtained by adding the corresponding subscripts. Due to the symmetry of Eq.~(\ref{efde}), it is convenient to solve it using a linear combination of odd and even functions:
\begin{subequations}
\begin{align}
u^o(|\widetilde{x}| <\widetilde{L}/2,\lambda) &= A^o \sin(\sqrt{\lambda} \widetilde{x})\\
u^o(|\widetilde{x}|>\widetilde{L}/2,\lambda) &= \e^{-\frac{|\widetilde{x}|}{2}}[B^o_1 \sin(\widetilde{x}\beta_{\lambda}) + B^o_2  \text{sgn}(\widetilde{x})\cos(\widetilde{x}\beta_{\lambda})]\\
u^e(|\widetilde{x}|<\widetilde{L}/2,\lambda) &=  A^e  \cos(\sqrt{\lambda} \widetilde{x}) \\
u^e(|\widetilde{x}|>\widetilde{L}/2,\lambda) &= \e^{-\frac{|\widetilde{x}|}{2}}[B^e_1\text{sgn}(\widetilde{x}) \sin(\widetilde{x}\beta_{\lambda}) + B^e_2  \cos(\widetilde{x}\beta_{\lambda})],
\end{align}
\end{subequations}
and for the discrete spectrum:
\begin{subequations}
\begin{align}
\label{u_discrete_a}
u^o_{k}(|\widetilde{x}|<\widetilde{L}/2) &= C^o \sin(\widetilde{x}\sqrt{\lambda_k^{o}})\\
u^o_{k}(|\widetilde{x}|>\widetilde{L}/2) &= D^o\text{sgn}(\widetilde{x})\e^{-|\widetilde{x}|\left(1+\sqrt{1-4\lambda_k^{o}}\right)/2}\\
u^e_{\lambda_k^{e}}(|\widetilde{x}|<\widetilde{L}/2) &=  C^e  \cos( \widetilde{x}\sqrt{\lambda_k^{e}}) \\
u^e_{k}(|\widetilde{x}|>\widetilde{L}/2) &=  D^e\e^{-|\widetilde{x}|\left(1+\sqrt{1-4\lambda_k^{e}}\right)/2},
\label{u_discrete_d}
\end{align}
\end{subequations}
where $\beta_{\lambda} = \sqrt{\lambda-1/4}$. The continuous eigenvalues are $\lambda>1/4$ and the discrete ones are $0\leq \lambda_k^{\alpha} < 1/4$. 
The superscripts ``$o$'' and ``$e$'' give the parity of the eigenfunctions. The constants are found at the end of this appendix.
The eigenvalue $\lambda_k^{\alpha}$ belongs either to an odd ($\alpha = o$) or even ($\alpha = e$) eigenfunction, and it is a solution to Eq.~\eqref{eodd} (if odd) or Eq.~\eqref{eeven} (if even) in the domain $[0,1/4)$. These equations result from imposing a vanishing determinant of the linear system made by Eqs.~\eqref{cef}-\eqref{cpc} for the eigenfunctions (Eqs.~\eqref{u_discrete_a}-\eqref{u_discrete_d}). This ensures a non-trivial solution for the discrete family of constants ($C^e,C^o,D^e$ and $D^o$). 

The normalization constants  $Z^{\alpha}(\lambda)$ and $Z^{\alpha}_{k}$ are defined by the relations:
\begin{align}
Z^{\alpha}(\lambda) \delta(\lambda-\lambda') \delta_{\alpha \alpha'}&= \int_{\mathbb{R}}  d\widetilde{x}\; u^{\alpha}(\widetilde{x},\lambda)\nu^{\alpha'}(\widetilde{x},\lambda') \label{cz}  \\
Z_{k}^{\alpha} \delta_{\lambda_k^{\alpha}\lambda_{k'}^{\alpha'}} &= \int_{\mathbb{R}}  d\widetilde{x}\; u_{k}^{\alpha}(\widetilde{x}) \nu^{\alpha'}_{k'}(\widetilde{x}) \label{dz},
\end{align}
which yield the following expressions:
\begin{subequations}
\begin{align}
&Z^o(\lambda) = 2 \pi \beta_{\lambda}({B^o_1}^2 + {B^o_2}^2)\\
&Z^e(\lambda) = 2 \pi \beta_{\lambda}({B^e_1}^2 + {B^e_2}^2)\\
&Z_{k}^o =  {C^o}^2 \Big(\frac{\widetilde{L}}{2}- \frac{\sin(\sqrt{\lambda_k^{o}}\widetilde{L})}{2 \sqrt{\lambda_k^{o}}}\Big)\e^{\frac{\widetilde{L}}{2}}+ \frac{{D^o}^2\e^{-i\beta_{\lambda_k^{o}} \widetilde{L}}}{i\beta_{\lambda_k^{o}}}\\
&Z_{k}^e =  {C^e}^2 \Big(\frac{\widetilde{L}}{2} + \frac{\sin(\sqrt{\lambda_k^{e}}\widetilde{L})}{2 \sqrt{\lambda_k^{e}}}\Big)\e^{\frac{\widetilde{L}}{2}} + \frac{{D^e}^2\e^{-i\beta_{\lambda_k^{e}} \widetilde{L}}}{i\beta_{\lambda_k^{e}}}\\
&Z_{\infty} =(\widetilde{L} + 2) \e^{-\widetilde{L}/2},
\end{align}
\end{subequations}
where $i\beta_{\lambda_k} = \sqrt{1/4-\lambda_k}$, and $Z_{\infty}$ is 
the partition function 
for the equilibrium distribution ($\lambda_0 = 0$), which is given by Eq.~\eqref{eq_dist}.

Finally, the following constants are determined by the vanishing boundary conditions at infinity and  Eqs.~\eqref{cef}-\eqref{cpc}. Additionally, for the discrete case, the determinant for the linear system that rules the family of constants must vanish (Eqs.~\eqref{eodd}-\eqref{eeven}) 
\begin{align}
	\frac{B_1^{o}}{B_2^{o}} &= \frac{2\beta_{\lambda}+\tan \big( \frac{\widetilde{L}}{2}\beta_{\lambda} \big) \big[1-2 \sqrt{\lambda } \cot \big( \frac{\widetilde{L}}{2}\sqrt{\lambda }\big)\big]  }{2\beta_{\lambda} \tan \big( \frac{\widetilde{L}}{2}\beta_{\lambda} \big)+2 \sqrt{\lambda } \cot \big( \frac{\widetilde{L}}{2}\sqrt{\lambda }\big)-1} \\
	\frac{B_1^{e}}{B_2^{e}} &= \frac{2\beta_{\lambda}  +\tan \big( \frac{\widetilde{L}}{2}\beta_{\lambda} \big) \big[2 \sqrt{\lambda } \tan \big( \frac{\widetilde{L}}{2}\sqrt{\lambda }\big)+1\big]}{2 \sqrt{\lambda } \tan \big( \frac{\widetilde{L}}{2}\sqrt{\lambda }\big)-2\beta_{\lambda} \tan \big( \frac{\widetilde{L}}{2}\beta_{\lambda} \big) +1}
	\\
	\frac{A^{o}}{B_2^{o}} &=
	\frac{2\beta_{\lambda} \e^{-\widetilde{L}/4} \csc \big( \frac{\widetilde{L}}{2}\sqrt{\lambda }\big) \sec \big( \frac{\widetilde{L}}{2}\beta_{\lambda} \big)}{2\beta_{\lambda} \tan \big( \frac{\widetilde{L}}{2}\beta_{\lambda} \big)+2 \sqrt{\lambda } \cot \big( \frac{\widetilde{L}}{2}\sqrt{\lambda }\big)-1}
	\\
	\frac{A^{e}}{B_2^{o}} &=\frac{2\beta_{\lambda} \e^{-\widetilde{L}/4} \sec \big( \frac{\widetilde{L}}{2}\sqrt{\lambda }\big) \sec\big( \frac{\widetilde{L}}{2}\beta_{\lambda} \big) }{2 \sqrt{\lambda } \tan \big( \frac{\widetilde{L}}{2}\sqrt{\lambda }\big)+1-2\beta_{\lambda} \tan \big( \frac{\widetilde{L}}{2}\beta_{\lambda} \big)}
	\\
\frac{D^o}{C^o} &= \e^{(2i\beta_{\lambda_k^{o}}+1) {\widetilde{L}}/{4}} \sin \big(\widetilde{L}\sqrt{\lambda_k^{o} } /2\big)
		\\
\frac{D^e}{C^e} &= \e^{ (2i\beta_{\lambda_k^{e}}+1) {\widetilde{L}}/{4}} \cos \big(\widetilde{L}\sqrt{\lambda_k^{e} } /{2}\big).
	\end{align}

\section{Asymptotic Kullback-Leibler Divergence for  \texorpdfstring{$N=1$}{} }
\label{appx_KLD}

This appendix computes the asymptotic behavior of the Kullback-Leibler divergence (KLD) when $N=1$ (Eq.~\eqref{DKL_epsilon}). For this purpose it will be useful to realize that $\delta p$ is the sum of a term that comes from the continuous spectrum ($\varepsilon_{\lambda}$) and  another from the discrete one ($\sum_{\alpha,k}\varepsilon_{\lambda_k^{\alpha}}$): 
\begin{equation}
\delta p(\widetilde{x},\widetilde{t},\widetilde{x}_0) = \varepsilon_{\lambda} + \sum_{\alpha,k} \varepsilon_{\lambda_k^{\alpha}},
\label{pepsilon}
\end{equation}
where $\varepsilon_{\lambda}$ and  $\sum_{\alpha,k} \varepsilon_{\lambda_k^{\alpha}}$ are the terms of the ionic density (Eq.~\eqref{1c_prop}) associated to the continuous ($\lambda\geq 1/4$) and non-zero discrete ($0<\lambda_k^{\alpha}<1/4$) parts of the spectrum  respectively. The contribution of $\varepsilon_{\lambda}$ is always non-zero, while $\sum_{\alpha,k}\varepsilon_{\lambda_k^{\alpha}}$ can vanish completely depending on $\widetilde{L}$ and $\widetilde{x}_0$. 

We start by analyzing the large time behavior of the KLD when the only contribution is due to the continuous spectrum term $\varepsilon_{\lambda}$ and therefore we have:  
\begin{equation}
D_{\text{KL}}(p||p_{\infty}) 
\sim \int_{\mathbb{R}} d \widetilde{x}\; \frac{(\varepsilon_{\lambda})^2}{p_{\infty}} =\frac{ Z_{\infty}  \e^{\frac{\widetilde{L}}{2}}}{2\pi} I(\widetilde{x}_0,\widetilde{L},t),
\end{equation}
where $I$ is the following integral:
\begin{equation}
I = \int_{\frac{1}{4}}^{\infty} d\lambda\frac{ \left(4 \sqrt{\lambda }-2\sin \left( \sqrt{\lambda } \widetilde{L} \right)\cos \left(2 \sqrt{\lambda }\widetilde{x}_0\right) \right)\e^{-2\lambda{\widetilde{t}}}}{(4\lambda-1)^{-\frac{1}{2}}\sqrt{\lambda } \left(8 \lambda +\cos \left(2 \sqrt{\lambda } \widetilde{L}\right)-1\right)}.
\end{equation}
This term is obtained by exchanging the integration order of $\lambda$ and $\widetilde{x}$, and then calculating the spatial integral.
After some algebra, the remaining expression is $I$, which unfortunately has no closed form. However, it is possible to find lower and upper bounds. Since $I$ is an integral of a positive function, the integral of a function that bounds the integrand is a bound of $I$. Following this reasoning we find the following inequality   
\begin{equation}
 \e^{-\frac{\widetilde{t}}{2}} \Big[ \frac{1}{\widetilde{t}^{\frac{5}{2}}} +\mathcal{O}\Big(\frac{1}{\widetilde{t}^{\frac{7}{2}}}\Big) \Big]\leq \frac{\sqrt{8}I}{3\sqrt{\pi}} \leq
 \e^{-\frac{\widetilde{t}}{2}} \Big[\frac{1}{\widetilde{t}^{\frac{1}{2}}} +\mathcal{O}\Big(\frac{1}{\widetilde{t}^{\frac{3}{2}}}\Big)\Big].
\end{equation}
This implies that 
the KLD is dominantly of the form $\exp(-\widetilde{t}/2)$ for asymptotically
long times.
This behavior is not a surprise, since $I$ has an integrand dominated by $\exp(-2\lambda \widetilde{t})$ when $\widetilde{t} \to \infty$, which is maximum when $\lambda$ takes its minimum value $1/4$. The next order correction is a power function term $\widetilde{t}^{-s}$, where $1/2 \leq s\leq 5/2$.  For $\widetilde{L}=0$, the explicit calculation is possible, which yields $s=3/2$. The subdominant term $\widetilde{t}^{-s}$ is relevant in practice, 
 due to the limited time domain accessible in the simulations.       

We now focus on the situation where $\sum_{\alpha,k}\varepsilon_{\lambda_k^{\alpha}}$ is non-zero. 
Discrete branches, featuring smaller eigenvalues, do dominate at long times 
over their continuous counterpart. 
Which eigenvalue it is depends on $\widetilde{L}$ (as seen in Fig.~\ref{spectralgap}) and also on the symmetry of the initial conditions. 
We summarize all the possible cases, including the previous continuous result, in the following relation:
\begin{equation}
D_{\text{KL}}(\rho||p_{\infty}) 
\sim
\begin{cases}
\frac{Z_{\infty}[\nu_1^o(\widetilde{x}_0)]^2}{Z^{o}_1}\e^{-2\lambda_1^o \widetilde{t}}, & \widetilde{L}>\pi, \widetilde{x}_0 \neq 0 \\
\frac{Z_{\infty}[\nu_1^e(\widetilde{x}_0)]^2}{Z^{e}_1}\e^{-2\lambda_1^e \widetilde{t}}, & \widetilde{L}>3\pi, \widetilde{x}_0 = 0 \\
\frac{ 3Z_{\infty}  \e^{\widetilde{L}/2}}{4\sqrt{2\pi}}\e^{-\widetilde{t}/2} \widetilde{t}^{-s},   & \text{elsewhere}.
\end{cases}
\label{DKL_N1_asymptotics}
\end{equation}

We conclude this appendix with a note on the behavior of these asymptotic relations when used on time domains which do not take large enough values. In this situation it can happen that some subdominant terms are not negligible, as previously stated for the continuous spectrum term $\varepsilon_{\lambda}$. It can also lead to problems in the scheme described in Sec. \ref{sec:extrapolationmethod}, for the cross-over regions where the dominant term in $\delta p$ changes from $\varepsilon_{\lambda}$ to  $\sum_{\alpha,k}\varepsilon_{\lambda_k^{\alpha}}$. 

    

\section{Derivation of the simplified transport time distribution}
\label{appx:transport_time}

The distribution of $\tau_l$ as defined in Eq.~\eqref{eq:tau_l} can be derived following the ideas in \cite[Ch. XII]{VanKampen}, and we summarize the method here. The probability density function  $p_l(t,x)$ for the free Brownian motion we consider, obeys the Fokker-Planck equation
\begin{equation}
\label{eq:tfp}
\frac{\partial}{\partial t}p_l(t,x)=\frac{\partial^2}{\partial x^2}p_l(t,x).
\end{equation}
The boundary conditions of the problem are that of a reflecting wall at the origin and an absorbing wall at $x=l$, but because Brownian motion has left-right symmetry, we can replace the reflecting wall by a second absorbing wall at $x=-l$. Then, we have
\begin{equation}
\begin{split}
\text{Initial condition: }&p_l(0,x)=\delta(x),\\
\text{Boundary condition: }&p_l(t,-l)=p_l(t,l)=0 .
\label{eq:conds}
\end{split}
\end{equation}
The total (integrated) probability is not conserved, and the rate at which it decreases due to adsorption gives the transport time distribution,
\begin{equation}\label{eq:pqrel}
q_l(t)=
-\frac{\partial}{\partial x}p_l(t,x)\Big|_{x=-l}^{x=l}.
\end{equation}
To solve \eqref{eq:tfp}, we introduce the Laplace transforms
\begin{equation}
    \begin{split}
\varphi_l(s,x)&=\int_0^\infty \e^{-st}p_l(t,x)\>dt\ \\ 
\phi_l(s)&=\int_0^\infty \e^{-st}q_l(t)\>dt,
    \end{split}
\end{equation}
with $\operatorname{Re}(s)>0$. Then, the transform of \eqref{eq:pqrel} reads
\begin{equation}\label{eq:ltrel}
\phi_l(s)=-\frac{\partial}{\partial x}\varphi_l(t,x)\Big|_{x=-l}^{x=l}.
\end{equation}
Now, the transform of \eqref{eq:tfp} becomes an ordinary differential equation,
which is solved using \eqref{eq:conds} to obtain
\begin{equation}
\varphi_l(s,x)=\frac{\tanh(\sqrt{s}l)\cosh(\sqrt{s}x)-|\sinh(\sqrt{s}x)|}{2\sqrt{s}},
\end{equation}
and due to \eqref{eq:ltrel} we get
\begin{equation}
\phi_l(s)=1/\cosh(\sqrt{s}l).
\end{equation}
This transform is inverted using the Mittag-Leffler expansion of $\cosh$,
\begin{equation}
\phi_l(s)=\sum_{n=0}^\infty(-1)^n\frac{4\pi(2n+1)}{(2n+1)^2\pi^2+4 s l^2},
\end{equation}
and \eqref{eq:trtimedist} follows from tables of Laplace transforms.

\bibliography{refs}

\begin{thebibliography}{67}%
\makeatletter
\providecommand \@ifxundefined [1]{%
 \@ifx{#1\undefined}
}%
\providecommand \@ifnum [1]{%
 \ifnum #1\expandafter \@firstoftwo
 \else \expandafter \@secondoftwo
 \fi
}%
\providecommand \@ifx [1]{%
 \ifx #1\expandafter \@firstoftwo
 \else \expandafter \@secondoftwo
 \fi
}%
\providecommand \natexlab [1]{#1}%
\providecommand \enquote  [1]{``#1''}%
\providecommand \bibnamefont  [1]{#1}%
\providecommand \bibfnamefont [1]{#1}%
\providecommand \citenamefont [1]{#1}%
\providecommand \href@noop [0]{\@secondoftwo}%
\providecommand \href [0]{\begingroup \@sanitize@url \@href}%
\providecommand \@href[1]{\@@startlink{#1}\@@href}%
\providecommand \@@href[1]{\endgroup#1\@@endlink}%
\providecommand \@sanitize@url [0]{\catcode `\\12\catcode `\$12\catcode
  `\&12\catcode `\#12\catcode `\^12\catcode `\_12\catcode `\%12\relax}%
\providecommand \@@startlink[1]{}%
\providecommand \@@endlink[0]{}%
\providecommand \url  [0]{\begingroup\@sanitize@url \@url }%
\providecommand \@url [1]{\endgroup\@href {#1}{\urlprefix }}%
\providecommand \urlprefix  [0]{URL }%
\providecommand \Eprint [0]{\href }%
\providecommand \doibase [0]{https://doi.org/}%
\providecommand \selectlanguage [0]{\@gobble}%
\providecommand \bibinfo  [0]{\@secondoftwo}%
\providecommand \bibfield  [0]{\@secondoftwo}%
\providecommand \translation [1]{[#1]}%
\providecommand \BibitemOpen [0]{}%
\providecommand \bibitemStop [0]{}%
\providecommand \bibitemNoStop [0]{.\EOS\space}%
\providecommand \EOS [0]{\spacefactor3000\relax}%
\providecommand \BibitemShut  [1]{\csname bibitem#1\endcsname}%
\let\auto@bib@innerbib\@empty
\bibitem [{\citenamefont {Hunter}(2001)}]{hunterbook}%
  \BibitemOpen
  \bibfield  {author} {\bibinfo {author} {\bibfnamefont {R.~J.}\ \bibnamefont
  {Hunter}},\ }\href {https://nla.gov.au/nla.cat-vn1888411} {\emph {\bibinfo
  {title} {Foundations of colloid science}}},\ \bibinfo {edition} {2nd}\ ed.\
  (\bibinfo  {publisher} {Oxford University Press Oxford ; New York},\ \bibinfo
  {year} {2001})\BibitemShut {NoStop}%
\bibitem [{\citenamefont {Levin}(2002)}]{Levin2002}%
  \BibitemOpen
  \bibfield  {author} {\bibinfo {author} {\bibfnamefont {Y.}~\bibnamefont
  {Levin}},\ }\bibfield  {title} {\bibinfo {title} {Electrostatic correlations:
  from plasma to biology},\ }\href
  {https://doi.org/10.1088/0034-4885/65/11/201} {\bibfield  {journal} {\bibinfo
   {journal} {Reports on Progress in Physics}\ }\textbf {\bibinfo {volume}
  {65}},\ \bibinfo {pages} {1577} (\bibinfo {year} {2002})}\BibitemShut
  {NoStop}%
\bibitem [{\citenamefont {J{\"o}nsson}\ and\ \citenamefont
  {Wennerstr{\"o}m}(2001)}]{Jonsson}%
  \BibitemOpen
  \bibfield  {author} {\bibinfo {author} {\bibfnamefont {B.}~\bibnamefont
  {J{\"o}nsson}}\ and\ \bibinfo {author} {\bibfnamefont {H.}~\bibnamefont
  {Wennerstr{\"o}m}},\ }\bibfield  {title} {\bibinfo {title} {When ion-ion
  correlations are important in charged colloidal systems},\ }in\ \href
  {https://doi.org/10.1007/978-94-010-0577-7_7} {\emph {\bibinfo {booktitle}
  {Electrostatic Effects in Soft Matter and Biophysics}}},\ \bibinfo {editor}
  {edited by\ \bibinfo {editor} {\bibfnamefont {C.}~\bibnamefont {Holm}},
  \bibinfo {editor} {\bibfnamefont {P.}~\bibnamefont {K{\'e}kicheff}},\ and\
  \bibinfo {editor} {\bibfnamefont {R.}~\bibnamefont {Podgornik}}}\ (\bibinfo
  {publisher} {Kluwer Academic, Dordrecht},\ \bibinfo {year} {2001})\ pp.\
  \bibinfo {pages} {171--204}\BibitemShut {NoStop}%
\bibitem [{\citenamefont {Levin}(2005)}]{Levin05}%
  \BibitemOpen
  \bibfield  {author} {\bibinfo {author} {\bibfnamefont {Y.}~\bibnamefont
  {Levin}},\ }\bibfield  {title} {\bibinfo {title} {Strange electrostatics in
  physics, chemistry, and biology},\ }\href@noop {} {\bibfield  {journal}
  {\bibinfo  {journal} {Physica A}\ }\textbf {\bibinfo {volume} {353}},\
  \bibinfo {pages} {43} (\bibinfo {year} {2005})}\BibitemShut {NoStop}%
\bibitem [{\citenamefont {Caffrey}(2001)}]{Caffrey}%
  \BibitemOpen
  \bibfield  {author} {\bibinfo {author} {\bibfnamefont {M.}~\bibnamefont
  {Caffrey}},\ }\bibfield  {title} {\bibinfo {title} {Structure and dynamic
  properties of membrane lipid and protein},\ }in\ \href
  {https://doi.org/10.1007/978-94-010-0577-7_1} {\emph {\bibinfo {booktitle}
  {Electrostatic Effects in Soft Matter and Biophysics}}},\ \bibinfo {editor}
  {edited by\ \bibinfo {editor} {\bibfnamefont {C.}~\bibnamefont {Holm}},
  \bibinfo {editor} {\bibfnamefont {P.}~\bibnamefont {K{\'e}kicheff}},\ and\
  \bibinfo {editor} {\bibfnamefont {R.}~\bibnamefont {Podgornik}}}\ (\bibinfo
  {publisher} {Kluwer Academic, Dordrecht},\ \bibinfo {year}
  {2001})\BibitemShut {NoStop}%
\bibitem [{\citenamefont {Schoch}\ \emph {et~al.}(2008)\citenamefont {Schoch},
  \citenamefont {Han},\ and\ \citenamefont {Renaud}}]{SchochHanRenaud}%
  \BibitemOpen
  \bibfield  {author} {\bibinfo {author} {\bibfnamefont {R.~B.}\ \bibnamefont
  {Schoch}}, \bibinfo {author} {\bibfnamefont {J.}~\bibnamefont {Han}},\ and\
  \bibinfo {author} {\bibfnamefont {P.}~\bibnamefont {Renaud}},\ }\bibfield
  {title} {\bibinfo {title} {Transport phenomena in nanofluidics},\ }\href
  {https://doi.org/https://doi.org/10.1103/RevModPhys.80.839} {\bibfield
  {journal} {\bibinfo  {journal} {Rev. Mod. Phys.}\ }\textbf {\bibinfo {volume}
  {80}},\ \bibinfo {pages} {839} (\bibinfo {year} {2008})}\BibitemShut
  {NoStop}%
\bibitem [{\citenamefont {Jubin}\ \emph {et~al.}(2018)\citenamefont {Jubin},
  \citenamefont {Poggioli}, \citenamefont {Siria},\ and\ \citenamefont
  {Bocquet}}]{Jubin2018}%
  \BibitemOpen
  \bibfield  {author} {\bibinfo {author} {\bibfnamefont {L.}~\bibnamefont
  {Jubin}}, \bibinfo {author} {\bibfnamefont {A.}~\bibnamefont {Poggioli}},
  \bibinfo {author} {\bibfnamefont {A.}~\bibnamefont {Siria}},\ and\ \bibinfo
  {author} {\bibfnamefont {L.}~\bibnamefont {Bocquet}},\ }\bibfield  {title}
  {\bibinfo {title} {Dramatic pressure-sensitive ion conduction in conical
  nanopores},\ }\href {https://doi.org/10.1073/pnas.1721987115} {\bibfield
  {journal} {\bibinfo  {journal} {Proceedings of the National Academy of
  Sciences}\ }\textbf {\bibinfo {volume} {115}},\ \bibinfo {pages} {4063}
  (\bibinfo {year} {2018})}\BibitemShut {NoStop}%
\bibitem [{\citenamefont {Kavokine}\ \emph {et~al.}(2019)\citenamefont
  {Kavokine}, \citenamefont {Marbach}, \citenamefont {Siria},\ and\
  \citenamefont {Bocquet}}]{Kavokine2019}%
  \BibitemOpen
  \bibfield  {author} {\bibinfo {author} {\bibfnamefont {N.}~\bibnamefont
  {Kavokine}}, \bibinfo {author} {\bibfnamefont {S.}~\bibnamefont {Marbach}},
  \bibinfo {author} {\bibfnamefont {A.}~\bibnamefont {Siria}},\ and\ \bibinfo
  {author} {\bibfnamefont {L.}~\bibnamefont {Bocquet}},\ }\bibfield  {title}
  {\bibinfo {title} {Ionic {Coulomb} blockade as a fractional {Wien} effect},\
  }\href {https://doi.org/10.1038/s41565-019-0425-y} {\bibfield  {journal}
  {\bibinfo  {journal} {Nature Nanotechnology}\ }\textbf {\bibinfo {volume}
  {14}},\ \bibinfo {pages} {573} (\bibinfo {year} {2019})}\BibitemShut
  {NoStop}%
\bibitem [{\citenamefont {Drummond}\ \emph {et~al.}(2003)\citenamefont
  {Drummond}, \citenamefont {Hill},\ and\ \citenamefont
  {Barton}}]{Drummond2003}%
  \BibitemOpen
  \bibfield  {author} {\bibinfo {author} {\bibfnamefont {T.~G.}\ \bibnamefont
  {Drummond}}, \bibinfo {author} {\bibfnamefont {M.~G.}\ \bibnamefont {Hill}},\
  and\ \bibinfo {author} {\bibfnamefont {J.~K.}\ \bibnamefont {Barton}},\
  }\bibfield  {title} {\bibinfo {title} {Electrochemical {DNA} sensors},\
  }\href {https://doi.org/10.1038/nbt873} {\bibfield  {journal} {\bibinfo
  {journal} {Nature Biotechnology}\ }\textbf {\bibinfo {volume} {21}},\
  \bibinfo {pages} {1192} (\bibinfo {year} {2003})}\BibitemShut {NoStop}%
\bibitem [{\citenamefont {Burt}\ \emph {et~al.}(2014)\citenamefont {Burt},
  \citenamefont {Birkett},\ and\ \citenamefont {Zhao}}]{BurtRyanZhao}%
  \BibitemOpen
  \bibfield  {author} {\bibinfo {author} {\bibfnamefont {R.}~\bibnamefont
  {Burt}}, \bibinfo {author} {\bibfnamefont {G.}~\bibnamefont {Birkett}},\ and\
  \bibinfo {author} {\bibfnamefont {X.~S.}\ \bibnamefont {Zhao}},\ }\bibfield
  {title} {\bibinfo {title} {A review of molecular modelling of electric double
  layer capacitors},\ }\href {http://dx.doi.org/10.1039/C3CP55186E} {\bibfield
  {journal} {\bibinfo  {journal} {Phys. Chem. Chem. Phys.}\ }\textbf {\bibinfo
  {volume} {16}},\ \bibinfo {pages} {6519} (\bibinfo {year}
  {2014})}\BibitemShut {NoStop}%
\bibitem [{\citenamefont {Simon}\ and\ \citenamefont
  {Gogotsi}(2008)}]{Simon2008}%
  \BibitemOpen
  \bibfield  {author} {\bibinfo {author} {\bibfnamefont {P.}~\bibnamefont
  {Simon}}\ and\ \bibinfo {author} {\bibfnamefont {Y.}~\bibnamefont
  {Gogotsi}},\ }\bibfield  {title} {\bibinfo {title} {Materials for
  electrochemical capacitors},\ }\href {https://doi.org/10.1038/nmat2297}
  {\bibfield  {journal} {\bibinfo  {journal} {Nature Materials}\ }\textbf
  {\bibinfo {volume} {7}},\ \bibinfo {pages} {845} (\bibinfo {year}
  {2008})}\BibitemShut {NoStop}%
\bibitem [{\citenamefont {Sharma}\ and\ \citenamefont
  {Bhatti}(2010)}]{Sharma2010}%
  \BibitemOpen
  \bibfield  {author} {\bibinfo {author} {\bibfnamefont {P.}~\bibnamefont
  {Sharma}}\ and\ \bibinfo {author} {\bibfnamefont {T.}~\bibnamefont
  {Bhatti}},\ }\bibfield  {title} {\bibinfo {title} {A review on
  electrochemical double-layer capacitors},\ }\href
  {https://doi.org/https://doi.org/10.1016/j.enconman.2010.06.031} {\bibfield
  {journal} {\bibinfo  {journal} {Energy Conversion and Management}\ }\textbf
  {\bibinfo {volume} {51}},\ \bibinfo {pages} {2901 } (\bibinfo {year}
  {2010})}\BibitemShut {NoStop}%
\bibitem [{\citenamefont {Merlet}\ \emph {et~al.}(2012)\citenamefont {Merlet},
  \citenamefont {Rotenberg}, \citenamefont {Madden}, \citenamefont {Taberna},
  \citenamefont {Simon}, \citenamefont {Gogotsi},\ and\ \citenamefont
  {Salanne}}]{Merlet2012}%
  \BibitemOpen
  \bibfield  {author} {\bibinfo {author} {\bibfnamefont {C.}~\bibnamefont
  {Merlet}}, \bibinfo {author} {\bibfnamefont {B.}~\bibnamefont {Rotenberg}},
  \bibinfo {author} {\bibfnamefont {P.~A.}\ \bibnamefont {Madden}}, \bibinfo
  {author} {\bibfnamefont {P.~L.}\ \bibnamefont {Taberna}}, \bibinfo {author}
  {\bibfnamefont {P.}~\bibnamefont {Simon}}, \bibinfo {author} {\bibfnamefont
  {Y.}~\bibnamefont {Gogotsi}},\ and\ \bibinfo {author} {\bibfnamefont
  {M.}~\bibnamefont {Salanne}},\ }\bibfield  {title} {\bibinfo {title} {On the
  molecular origin of supercapacitance in nanoporous carbon electrodes},\
  }\href {https://doi.org/10.1038/nmat3260} {\bibfield  {journal} {\bibinfo
  {journal} {Nature Materials}\ }\textbf {\bibinfo {volume} {11}},\ \bibinfo
  {pages} {306} (\bibinfo {year} {2012})}\BibitemShut {NoStop}%
\bibitem [{\citenamefont {Xu}\ and\ \citenamefont
  {Fullerton-Shirey}(2020)}]{Xu2020}%
  \BibitemOpen
  \bibfield  {author} {\bibinfo {author} {\bibfnamefont {K.}~\bibnamefont
  {Xu}}\ and\ \bibinfo {author} {\bibfnamefont {S.~K.}\ \bibnamefont
  {Fullerton-Shirey}},\ }\bibfield  {title} {\bibinfo {title}
  {Electric-double-layer-gated transistors based on two-dimensional crystals:
  recent approaches and advances},\ }\href
  {https://doi.org/10.1088/2515-7639/ab8270} {\bibfield  {journal} {\bibinfo
  {journal} {Journal of Physics: Materials}\ }\textbf {\bibinfo {volume} {3}},\
  \bibinfo {pages} {032001} (\bibinfo {year} {2020})}\BibitemShut {NoStop}%
\bibitem [{\citenamefont {Anderson}\ \emph {et~al.}(2010)\citenamefont
  {Anderson}, \citenamefont {Cudero},\ and\ \citenamefont
  {Palma}}]{AndersonCuderoPalma}%
  \BibitemOpen
  \bibfield  {author} {\bibinfo {author} {\bibfnamefont {M.~A.}\ \bibnamefont
  {Anderson}}, \bibinfo {author} {\bibfnamefont {A.~L.}\ \bibnamefont
  {Cudero}},\ and\ \bibinfo {author} {\bibfnamefont {J.}~\bibnamefont
  {Palma}},\ }\bibfield  {title} {\bibinfo {title} {Capacitive deionization as
  an electrochemical means of saving energy and delivering clean water.
  {C}omparison to present desalination practices: Will it compete?},\ }\href
  {https://doi.org/https://doi.org/10.1016/j.electacta.2010.02.012} {\bibfield
  {journal} {\bibinfo  {journal} {Electrochimica Acta}\ }\textbf {\bibinfo
  {volume} {55}},\ \bibinfo {pages} {3845 } (\bibinfo {year}
  {2010})}\BibitemShut {NoStop}%
\bibitem [{\citenamefont {Grahame}(1947)}]{Grahame1947}%
  \BibitemOpen
  \bibfield  {author} {\bibinfo {author} {\bibfnamefont {D.~C.}\ \bibnamefont
  {Grahame}},\ }\bibfield  {title} {\bibinfo {title} {The electrical double
  layer and the theory of electrocapillarity},\ }\href
  {https://doi.org/10.1021/cr60130a002} {\bibfield  {journal} {\bibinfo
  {journal} {Chemical Reviews}\ }\textbf {\bibinfo {volume} {41}},\ \bibinfo
  {pages} {441} (\bibinfo {year} {1947})}\BibitemShut {NoStop}%
\bibitem [{\citenamefont {Verwey}\ \emph {et~al.}(1948)\citenamefont {Verwey},
  \citenamefont {Overbeek},\ and\ \citenamefont {van Nes}}]{VerweyOverbeek}%
  \BibitemOpen
  \bibfield  {author} {\bibinfo {author} {\bibfnamefont {E.~J.~W.}\
  \bibnamefont {Verwey}}, \bibinfo {author} {\bibfnamefont {J.~T.~G.}\
  \bibnamefont {Overbeek}},\ and\ \bibinfo {author} {\bibfnamefont
  {K.}~\bibnamefont {van Nes}},\ }\href
  {https://books.google.com.co/books?id=1inDjgEACAAJ} {\emph {\bibinfo {title}
  {Theory of the Stability of Lyophobic Colloids: The Interaction of Sol
  Particles Having an Electric Double Layer}}}\ (\bibinfo  {publisher}
  {Elsevier},\ \bibinfo {year} {1948})\BibitemShut {NoStop}%
\bibitem [{\citenamefont {Ohshima}(2006)}]{Ohshima}%
  \BibitemOpen
  \bibfield  {author} {\bibinfo {author} {\bibfnamefont {H.}~\bibnamefont
  {Ohshima}},\ }\href
  {https://www.elsevier.com/books/theory-of-colloid-and-interfacial-electric-phenomena/ohshima/978-0-12-370642-3}
  {\emph {\bibinfo {title} {Theory of colloid and interfacial electric
  phenomena}}},\ \bibinfo {edition} {1st}\ ed.\ (\bibinfo  {publisher}
  {Elsevier},\ \bibinfo {year} {2006})\BibitemShut {NoStop}%
\bibitem [{\citenamefont {Helmholtz}(1879)}]{Helmholtz}%
  \BibitemOpen
  \bibfield  {author} {\bibinfo {author} {\bibfnamefont {H.}~\bibnamefont
  {Helmholtz}},\ }\bibfield  {title} {\bibinfo {title} {Studien über
  electrische grenzschichten},\ }\href
  {https://doi.org/https://doi.org/10.1002/andp.18792430702} {\bibfield
  {journal} {\bibinfo  {journal} {Annalen der Physik}\ }\textbf {\bibinfo
  {volume} {243}},\ \bibinfo {pages} {337} (\bibinfo {year}
  {1879})}\BibitemShut {NoStop}%
\bibitem [{\citenamefont {Gouy}(1910)}]{Gouy}%
  \BibitemOpen
  \bibfield  {author} {\bibinfo {author} {\bibfnamefont {M.}~\bibnamefont
  {Gouy}},\ }\bibfield  {title} {\bibinfo {title} {Sur la constitution de la
  charge \'electrique \`a la surface d'un \'electrolyte},\ }\href
  {https://doi.org/10.1051/jphystap:019100090045700} {\bibfield  {journal}
  {\bibinfo  {journal} {J. Phys. Theor. Appl.}\ }\textbf {\bibinfo {volume}
  {9}},\ \bibinfo {pages} {457} (\bibinfo {year} {1910})}\BibitemShut {NoStop}%
\bibitem [{\citenamefont {Chapman}(1913)}]{Chapman}%
  \BibitemOpen
  \bibfield  {author} {\bibinfo {author} {\bibfnamefont {D.~L.}\ \bibnamefont
  {Chapman}},\ }\bibfield  {title} {\bibinfo {title} {{LI}. a contribution to
  the theory of electrocapillarity},\ }\href
  {https://doi.org/10.1080/14786440408634187} {\bibfield  {journal} {\bibinfo
  {journal} {The London, Edinburgh, and Dublin Philosophical Magazine and
  Journal of Science}\ }\textbf {\bibinfo {volume} {25}},\ \bibinfo {pages}
  {475} (\bibinfo {year} {1913})}\BibitemShut {NoStop}%
\bibitem [{\citenamefont {Stern}(1924)}]{Stern}%
  \BibitemOpen
  \bibfield  {author} {\bibinfo {author} {\bibfnamefont {O.}~\bibnamefont
  {Stern}},\ }\bibfield  {title} {\bibinfo {title} {Zur theorie der
  elektrolytischen doppelschicht},\ }\href
  {https://doi.org/https://doi.org/10.1002/bbpc.192400182} {\bibfield
  {journal} {\bibinfo  {journal} {Zeitschrift für Elektrochemie und angewandte
  physikalische Chemie}\ }\textbf {\bibinfo {volume} {30}},\ \bibinfo {pages}
  {508} (\bibinfo {year} {1924})}\BibitemShut {NoStop}%
\bibitem [{\citenamefont {Bikerman}(1942)}]{Bikerman}%
  \BibitemOpen
  \bibfield  {author} {\bibinfo {author} {\bibfnamefont {J.~J.}\ \bibnamefont
  {Bikerman}},\ }\bibfield  {title} {\bibinfo {title} {{XXXIX}. structure and
  capacity of electrical double layer},\ }\href
  {https://doi.org/10.1080/14786444208520813} {\bibfield  {journal} {\bibinfo
  {journal} {The London, Edinburgh, and Dublin Philosophical Magazine and
  Journal of Science}\ }\textbf {\bibinfo {volume} {33}},\ \bibinfo {pages}
  {384} (\bibinfo {year} {1942})}\BibitemShut {NoStop}%
\bibitem [{\citenamefont {Freise}(1952)}]{Freise}%
  \BibitemOpen
  \bibfield  {author} {\bibinfo {author} {\bibfnamefont {V.}~\bibnamefont
  {Freise}},\ }\bibfield  {title} {\bibinfo {title} {Zur theorie der diffusen
  doppelschicht},\ }\href
  {https://doi.org/https://doi.org/10.1002/bbpc.19520560826} {\bibfield
  {journal} {\bibinfo  {journal} {Zeitschrift für Elektrochemie, Berichte der
  Bunsengesellschaft für physikalische Chemie}\ }\textbf {\bibinfo {volume}
  {56}},\ \bibinfo {pages} {822} (\bibinfo {year} {1952})}\BibitemShut
  {NoStop}%
\bibitem [{\citenamefont {Kralj-Iglic}\ and\ \citenamefont
  {Iglic}(1996)}]{Kralj}%
  \BibitemOpen
  \bibfield  {author} {\bibinfo {author} {\bibfnamefont {V.}~\bibnamefont
  {Kralj-Iglic}}\ and\ \bibinfo {author} {\bibfnamefont {A.}~\bibnamefont
  {Iglic}},\ }\bibfield  {title} {\bibinfo {title} {A simple statistical
  mechanical approach to the free energy of the electric double layer including
  the excluded volume effect},\ }\href {https://doi.org/10.1051/jp2:1996193}
  {\bibfield  {journal} {\bibinfo  {journal} {J. Phys. II France}\ }\textbf
  {\bibinfo {volume} {6}},\ \bibinfo {pages} {477} (\bibinfo {year}
  {1996})}\BibitemShut {NoStop}%
\bibitem [{\citenamefont {Borukhov}\ \emph {et~al.}(1997)\citenamefont
  {Borukhov}, \citenamefont {Andelman},\ and\ \citenamefont
  {Orland}}]{Borukhov}%
  \BibitemOpen
  \bibfield  {author} {\bibinfo {author} {\bibfnamefont {I.}~\bibnamefont
  {Borukhov}}, \bibinfo {author} {\bibfnamefont {D.}~\bibnamefont {Andelman}},\
  and\ \bibinfo {author} {\bibfnamefont {H.}~\bibnamefont {Orland}},\
  }\bibfield  {title} {\bibinfo {title} {Steric effects in electrolytes: A
  modified {Poisson-Boltzmann} equation},\ }\href
  {https://doi.org/10.1103/PhysRevLett.79.435} {\bibfield  {journal} {\bibinfo
  {journal} {Phys. Rev. Lett.}\ }\textbf {\bibinfo {volume} {79}},\ \bibinfo
  {pages} {435} (\bibinfo {year} {1997})}\BibitemShut {NoStop}%
\bibitem [{\citenamefont {Kornyshev}(2007)}]{Kornyshev2007}%
  \BibitemOpen
  \bibfield  {author} {\bibinfo {author} {\bibfnamefont {A.~A.}\ \bibnamefont
  {Kornyshev}},\ }\bibfield  {title} {\bibinfo {title} {Double-layer in ionic
  liquids:{\thinspace} paradigm change?},\ }\href
  {https://doi.org/10.1021/jp067857o} {\bibfield  {journal} {\bibinfo
  {journal} {The Journal of Physical Chemistry B}\ }\textbf {\bibinfo {volume}
  {111}},\ \bibinfo {pages} {5545} (\bibinfo {year} {2007})}\BibitemShut
  {NoStop}%
\bibitem [{\citenamefont {Frydel}\ and\ \citenamefont {Levin}(2012)}]{FrLe12}%
  \BibitemOpen
  \bibfield  {author} {\bibinfo {author} {\bibfnamefont {D.}~\bibnamefont
  {Frydel}}\ and\ \bibinfo {author} {\bibfnamefont {Y.}~\bibnamefont {Levin}},\
  }\bibfield  {title} {\bibinfo {title} {A close look into the excluded volume
  effects within a double layer},\ }\href {https://doi.org/10.1063/1.4761938}
  {\bibfield  {journal} {\bibinfo  {journal} {The Journal of Chemical Physics}\
  }\textbf {\bibinfo {volume} {137}},\ \bibinfo {pages} {164703} (\bibinfo
  {year} {2012})}\BibitemShut {NoStop}%
\bibitem [{\citenamefont {Huang}\ \emph {et~al.}(2016)\citenamefont {Huang},
  \citenamefont {Liu}, \citenamefont {Li},\ and\ \citenamefont
  {Yan}}]{Huang2016}%
  \BibitemOpen
  \bibfield  {author} {\bibinfo {author} {\bibfnamefont {Y.}~\bibnamefont
  {Huang}}, \bibinfo {author} {\bibfnamefont {X.}~\bibnamefont {Liu}}, \bibinfo
  {author} {\bibfnamefont {S.}~\bibnamefont {Li}},\ and\ \bibinfo {author}
  {\bibfnamefont {T.}~\bibnamefont {Yan}},\ }\bibfield  {title} {\bibinfo
  {title} {Development of mean-field electrical double layer theory},\ }\href
  {https://doi.org/10.1088%2F1674-1056%2F25%2F1%2F016801} {\bibfield  {journal}
  {\bibinfo  {journal} {Chinese Physics B}\ }\textbf {\bibinfo {volume} {25}},\
  \bibinfo {pages} {016801} (\bibinfo {year} {2016})}\BibitemShut {NoStop}%
\bibitem [{\citenamefont {Guldbrand}\ \emph {et~al.}(1984)\citenamefont
  {Guldbrand}, \citenamefont {Jönsson}, \citenamefont {Wennerström},\ and\
  \citenamefont {Linse}}]{Guldbrand1984}%
  \BibitemOpen
  \bibfield  {author} {\bibinfo {author} {\bibfnamefont {L.}~\bibnamefont
  {Guldbrand}}, \bibinfo {author} {\bibfnamefont {B.}~\bibnamefont {Jönsson}},
  \bibinfo {author} {\bibfnamefont {H.}~\bibnamefont {Wennerström}},\ and\
  \bibinfo {author} {\bibfnamefont {P.}~\bibnamefont {Linse}},\ }\bibfield
  {title} {\bibinfo {title} {Electrical double layer forces. a {Monte Carlo}
  study},\ }\href {https://doi.org/10.1063/1.446912} {\bibfield  {journal}
  {\bibinfo  {journal} {The Journal of Chemical Physics}\ }\textbf {\bibinfo
  {volume} {80}},\ \bibinfo {pages} {2221} (\bibinfo {year}
  {1984})}\BibitemShut {NoStop}%
\bibitem [{\citenamefont {Allahyarov}\ \emph {et~al.}(1999)\citenamefont
  {Allahyarov}, \citenamefont {D'Amico},\ and\ \citenamefont
  {L\"owen}}]{Allahyarov1999}%
  \BibitemOpen
  \bibfield  {author} {\bibinfo {author} {\bibfnamefont {E.}~\bibnamefont
  {Allahyarov}}, \bibinfo {author} {\bibfnamefont {I.}~\bibnamefont
  {D'Amico}},\ and\ \bibinfo {author} {\bibfnamefont {H.}~\bibnamefont
  {L\"owen}},\ }\bibfield  {title} {\bibinfo {title} {Effect of geometrical
  confinement on the interaction between charged colloidal suspensions},\
  }\href {https://doi.org/https://doi.org/10.1103/PhysRevE.60.3199} {\bibfield
  {journal} {\bibinfo  {journal} {Phys. Rev. E}\ }\textbf {\bibinfo {volume}
  {60}},\ \bibinfo {pages} {3199} (\bibinfo {year} {1999})}\BibitemShut
  {NoStop}%
\bibitem [{\citenamefont {Moreira}\ and\ \citenamefont
  {Netz}(2002)}]{Moreira2002MC}%
  \BibitemOpen
  \bibfield  {author} {\bibinfo {author} {\bibfnamefont {A.~G.}\ \bibnamefont
  {Moreira}}\ and\ \bibinfo {author} {\bibfnamefont {R.~R.}\ \bibnamefont
  {Netz}},\ }\bibfield  {title} {\bibinfo {title} {Simulations of counterions
  at charged plates},\ }\href {https://doi.org/10.1140/epje/i2001-10091-9}
  {\bibfield  {journal} {\bibinfo  {journal} {The European Physical Journal E}\
  }\textbf {\bibinfo {volume} {8}},\ \bibinfo {pages} {33} (\bibinfo {year}
  {2002})}\BibitemShut {NoStop}%
\bibitem [{\citenamefont {Moreira}\ and\ \citenamefont
  {Netz}(2001)}]{MoreiraNetz2001}%
  \BibitemOpen
  \bibfield  {author} {\bibinfo {author} {\bibfnamefont {A.~G.}\ \bibnamefont
  {Moreira}}\ and\ \bibinfo {author} {\bibfnamefont {R.~R.}\ \bibnamefont
  {Netz}},\ }\bibfield  {title} {\bibinfo {title} {Field-theoretic approaches
  to classical charged systems},\ }in\ \href
  {https://doi.org/10.1007/978-94-010-0577-7_11} {\emph {\bibinfo {booktitle}
  {Electrostatic Effects in Soft Matter and Biophysics}}},\ \bibinfo {editor}
  {edited by\ \bibinfo {editor} {\bibfnamefont {C.}~\bibnamefont {Holm}},
  \bibinfo {editor} {\bibfnamefont {P.}~\bibnamefont {K{\'e}kicheff}},\ and\
  \bibinfo {editor} {\bibfnamefont {R.}~\bibnamefont {Podgornik}}}\ (\bibinfo
  {publisher} {Kluwer Academic, Dordrecht},\ \bibinfo {address} {Dordrecht},\
  \bibinfo {year} {2001})\ pp.\ \bibinfo {pages} {367--408}\BibitemShut
  {NoStop}%
\bibitem [{\citenamefont {Boroudjerdi}\ \emph {et~al.}(2005)\citenamefont
  {Boroudjerdi}, \citenamefont {Kim}, \citenamefont {Naji}, \citenamefont
  {Netz}, \citenamefont {Schlagberger},\ and\ \citenamefont
  {Serr}}]{Boroudjerdi2005}%
  \BibitemOpen
  \bibfield  {author} {\bibinfo {author} {\bibfnamefont {H.}~\bibnamefont
  {Boroudjerdi}}, \bibinfo {author} {\bibfnamefont {Y.-W.}\ \bibnamefont
  {Kim}}, \bibinfo {author} {\bibfnamefont {A.}~\bibnamefont {Naji}}, \bibinfo
  {author} {\bibfnamefont {R.~R.}\ \bibnamefont {Netz}}, \bibinfo {author}
  {\bibfnamefont {X.}~\bibnamefont {Schlagberger}},\ and\ \bibinfo {author}
  {\bibfnamefont {A.}~\bibnamefont {Serr}},\ }\bibfield  {title} {\bibinfo
  {title} {Statics and dynamics of strongly charged soft matter},\ }\href
  {https://doi.org/https://doi.org/10.1016/j.physrep.2005.06.006} {\bibfield
  {journal} {\bibinfo  {journal} {Physics Reports}\ }\textbf {\bibinfo {volume}
  {416}},\ \bibinfo {pages} {129 } (\bibinfo {year} {2005})}\BibitemShut
  {NoStop}%
\bibitem [{\citenamefont {Naji}\ \emph {et~al.}(2005)\citenamefont {Naji},
  \citenamefont {Jungblut}, \citenamefont {Moreira},\ and\ \citenamefont
  {Netz}}]{Naji2005}%
  \BibitemOpen
  \bibfield  {author} {\bibinfo {author} {\bibfnamefont {A.}~\bibnamefont
  {Naji}}, \bibinfo {author} {\bibfnamefont {S.}~\bibnamefont {Jungblut}},
  \bibinfo {author} {\bibfnamefont {A.~G.}\ \bibnamefont {Moreira}},\ and\
  \bibinfo {author} {\bibfnamefont {R.~R.}\ \bibnamefont {Netz}},\ }\bibfield
  {title} {\bibinfo {title} {Electrostatic interactions in strongly coupled
  soft matter},\ }\href
  {https://doi.org/https://doi.org/10.1016/j.physa.2004.12.029} {\bibfield
  {journal} {\bibinfo  {journal} {Physica A: Statistical Mechanics and its
  Applications}\ }\textbf {\bibinfo {volume} {352}},\ \bibinfo {pages} {131 }
  (\bibinfo {year} {2005})},\ \bibinfo {note} {physics Applied to Biological
  Systems}\BibitemShut {NoStop}%
\bibitem [{\citenamefont {{{\v{S}}amaj}}\ and\ \citenamefont
  {Trizac}(2011)}]{SaTr11PRL}%
  \BibitemOpen
  \bibfield  {author} {\bibinfo {author} {\bibfnamefont {L.}~\bibnamefont
  {{{\v{S}}amaj}}}\ and\ \bibinfo {author} {\bibfnamefont {E.}~\bibnamefont
  {Trizac}},\ }\bibfield  {title} {\bibinfo {title} {Counterions at highly
  charged interfaces: From one plate to like-charge attraction},\ }\href
  {https://doi.org/https://doi.org/10.1103/PhysRevLett.106.078301} {\bibfield
  {journal} {\bibinfo  {journal} {Phys. Rev. Lett.}\ }\textbf {\bibinfo
  {volume} {106}},\ \bibinfo {pages} {078301} (\bibinfo {year}
  {2011})}\BibitemShut {NoStop}%
\bibitem [{\citenamefont {{\v{S}}amaj}\ \emph {et~al.}(2018)\citenamefont
  {{\v{S}}amaj}, \citenamefont {Trulsson},\ and\ \citenamefont
  {Trizac}}]{SaTT18}%
  \BibitemOpen
  \bibfield  {author} {\bibinfo {author} {\bibfnamefont {L.}~\bibnamefont
  {{\v{S}}amaj}}, \bibinfo {author} {\bibfnamefont {M.}~\bibnamefont
  {Trulsson}},\ and\ \bibinfo {author} {\bibfnamefont {E.}~\bibnamefont
  {Trizac}},\ }\bibfield  {title} {\bibinfo {title} {Strong-coupling theory of
  counterions between symmetrically charged walls: From crystal to fluid
  phases},\ }\href {https://doi.org/https://doi.org/10.1039/c8sm00571k}
  {\bibfield  {journal} {\bibinfo  {journal} {Soft Matter}\ }\textbf {\bibinfo
  {volume} {14}},\ \bibinfo {pages} {4040} (\bibinfo {year}
  {2018})}\BibitemShut {NoStop}%
\bibitem [{\citenamefont {Golestanian}(2000)}]{Golestanian2000}%
  \BibitemOpen
  \bibfield  {author} {\bibinfo {author} {\bibfnamefont {R.}~\bibnamefont
  {Golestanian}},\ }\bibfield  {title} {\bibinfo {title} {Dynamics of
  counterion condensation},\ }\href {https://doi.org/10.1209/epl/i2000-00402-x}
  {\bibfield  {journal} {\bibinfo  {journal} {Europhysics Letters ({EPL})}\
  }\textbf {\bibinfo {volume} {52}},\ \bibinfo {pages} {47} (\bibinfo {year}
  {2000})}\BibitemShut {NoStop}%
\bibitem [{\citenamefont {Bazant}\ \emph {et~al.}(2004)\citenamefont {Bazant},
  \citenamefont {Thornton},\ and\ \citenamefont {Ajdari}}]{Bazant2004}%
  \BibitemOpen
  \bibfield  {author} {\bibinfo {author} {\bibfnamefont {M.~Z.}\ \bibnamefont
  {Bazant}}, \bibinfo {author} {\bibfnamefont {K.}~\bibnamefont {Thornton}},\
  and\ \bibinfo {author} {\bibfnamefont {A.}~\bibnamefont {Ajdari}},\
  }\bibfield  {title} {\bibinfo {title} {Diffuse-charge dynamics in
  electrochemical systems},\ }\href
  {https://doi.org/https://doi.org/10.1103/PhysRevE.70.021506} {\bibfield
  {journal} {\bibinfo  {journal} {Phys. Rev. E}\ }\textbf {\bibinfo {volume}
  {70}},\ \bibinfo {pages} {021506} (\bibinfo {year} {2004})}\BibitemShut
  {NoStop}%
\bibitem [{\citenamefont {Morrow}\ \emph {et~al.}(2006)\citenamefont {Morrow},
  \citenamefont {McKenzie},\ and\ \citenamefont {Bilek}}]{Morrow2006}%
  \BibitemOpen
  \bibfield  {author} {\bibinfo {author} {\bibfnamefont {R.}~\bibnamefont
  {Morrow}}, \bibinfo {author} {\bibfnamefont {D.~R.}\ \bibnamefont
  {McKenzie}},\ and\ \bibinfo {author} {\bibfnamefont {M.~M.~M.}\ \bibnamefont
  {Bilek}},\ }\bibfield  {title} {\bibinfo {title} {The time-dependent
  development of electric double-layers in saline solutions},\ }\href
  {https://doi.org/https://doi.org/10.1088/0022-3727/39/5/007} {\bibfield
  {journal} {\bibinfo  {journal} {Journal of Physics D: Applied Physics}\
  }\textbf {\bibinfo {volume} {39}},\ \bibinfo {pages} {937} (\bibinfo {year}
  {2006})}\BibitemShut {NoStop}%
\bibitem [{\citenamefont {Alexe‐Ionescu}\ \emph {et~al.}(2006)\citenamefont
  {Alexe‐Ionescu}, \citenamefont {Barbero}, \citenamefont {Freire},\ and\
  \citenamefont {Scalerandi}}]{AlexeIonescu2006}%
  \BibitemOpen
  \bibfield  {author} {\bibinfo {author} {\bibfnamefont {A.~L.}\ \bibnamefont
  {Alexe‐Ionescu}}, \bibinfo {author} {\bibfnamefont {G.}~\bibnamefont
  {Barbero}}, \bibinfo {author} {\bibfnamefont {F.~C.~M.}\ \bibnamefont
  {Freire}},\ and\ \bibinfo {author} {\bibfnamefont {M.}~\bibnamefont
  {Scalerandi}},\ }\bibfield  {title} {\bibinfo {title} {Transient effects in
  electrolytic cells submitted to an external electric field},\ }\href
  {https://doi.org/10.1080/02678290601010972} {\bibfield  {journal} {\bibinfo
  {journal} {Liquid Crystals}\ }\textbf {\bibinfo {volume} {33}},\ \bibinfo
  {pages} {1177} (\bibinfo {year} {2006})}\BibitemShut {NoStop}%
\bibitem [{\citenamefont {Lim}\ \emph {et~al.}(2007)\citenamefont {Lim},
  \citenamefont {Whitcomb}, \citenamefont {Boyd},\ and\ \citenamefont
  {Varghese}}]{Lim2007}%
  \BibitemOpen
  \bibfield  {author} {\bibinfo {author} {\bibfnamefont {J.}~\bibnamefont
  {Lim}}, \bibinfo {author} {\bibfnamefont {J.}~\bibnamefont {Whitcomb}},
  \bibinfo {author} {\bibfnamefont {J.}~\bibnamefont {Boyd}},\ and\ \bibinfo
  {author} {\bibfnamefont {J.}~\bibnamefont {Varghese}},\ }\bibfield  {title}
  {\bibinfo {title} {Transient finite element analysis of electric double layer
  using {Nernst–Planck–Poisson} equations with a modified {Stern} layer},\
  }\href {https://doi.org/https://doi.org/10.1016/j.jcis.2006.08.049}
  {\bibfield  {journal} {\bibinfo  {journal} {Journal of Colloid and Interface
  Science}\ }\textbf {\bibinfo {volume} {305}},\ \bibinfo {pages} {159 }
  (\bibinfo {year} {2007})}\BibitemShut {NoStop}%
\bibitem [{\citenamefont {H\o{}jgaard~Olesen}\ \emph
  {et~al.}(2010)\citenamefont {H\o{}jgaard~Olesen}, \citenamefont {Bazant},\
  and\ \citenamefont {Bruus}}]{Hojgaard2010}%
  \BibitemOpen
  \bibfield  {author} {\bibinfo {author} {\bibfnamefont {L.}~\bibnamefont
  {H\o{}jgaard~Olesen}}, \bibinfo {author} {\bibfnamefont {M.~Z.}\ \bibnamefont
  {Bazant}},\ and\ \bibinfo {author} {\bibfnamefont {H.}~\bibnamefont
  {Bruus}},\ }\bibfield  {title} {\bibinfo {title} {Strongly nonlinear dynamics
  of electrolytes in large ac voltages},\ }\href
  {https://doi.org/https://doi.org/10.1103/PhysRevE.82.011501} {\bibfield
  {journal} {\bibinfo  {journal} {Phys. Rev. E}\ }\textbf {\bibinfo {volume}
  {82}},\ \bibinfo {pages} {011501} (\bibinfo {year} {2010})}\BibitemShut
  {NoStop}%
\bibitem [{\citenamefont {Janssen}\ and\ \citenamefont
  {Bier}(2018)}]{JanssenBier2018}%
  \BibitemOpen
  \bibfield  {author} {\bibinfo {author} {\bibfnamefont {M.}~\bibnamefont
  {Janssen}}\ and\ \bibinfo {author} {\bibfnamefont {M.}~\bibnamefont {Bier}},\
  }\bibfield  {title} {\bibinfo {title} {Transient dynamics of electric
  double-layer capacitors: Exact expressions within the {Debye-Falkenhagen}
  approximation},\ }\href
  {https://doi.org/https://doi.org/10.1103/PhysRevE.97.052616} {\bibfield
  {journal} {\bibinfo  {journal} {Phys. Rev. E}\ }\textbf {\bibinfo {volume}
  {97}},\ \bibinfo {pages} {052616} (\bibinfo {year} {2018})}\BibitemShut
  {NoStop}%
\bibitem [{\citenamefont {Palaia}\ \emph {et~al.}(2020)\citenamefont {Palaia},
  \citenamefont {Telles}, \citenamefont {dos Santos},\ and\ \citenamefont
  {Trizac}}]{Palaia2020}%
  \BibitemOpen
  \bibfield  {author} {\bibinfo {author} {\bibfnamefont {I.}~\bibnamefont
  {Palaia}}, \bibinfo {author} {\bibfnamefont {I.~M.}\ \bibnamefont {Telles}},
  \bibinfo {author} {\bibfnamefont {A.~P.}\ \bibnamefont {dos Santos}},\ and\
  \bibinfo {author} {\bibfnamefont {E.}~\bibnamefont {Trizac}},\ }\bibfield
  {title} {\bibinfo {title} {Electroosmosis as a probe for electrostatic
  correlations},\ }\href {https://doi.org/10.1039/D0SM01523G} {\bibfield
  {journal} {\bibinfo  {journal} {Soft Matter}\ }\textbf {\bibinfo {volume}
  {16}},\ \bibinfo {pages} {10688} (\bibinfo {year} {2020})}\BibitemShut
  {NoStop}%
\bibitem [{\citenamefont {Telles}\ and\ \citenamefont {dos
  Santos}(2021)}]{telles2021}%
  \BibitemOpen
  \bibfield  {author} {\bibinfo {author} {\bibfnamefont {I.~M.}\ \bibnamefont
  {Telles}}\ and\ \bibinfo {author} {\bibfnamefont {A.~P.}\ \bibnamefont {dos
  Santos}},\ }\bibfield  {title} {\bibinfo {title} {Electroosmotic flow grows
  with electrostatic coupling in confining charged dielectric surfaces},\
  }\href {https://doi.org/10.1021/acs.langmuir.0c03116} {\bibfield  {journal}
  {\bibinfo  {journal} {Langmuir}\ }\textbf {\bibinfo {volume} {37}},\ \bibinfo
  {pages} {2104} (\bibinfo {year} {2021})}\BibitemShut {NoStop}%
\bibitem [{\citenamefont {Bazant}\ \emph {et~al.}(2009)\citenamefont {Bazant},
  \citenamefont {Kilic}, \citenamefont {Storey},\ and\ \citenamefont
  {Ajdari}}]{Bazant2009}%
  \BibitemOpen
  \bibfield  {author} {\bibinfo {author} {\bibfnamefont {M.~Z.}\ \bibnamefont
  {Bazant}}, \bibinfo {author} {\bibfnamefont {M.~S.}\ \bibnamefont {Kilic}},
  \bibinfo {author} {\bibfnamefont {B.~D.}\ \bibnamefont {Storey}},\ and\
  \bibinfo {author} {\bibfnamefont {A.}~\bibnamefont {Ajdari}},\ }\bibfield
  {title} {\bibinfo {title} {Towards an understanding of induced-charge
  electrokinetics at large applied voltages in concentrated solutions},\ }\href
  {https://doi.org/https://doi.org/10.1016/j.cis.2009.10.001} {\bibfield
  {journal} {\bibinfo  {journal} {Advances in Colloid and Interface Science}\
  }\textbf {\bibinfo {volume} {152}},\ \bibinfo {pages} {48 } (\bibinfo {year}
  {2009})}\BibitemShut {NoStop}%
\bibitem [{\citenamefont {Storey}\ and\ \citenamefont
  {Bazant}(2012)}]{Storey2012}%
  \BibitemOpen
  \bibfield  {author} {\bibinfo {author} {\bibfnamefont {B.~D.}\ \bibnamefont
  {Storey}}\ and\ \bibinfo {author} {\bibfnamefont {M.~Z.}\ \bibnamefont
  {Bazant}},\ }\bibfield  {title} {\bibinfo {title} {Effects of electrostatic
  correlations on electrokinetic phenomena},\ }\href
  {https://doi.org/10.1103/PhysRevE.86.056303} {\bibfield  {journal} {\bibinfo
  {journal} {Phys. Rev. E}\ }\textbf {\bibinfo {volume} {86}},\ \bibinfo
  {pages} {056303} (\bibinfo {year} {2012})}\BibitemShut {NoStop}%
\bibitem [{\citenamefont {Dean}\ \emph {et~al.}(1998)\citenamefont {Dean},
  \citenamefont {Horgan},\ and\ \citenamefont {Sentenac}}]{Dean1998}%
  \BibitemOpen
  \bibfield  {author} {\bibinfo {author} {\bibfnamefont {D.~S.}\ \bibnamefont
  {Dean}}, \bibinfo {author} {\bibfnamefont {R.~R.}\ \bibnamefont {Horgan}},\
  and\ \bibinfo {author} {\bibfnamefont {D.}~\bibnamefont {Sentenac}},\
  }\bibfield  {title} {\bibinfo {title} {Boundary effects in the
  one-dimensional {Coulomb} gas},\ }\href
  {https://doi.org/10.1023/A:1023241407140} {\bibfield  {journal} {\bibinfo
  {journal} {Journal of Statistical Physics}\ }\textbf {\bibinfo {volume}
  {90}},\ \bibinfo {pages} {899} (\bibinfo {year} {1998})}\BibitemShut
  {NoStop}%
\bibitem [{\citenamefont {Dean}\ \emph {et~al.}(2009)\citenamefont {Dean},
  \citenamefont {Horgan},\ and\ \citenamefont {Podgornik}}]{dean}%
  \BibitemOpen
  \bibfield  {author} {\bibinfo {author} {\bibfnamefont {D.~S.}\ \bibnamefont
  {Dean}}, \bibinfo {author} {\bibfnamefont {R.}~\bibnamefont {Horgan}},\ and\
  \bibinfo {author} {\bibfnamefont {R.}~\bibnamefont {Podgornik}},\ }\bibfield
  {title} {\bibinfo {title} {One-dimensional counterion gas between charged
  surfaces: Exact results compared with weak- and strong-coupling analyses},\
  }\href {https://doi.org/10.1063/1.3078492} {\bibfield  {journal} {\bibinfo
  {journal} {The Journal of Chemical Physics}\ }\textbf {\bibinfo {volume}
  {130}},\ \bibinfo {pages} {094504} (\bibinfo {year} {2009})}\BibitemShut
  {NoStop}%
\bibitem [{\citenamefont {T\'ellez}\ and\ \citenamefont
  {Trizac}(2015)}]{TellezTrizac2015}%
  \BibitemOpen
  \bibfield  {author} {\bibinfo {author} {\bibfnamefont {G.}~\bibnamefont
  {T\'ellez}}\ and\ \bibinfo {author} {\bibfnamefont {E.}~\bibnamefont
  {Trizac}},\ }\bibfield  {title} {\bibinfo {title} {Screening like charges in
  one-dimensional {Coulomb} systems: Exact results},\ }\href
  {https://doi.org/10.1103/PhysRevE.92.042134} {\bibfield  {journal} {\bibinfo
  {journal} {Phys. Rev. E}\ }\textbf {\bibinfo {volume} {92}},\ \bibinfo
  {pages} {042134} (\bibinfo {year} {2015})}\BibitemShut {NoStop}%
\bibitem [{\citenamefont {Varela}\ \emph {et~al.}(2017)\citenamefont {Varela},
  \citenamefont {T\'ellez},\ and\ \citenamefont {Trizac}}]{vtt2016}%
  \BibitemOpen
  \bibfield  {author} {\bibinfo {author} {\bibfnamefont {L.}~\bibnamefont
  {Varela}}, \bibinfo {author} {\bibfnamefont {G.}~\bibnamefont {T\'ellez}},\
  and\ \bibinfo {author} {\bibfnamefont {E.}~\bibnamefont {Trizac}},\
  }\bibfield  {title} {\bibinfo {title} {Configurational and energy landscape
  in one-dimensional {Coulomb} systems},\ }\href
  {https://doi.org/https://doi.org/10.1103/PhysRevE.95.022112} {\bibfield
  {journal} {\bibinfo  {journal} {Phys. Rev. E}\ }\textbf {\bibinfo {volume}
  {95}},\ \bibinfo {pages} {022112} (\bibinfo {year} {2017})}\BibitemShut
  {NoStop}%
\bibitem [{\citenamefont {Trizac}\ and\ \citenamefont
  {T{\'{e}}llez}(2018)}]{Trizac2018}%
  \BibitemOpen
  \bibfield  {author} {\bibinfo {author} {\bibfnamefont {E.}~\bibnamefont
  {Trizac}}\ and\ \bibinfo {author} {\bibfnamefont {G.}~\bibnamefont
  {T{\'{e}}llez}},\ }\bibfield  {title} {\bibinfo {title} {Like-charge
  attraction in a one-dimensional setting: the importance of being odd},\
  }\href {https://doi.org/10.1088/1361-6404/aa9e80} {\bibfield  {journal}
  {\bibinfo  {journal} {European Journal of Physics}\ }\textbf {\bibinfo
  {volume} {39}},\ \bibinfo {pages} {025102} (\bibinfo {year}
  {2018})}\BibitemShut {NoStop}%
\bibitem [{\citenamefont {Netz}(2001)}]{Netz2001}%
  \BibitemOpen
  \bibfield  {author} {\bibinfo {author} {\bibfnamefont {R.~R.}\ \bibnamefont
  {Netz}},\ }\bibfield  {title} {\bibinfo {title} {Electrostatistics of
  counter-ions at and between planar charged walls: From {Poisson-Boltzmann} to
  the strong-coupling theory},\ }\href {https://doi.org/epje/v5/p557(e01021)}
  {\bibfield  {journal} {\bibinfo  {journal} {Eur. Phys. J. E}\ }\textbf
  {\bibinfo {volume} {5}},\ \bibinfo {pages} {557} (\bibinfo {year}
  {2001})}\BibitemShut {NoStop}%
\bibitem [{\citenamefont {Gennes}(2005)}]{Gennes2005}%
  \BibitemOpen
  \bibfield  {author} {\bibinfo {author} {\bibfnamefont {P.~G.~d.}\
  \bibnamefont {Gennes}},\ }\bibfield  {title} {\bibinfo {title} {Brownian
  motion with dry friction},\ }\href
  {https://doi.org/10.1007/s10955-005-4650-4} {\bibfield  {journal} {\bibinfo
  {journal} {Journal of Statistical Physics}\ }\textbf {\bibinfo {volume}
  {119}},\ \bibinfo {pages} {953} (\bibinfo {year} {2005})}\BibitemShut
  {NoStop}%
\bibitem [{\citenamefont {Touchette}\ \emph {et~al.}(2010)\citenamefont
  {Touchette}, \citenamefont {d.~Straeten},\ and\ \citenamefont
  {Just}}]{dryfriction}%
  \BibitemOpen
  \bibfield  {author} {\bibinfo {author} {\bibfnamefont {H.}~\bibnamefont
  {Touchette}}, \bibinfo {author} {\bibfnamefont {E.~V.}\ \bibnamefont
  {d.~Straeten}},\ and\ \bibinfo {author} {\bibfnamefont {W.}~\bibnamefont
  {Just}},\ }\bibfield  {title} {\bibinfo {title} {Brownian motion with dry
  friction: {Fokker–Planck} approach},\ }\href
  {http://dx.doi.org/10.1088/1751-8113/43/44/445002} {\bibfield  {journal}
  {\bibinfo  {journal} {Journal of Physics A: Mathematical and Theoretical}\
  }\textbf {\bibinfo {volume} {43}},\ \bibinfo {pages} {445002} (\bibinfo
  {year} {2010})}\BibitemShut {NoStop}%
\bibitem [{\citenamefont {Risken}(1984)}]{risken}%
  \BibitemOpen
  \bibfield  {author} {\bibinfo {author} {\bibfnamefont {H.}~\bibnamefont
  {Risken}},\ }\href {https://doi.org/10.1007/978-3-642-61544-3} {\emph
  {\bibinfo {title} {The {Fokker-Planck} Equation}}}\ (\bibinfo  {publisher}
  {Springer, Berlin, Heidelberg},\ \bibinfo {year} {1984})\BibitemShut
  {NoStop}%
\bibitem [{sch()}]{schrodinger}%
  \BibitemOpen
  \href@noop {} {\bibinfo {title} {{More precisely, the {F}okker-{P}lanck
  equation is formally equivalent to the time-independent {S}chr{\"o}dinger
  equation with a square well potential of depth $1/4$ with attractive deltas
  at the colloid positions, for an imaginary time $t_{S} = -i \hbar t $ and
  mass $m_{S} = \hbar^2/2{D}$ \cite{risken} }}}\BibitemShut {NoStop}%
\bibitem [{\citenamefont {Kullback}\ and\ \citenamefont
  {Leibler}(1951)}]{kullback1951}%
  \BibitemOpen
  \bibfield  {author} {\bibinfo {author} {\bibfnamefont {S.}~\bibnamefont
  {Kullback}}\ and\ \bibinfo {author} {\bibfnamefont {R.~A.}\ \bibnamefont
  {Leibler}},\ }\bibfield  {title} {\bibinfo {title} {On information and
  sufficiency},\ }\href {https://doi.org/10.1214/aoms/1177729694} {\bibfield
  {journal} {\bibinfo  {journal} {Ann. Math. Statist.}\ }\textbf {\bibinfo
  {volume} {22}},\ \bibinfo {pages} {79} (\bibinfo {year} {1951})}\BibitemShut
  {NoStop}%
\bibitem [{\citenamefont {Maruyama}(1955)}]{Maruyama}%
  \BibitemOpen
  \bibfield  {author} {\bibinfo {author} {\bibfnamefont {G.}~\bibnamefont
  {Maruyama}},\ }\bibfield  {title} {\bibinfo {title} {Continuous {Markov}
  processes and stochastic equations},\ }\href
  {https://doi.org/https://doi.org/10.1007/BF02846028} {\bibfield  {journal}
  {\bibinfo  {journal} {Rend. Circ. Mat. Palermo}\ }\textbf {\bibinfo {volume}
  {4}},\ \bibinfo {pages} {48} (\bibinfo {year} {1955})}\BibitemShut {NoStop}%
\bibitem [{\citenamefont {Milshtein}(1978)}]{Milstein}%
  \BibitemOpen
  \bibfield  {author} {\bibinfo {author} {\bibfnamefont {G.~N.}\ \bibnamefont
  {Milshtein}},\ }\bibfield  {title} {\bibinfo {title} {A method of second
  order accuracy integration of stochastic differential equations},\ }\href
  {https://doi.org/https://doi.org/10.1137/1123045} {\bibfield  {journal}
  {\bibinfo  {journal} {Theory Probab. Appl.}\ }\textbf {\bibinfo {volume}
  {23}},\ \bibinfo {pages} {396} (\bibinfo {year} {1978})}\BibitemShut
  {NoStop}%
\bibitem [{com()}]{comment10}%
  \BibitemOpen
  \href@noop {} {\bibinfo {title} {{For every polynomial function $g$ of the
  final counterion positions, and every small time step $\Delta t$, there
  exists a positive constant $C$ such that $|\langle
  g\rangle_\text{exact}-\langle g\rangle_\text{num}|\leq {C}\, \Delta
  t^{1.0}$}}}\BibitemShut {NoStop}%
\bibitem [{\citenamefont {Kloeden}\ and\ \citenamefont
  {Platen}(1992)}]{KloedenPlaten}%
  \BibitemOpen
  \bibfield  {author} {\bibinfo {author} {\bibfnamefont {P.~E.}\ \bibnamefont
  {Kloeden}}\ and\ \bibinfo {author} {\bibfnamefont {E.}~\bibnamefont
  {Platen}},\ }\href {https://www.springer.com/gp/book/9783540540625} {\emph
  {\bibinfo {title} {Numerical Solution of Stochastic Differential
  Equations}}}\ (\bibinfo  {publisher} {Springer-Verlag},\ \bibinfo {year}
  {1992})\BibitemShut {NoStop}%
\bibitem [{\citenamefont {Pal}\ and\ \citenamefont {Pitman}(2008)}]{pal2008}%
  \BibitemOpen
  \bibfield  {author} {\bibinfo {author} {\bibfnamefont {S.}~\bibnamefont
  {Pal}}\ and\ \bibinfo {author} {\bibfnamefont {J.}~\bibnamefont {Pitman}},\
  }\bibfield  {title} {\bibinfo {title} {One-dimensional {Brownian} particle
  systems with rank-dependent drifts},\ }\href
  {https://doi.org/10.1214/08-AAP516} {\bibfield  {journal} {\bibinfo
  {journal} {Ann. Appl. Probab.}\ }\textbf {\bibinfo {volume} {18}},\ \bibinfo
  {pages} {2179} (\bibinfo {year} {2008})}\BibitemShut {NoStop}%
\bibitem [{\citenamefont {Andelman}(2006)}]{andelman2006}%
  \BibitemOpen
  \bibfield  {author} {\bibinfo {author} {\bibfnamefont {D.}~\bibnamefont
  {Andelman}},\ }\bibinfo {title} {Introduction to electrostatics in soft and
  biological matter},\ in\ \href
  {https://books.google.com.co/books?id=hk9UbvyKZE0C} {\emph {\bibinfo
  {booktitle} {Soft Condensed Matter Physics in Molecular and Cell Biology}}},\
  \bibinfo {editor} {edited by\ \bibinfo {editor} {\bibfnamefont {W.~C.~K.}\
  \bibnamefont {Poon}}\ and\ \bibinfo {editor} {\bibfnamefont {D.}~\bibnamefont
  {Andelman}}}\ (\bibinfo  {publisher} {Taylor \& Francis, New York},\ \bibinfo
  {year} {2006})\ pp.\ \bibinfo {pages} {97--122}\BibitemShut {NoStop}%
\bibitem [{lav()}]{lavgforlargeL}%
  \BibitemOpen
  \href@noop {} {\bibinfo {title} {{Since Fig.~\ref{fig:DiffL} shows that
  $\widetilde{L}-\widetilde{l}_\text{avg}(N,\widetilde{L})$ is close to the
  limiting value at $\widetilde{L}=9$, we use
  $\widetilde{l}_\text{avg}(N,\widetilde{L})\approx
  \widetilde{l}_\text{avg}(N,9)+\widetilde{L}-9$ for
  $\widetilde{L}>9$}}}\BibitemShut {NoStop}%
\bibitem [{\citenamefont {Van~Kampen}(2007)}]{VanKampen}%
  \BibitemOpen
  \bibfield  {author} {\bibinfo {author} {\bibfnamefont {N.~G.}\ \bibnamefont
  {Van~Kampen}},\ }\href {https://doi.org/10.1016/B978-0-444-52965-7.X5000-4}
  {\emph {\bibinfo {title} {Stochastic Processes in Physics and Chemistry, 3rd
  ed.}}}\ (\bibinfo  {publisher} {Elsevier},\ \bibinfo {address} {Amsterdam},\
  \bibinfo {year} {2007})\BibitemShut {NoStop}%
\end{thebibliography}%
\end{document}